\documentclass[a4paper,titlepage,12pt]{article}
\pdfoutput=1
\usepackage{graphicx} 
\usepackage[utf8]{inputenc}
\usepackage{geometry}
\usepackage[english]{babel}
\usepackage{amsmath, amsfonts, graphicx}
\usepackage{mathtools}
\usepackage{graphicx}
\usepackage{float}
\usepackage{mciteplus}
\usepackage{bbold}
\usepackage{pstricks}
\usepackage{xcolor}
\usepackage[labelfont=bf,font={small}]{caption}
\usepackage{pgfplots}
\usepackage{empheq}
\usepackage{tikz}
\usetikzlibrary{positioning, arrows}
\usetikzlibrary{external}
\usepackage{footmisc}
\usepackage{multirow}
\usepackage{url}
\usepackage[breakable, theorems, skins]{tcolorbox}
\usepackage{enumerate}
\usepackage{physics}
\usepackage{bm}
\usepackage{braket}
\usepackage{epstopdf}
\usetikzlibrary{calc}
\usepackage{hyperref}
\usepackage{graphicx} 
\usepackage{tikz}     
\usepackage[OT2,T1]{fontenc}
\DeclareSymbolFont{cyrletters}{OT2}{wncyr}{m}{n}
\DeclareMathSymbol{\Sha}{\mathalpha}{cyrletters}{"58}
\date{June 2024}

\definecolor{darkblue}{RGB}{0,0,139}

\hypersetup{
    colorlinks,
    citecolor=blue,
    filecolor=blue,
    linkcolor=blue,
    urlcolor=blue
}

	\numberwithin{equation}{section}

\begin{document}

\begin{titlepage}

\setcounter{page}{1} \baselineskip=15.5pt \thispagestyle{empty}

\vfil

${}$

\begin{center}

\def\thefootnote{\fnsymbol{footnote}}
\begin{changemargin}{0.05cm}{0.05cm} 
\begin{center}
{\Large \bf A new perspective on dilaton gravity at finite cutoff}  
\end{center} 
\end{changemargin}

~\\[0.5cm]
{Luca Griguolo,\footnote{{\protect\path{luca.griguolo@unipr.it}}} Jacopo Papalini,\footnote{{\protect\path{jacopo.papalini@ugent.be}}} Lorenzo Russo,\footnote{{\protect\path{lorenzo.russo@unifi.it}}}}
Domenico Seminara,\footnote{{\protect\path{domenico.seminara@unifi.it}}}
\\[0.5cm]
{\normalsize { $^\ast$\sl Dipartimento SMFI, Università di Parma and INFN Gruppo Collegato di Parma,
\\[1.0mm]
Viale G.P.Usberti 7/A, 43100 Parma}}\\[3mm]
{\normalsize { $^\dagger$\sl Department of Physics and Astronomy
\\[1.0mm]
Ghent University, Krijgslaan, 281-S9, 9000 Gent, Belgium}}\\[3mm]
{\normalsize { $^\ddagger$ $^\S$\sl Dipartimento di Fisica, Università di Firenze and INFN Sezione di Firenze, \\[1.0mm]
via G. Sansone 1, 50019 Sesto Fiorentino, Italy}}

\end{center}

\begin{changemargin}{01cm}{1cm} 
{\small  \noindent 
\begin{center} 
\textbf{Abstract}
\end{center} 
The formulation of two-dimensional quantum gravity at finite cutoff remains an open problem.
We revisit this question in JT gravity from two perspectives: the closed-channel bulk path integral and the path integral over boundary curves. First, we study the radial evolution of a closed universe and derive the trumpet wavefunction as a transition amplitude between a geodesic boundary and a finite Dirichlet boundary.
Our analysis recovers the Hartle--Hawking wavefunction without imposing asymptotic boundary conditions, allowing the trumpet to be glued to a cap wavefunction to reconstruct the smooth disk. Second, we derive an exact Riccati equation for the extrinsic curvature of a finite-cutoff boundary curve in the Euclidean Poincar\'e disk.
A WKB expansion of this equation yields all perturbative corrections in the cutoff parameter and captures nonperturbative effects.
From this, we compute the quadratic boundary action and the one-loop partition function at finite cutoff, finding agreement with both the bulk approach and the expected one-loop effective action for the $T\bar{T}$ deformation of the Schwarzian theory. Extracting lessons from JT gravity, we then argue that similar relationships hold for general dilaton gravities with arbitrary potentials $V(\phi)$ and propose an exact expression for their finite cutoff partition functions. We finally investigate several signatures of UV completeness in these settings, introducing a canonical quantization approach within the finite cutoff framework.
}
\end{changemargin}
 \vspace{0.3cm}
\vfil

\end{titlepage}

\tableofcontents

\setcounter{footnote}{0}

\section{Introduction}

In order to make progress in understanding quantum gravity and holography, exactly solvable models provide powerful benchmarks. In this light, two-dimensional dilaton gravity models have attracted significant attention in recent years, with Jackiw-Teitelboim (JT) gravity \cite{Jackiw:1984je,Teitelboim:1983ux} emerging as the most renowned example. Its simplicity allows for exact computations while retaining a rich mathematical structure that captures essential features of black hole thermodynamics \cite{Mertens:2022irh}, connections to statistical models such as SYK \cite{Kitaev:2017awl,Maldacena:2016hyu,Saad:2018bqo}, and the physics of near-extremal black holes \cite{Iliesiu:2020qvm,Heydeman:2020hhw,Nayak:2018qej,Castro:2022cuo}. A cornerstone of its success is the correspondence between the bulk gravitational path integral and a boundary theory described by the Schwarzian action \cite{Almheiri:2014cka,Maldacena:2016upp,Engelsoy:2016xyb,Mertens:2018fds,Jensen:2016pah}, which governs the reparametrization modes of the asymptotic boundary. Remarkably, the disk partition function of JT gravity can be computed exactly in many different ways, matching the Schwarzian path-integral description \cite{Stanford:2017thb,Mertens:2017mtv,Kitaev:2018wpr,Harlow:2018tqv}. Among dilaton gravity models, JT gravity is not the only exactly solvable system; other remarkable examples include sinh dilaton gravity \cite{Mertens:2020hbs,Fan:2021bwt,Blommaert:2023wad,Belaey:2025kiu} and sine dilaton gravity \cite{Blommaert:2024ymv,Blommaert:2024whf,Blommaert:2025avl,Bossi:2024ffa}, the latter being particularly interesting due to its duality with the DSSYK model \cite{Berkooz:2018qkz,Berkooz:2018jqr,Berkooz:2024lgq}.\footnote{In recent years, the double-scaling limit of the SYK model and its connection with low dimensional holography has attracted considerable attention \cite{Berkooz:2022mfk,Lin:2022rbf,Goel:2023svz,Lin:2023trc,Narovlansky:2023lfz,Verlinde:2024zrh,Verlinde:2024znh,Almheiri:2024ayc,Okuyama:2023bch,Berkooz:2024evs,Belaey:2025ijg,vanderHeijden:2025zkr,Blommaert:2025eps,Schouten:2025tvn}.}

However, the dualities involving dilaton gravity models are often most clearly formulated in the limit where the holographic boundary is taken to asymptotic infinity. A conceptually important open question is therefore the formulation of dilaton gravity models with a finite bulk cutoff, that is, with a boundary placed at a finite distance in the bulk. Studying these setups is expected to provide a controlled way to improve the UV behavior of the bulk theories, naturally realizing integrable deformations of the boundary model. In the case of JT gravity, this scenario is concretely implemented by the $T\bar{T}$ deformation of the boundary quantum theory \cite{Smirnov:2016lqw}, which has been shown to correspond to moving the holographic boundary inward in the bulk spacetime \cite{McGough:2016lol,Gross:2019ach,Iliesiu:2020zld}.\footnote{Additional interesting related works in this context are \cite{Ebert:2022ehb, AliAhmad:2025kki, Hirano:2025cjg, Hirano:2025tkq, FarajiAstaneh:2024fig, Aguilar-Gutierrez:2024nst, Callebaut:2025thw, Blacker:2024rje, He:2025ppz, Ferrari:2024ndr, Tsolakidis:2024wut, Morone:2024ffm, Chaudhuri:2024yau, Ferrari:2024kpz}.}

A finite cutoff also allows one to define quasi-local gravitational observables and explore thermodynamics in a genuinely finite system. In this context, the connection to the SYK model \cite{Sachdev:1992fk,KitaevTalks} is particularly natural: the low-energy dynamics of SYK are governed by a Schwarzian action \cite{Kitaev:2017awl}, and since the full SYK model is characterized by a finite number of states and a bounded spectrum, the finite-cutoff JT setup is expected to mirror this structure, enabling controlled studies of entropy and UV energy bounds in a tractable holographic system.

Because of the above motivations, in this work we study the problem of dilaton gravity at finite cutoff, focusing in particular on JT gravity, in order to find universal rules for the more general case of arbitrary potentials $V(\phi)$ \cite{Grumiller:2002nm,Mertens:2022irh}.  More specifically, we investigate the finite-cutoff JT gravity framework from two complementary and self-contained approaches, which offer a new perspective on the topic.

Firstly we tackle the problem from the closed bulk channel: we compute the finite-cutoff "trumpet wavefunction" $\Psi_b(\phi, L)$ as a transition amplitude between a geodesic boundary of length $b$ and a finite Dirichlet boundary of length $L$ and fixed dilaton value $\phi$. We perform this path integral directly in the bulk formulation, by fixing a radial gauge and integrating over metric and dilaton fluctuations. This approach does not rely on boundary conditions at infinity nor on the Schwarzian boundary mode description. We then impose the Hartle-Hawking no-boundary condition by gluing this trumpet to a cap wavefunction, thereby reconstructing the smooth disk partition function at finite cutoff.

The second approach consists in the path integral over the wiggly boundaries: we provide a systematic analysis of the boundary path integral for a fluctuating curve at a finite distance in the Euclidean Poincaré disk. A key technical result is the derivation of an exact Riccati-type differential equation satisfied by the extrinsic curvature $\kappa$ of the boundary curve. This equation allows for a systematic WKB expansion in the cutoff parameter $\varepsilon$, generating all perturbative corrections to the boundary action and providing access to nonperturbative contributions. We use this framework to compute the one-loop partition function at finite cutoff from the boundary perspective.

A perfect agreement has been found between the disk partition functions obtained via the bulk trumpet gluing and the boundary one-loop calculation, as we substantiate in \eqref{summary}. Furthermore, a key feature of both derivations is that they naturally incorporate two saddle points (instantons) with a relative phase, which are essential for the nonperturbative completion of the theory. This structure aligns with the result obtained from the Borel resummation of the $T\bar{T}$-deformed partition function \cite{Griguolo:2021wgy}, placing it on a firmer gravitational foundation.

While deriving our results, we clarify several aspects of the finite-cutoff theory. From the bulk perspective, we elucidate the role of the Wheeler--DeWitt constraint in the bulk path integral and its relation to quantization, distinguishing in particular between the finite-cutoff trumpet wavefunction and eigenstates of the geodesic length operator \cite{Held:2024rmg,Blommaert:2025bgd}, and showing how different treatments of path-integral zero modes prepare distinct states in the gravitational Hilbert space.
From the boundary perspective, we verify that the quadratic action derived from the Riccati equation agrees with the \(T\bar{T}\)-deformed Schwarzian theory~\cite{Gross:2019ach}, thereby providing a robust consistency check between the bulk and boundary descriptions.

As anticipated, we then try to extend the study of finite-cutoff holography beyond JT gravity to a general class of two-dimensional dilaton gravities with arbitrary potential $V(\phi)$. We argue that imposing a finite radial cutoff in the bulk corresponds, in the dual boundary description, to a \(T\bar{T}\)-type deformation of the putative boundary quantum theory. 
To define the finite-cutoff partition function nonperturbatively, we propose a contour prescription that naturally incorporates the nonperturbative branch of the deformation, exploiting the lessons learned from JT gravity. This leads to an integral representation of the partition function for general $V(\phi)$, illustrated explicitly for Liouville/sinh-dilaton gravity \cite{Mertens:2020hbs,Fan:2021bwt,Blommaert:2023wad,Collier:2023cyw}. While the correspondence is most firmly established for potentials with JT-like asymptotics and that can be expressed via a gas of defects expansion \cite{Maxfield:2020ale,Turiaci:2020fjj,Kruthoff:2024gxc}, evidence suggests that our proposed exact expression may hold more generally, opening a path toward a unified understanding of finite-cutoff holography in two dimensions.

Finally, motivated by the discussions presented above, we explore possible signatures of UV completeness in the frameworks studied in this paper, highlighting close similarities with periodic dilaton gravity \cite{Blommaert:2024whf}. First, we propose how to incorporate higher-genus topologies in the finite-cutoff formulation of JT gravity \cite{Griguolo:2021wgy}, analyzing the energy spectrum of the theory and discussing the emergence of a natural UV cutoff. Second, we investigate the finite-cutoff boundary to boundary correlator of a matter probe and argue that it appears to resolve the usual UV singularities at coincident operator insertions. Third, we introduce a new framework for studying dilaton gravity at finite cutoff based on two-sided open channel quantization, and we speculate on how it might lead to a discretization of spacetime. A thorough analysis of this framework—which, to our knowledge, has never been applied to finite-cutoff holography before—will be presented in a future paper \cite{newpaper}.

The paper is organized as follows. In Section~\ref{tt}, we compute the trumpet wavefunction at finite cutoff from the bulk path integral. We show how this wavefunction leads to the disk partition function by imposing the no-boundary condition in Section~\ref{no}, and we introduce a variant of the path integral that prepares eigenstates of the geodesic length operator in Section~\ref{reduced}. In Section \ref{wiggly}, we analyze the wiggly boundary description, derive the Riccati equation for the extrinsic curvature, and study its classical solutions. In Section \ref{quantum}, we quantize the boundary theory and compute the one-loop partition function, demonstrating agreement with the bulk approach. A detailed comparison with earlier work is presented in Section \ref{onevs}, while in Section \ref{gene} we propose an extension of finite cut-off holography beyond the JT gravity paradigmatic case, considering a class of general dilaton potentials. The signatures of UV completeness in these models, as well as possible future directions, are instead discussed in Section \ref{toward}.
Several technical appendices complete the paper.

\section{Trumpet wavefunction at finite cutoff}\label{tt}

The closed-channel phase space of JT gravity, governed by the action
\begin{equation}\label{JT}
S_{\mathrm{JT}} = \frac{1}{2} \int_{\mathcal{M}} \mathrm{d}^2 x \sqrt{g} \, \Phi\left(R + 2\right) + \int_{\partial \mathcal{M}} \mathrm{d}\tau \sqrt{h} \, \Phi \kappa,
\end{equation}
can be characterized by two variables: the value of the dilaton $\phi$ and the size of the spatial  slice $h$. This reduction is often referred to as the minisuperspace approximation, where only the zero modes of the fields are considered and the universe wavefunction takes the form $\Psi = \Psi(\phi, h)$. Remarkably, in JT gravity and more generally in 2d dilaton gravity, this approximation turns out to be exact \cite{Iliesiu:2020zld}.
The goal of this section is to compute the trumpet wavefunction  $\Psi_{b}(\phi,L)$ in JT gravity at finite cutoff, interpreted as a transition amplitude from a geodesic boundary of fixed length $b$ to a Dirichlet boundary of fixed and finite length $L$, characterized by a fixed dilaton value $\phi$. Our approach is to evaluate this amplitude using the JT gravity path integral, without relying on the Schwarzian boundary mode description.\footnote{Similar calculations in different contexts were performed in \cite{Buchmuller:2024ksd,Fanaras:2021awm,Honda:2024hdr}.}

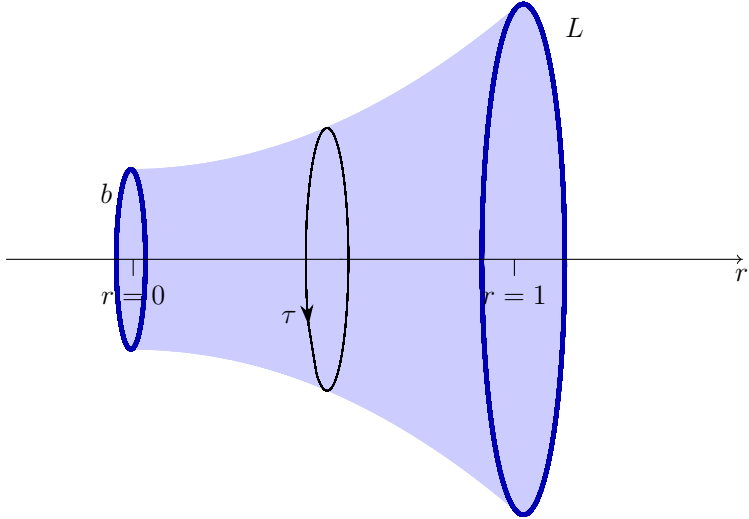
\begin{figure}

    \centering
    
    \begin{tikzpicture}[scale=0.88, transform shape]
        \node[inner sep=0] (img) {\includegraphics[width=0.52\textwidth, angle=270]{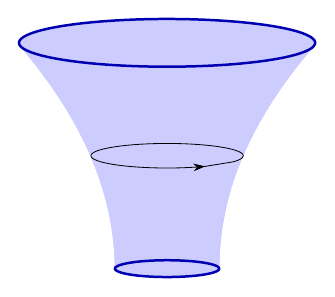}};

      \def\xL{0}      
\def\xR{6}      
\def\RLx{0.35}  
\def\RLy{1.30}  
\def\RRx{0.15}  
\def\RRy{0.55}  
\def\yaxis{0}

\draw[->] (\xL-5.0,\yaxis) -- (\xR,\yaxis) node[below] {$r$};
\draw (\xL - 3.1,\yaxis) -- ++(0,-0.25) node[below] {$r=0$};
\draw (\xR- 3.4,\yaxis) -- ++(0,-0.25) node[below] {$r=1$};

\draw[densely dotted] (\xL,0) -- (\xL,\yaxis);
\draw[densely dotted] (\xR,0) -- (\xR,\yaxis);

\begin{scope}[x={(img.south west)}, y={(img.north east)}]
    \node at (0.75,0.55) {$\tau$}; 
  \end{scope}

  \node at (-3.5, 1) {$b$};
  \node at (3.5, 3.5) {$L$};

    \end{tikzpicture}
    \caption{In this figure we represent the geometry of the spacetime \eqref{geom} that is a solution of the equation of motion and interpolates between the initial and final time circles, at $r=0$ and $r=1$, which are subject to the boundary conditions \eqref{bcc}.  The blue circle of length $b$ is a geodesic, the neck of the trumpet, while the other circle is the boundary of length $L$.}

    \label{trumpetonly}
\end{figure}

Specifically, we parametrize the class of metrics over which we perform the path integral as follows:
\begin{equation}\label{gauge}
\mathrm{d}s^2 = H(r) \, \mathrm{d}\tau^2 + \frac{\mathcal{N}^2}{H(r)} \, \mathrm{d}r^2
\end{equation}
where both $H(r)$ and the dilaton field $\Phi(r)$ depend only on the radial coordinate $r \in [0, 1]$ and $\tau$ is the Euclidean time coordinate. The function $H(r)^{\frac12 }$ encodes the radial profile of the proper length of the Euclidean time circle, i.e. the size of the universe as a function of the radial coordinate. We take $r = 0$ as the starting point of the radial evolution, where the geodesic boundary is located, and $r = 1$ as the endpoint, corresponding to the Dirichlet boundary at finite distance. 
See Figure \ref{trumpetonly} for a geometric visualization of our setup.

Expressing the JT action \eqref{JT} in the gauge \eqref{gauge}, we obtain
\begin{equation}\label{actionn}
S_{\mathrm{JT}}[H, \Phi; \mathcal{N}] =  \int_0^1 \mathrm{d}r \left( \frac{1}{2\mathcal{N}} \Phi'(r) H'(r) + \mathcal{N}  \Phi(r) \right).
\end{equation}
where the Gibbons–Hawking–York boundary term is exactly canceled by an integration by parts.\footnote{The scalar curvature and the extrinsic curvature of a boundary circle at radius $r$ in the gauge \eqref{gauge} are given by $R=-H''(r)/\mathcal{N}^2$ and $\kappa=\frac{1}{2\mathcal{N}} \frac{H'(r)}{\sqrt{H(r)}}$.}
As discussed in \cite{PhysRevD.38.2468}, the procedure that we are highlighting breaks the invariance under diffeomorphisms and requires the introduction of ghost fields in the path integral. Remarkably, with our choice of gauge fixing, the ghosts decouple and the  Euclidean transition amplitude we wish to compute simply reads
\begin{equation}\label{amplitude}
\Psi_{b}(\phi,L) = \int \mathrm{d}\mathcal{N} \  \mathcal{D}H \ \mathcal{D}\Phi  \ \exp \left( -S_{\mathrm{JT}}\left(H, \Phi, \mathcal{N}\right) \right),
\end{equation}
where we have to integrate over the lapse variable $\mathcal{N}$ and the two fields $H(r), \Phi(r)$ with boundary conditions 
\begin{equation}\label{bcc}
H(0) = b^2 \quad H(1) = L^2  \qquad \Phi(1) = \phi \qquad H'(0)=0.
\end{equation}
The first two conditions fix the initial and final size of the universe, respectively.  The third condition fixes the value of the dilaton on the Dirichlet boundary; this is physically meaningful, since it determines the coupling of the Schwarzian theory not only in the infinite-cutoff limit but also in its finite-cutoff deformation, as we will discuss in Section~\ref{quantum}.
Finally, the last condition requires that the initial circle be a geodesic boundary, since the extrinsic curvature in the gauge \eqref{gauge} is given by
\begin{equation}
\kappa = \frac{1}{2\mathcal{N}} \frac{H'(r)}{\sqrt{H(r)}} .
\end{equation} 
With the boundary conditions \eqref{bcc}, we define the path integral \eqref{amplitude} as the one that prepares the trumpet wavefunction $\Psi_{b}(\phi,L)$.

\paragraph{Semiclassical expansion} To proceed in the evaluation of the functional integral, we will perform a semiclassical expansion around the classical solutions of the system. The semiclassical picture will also provide us with a geometrical interpretation of the results.
By varying the action with respect to $\Phi$ and $H$, the following equations of motion are readily obtained:
\begin{equation}
H_{\mathrm{cl}}''(r) - 2 \mathcal{N}_{\mathrm{cl}}^2 = 0, \quad \Phi_{\mathrm{cl}}''(r) = 0.
\end{equation}
The classical solutions of the theory, compatible with the boundary conditions in \eqref{bcc}, are given by
\begin{equation}\label{solu}
H_{\mathrm{cl}}(r) = b^2 + (L^2-b^2) r^2, \quad \Phi_{\mathrm{cl}}(r) = \phi_0+(\phi-\phi_0) r, \qquad \mathcal{N}_{\mathrm{cl}}^2=L^2-b^2.
\end{equation}
At this level, $\phi_0$ can be regarded as a collective coordinate that parametrizes the moduli space of classical solutions.\footnote{The on shell action \eqref{saddles}, accordingly, does not depend on $\phi_0$.}
However, variation with respect to $\mathcal{N}$ yields the semiclassical Wheeler--DeWitt (WdW) constraint
\begin{equation}\label{wdw_contr}
- \frac{1}{2\mathcal{N}_{\mathrm{cl}}^2} \int_0^1 \mathrm{d}r \, \Phi_{\mathrm{cl}}'(r) H_{\mathrm{cl}}'(r) + \int_0^1 \mathrm{d}r \, \Phi_{\mathrm{cl}}(r) = 0,
\end{equation}
which, upon substituting the solutions~\eqref{solu}, imposes the condition
\begin{equation}\label{N}
\phi_0=0.
\end{equation}
Geometrically, this solution describes a trumpet-shaped spacetime, as we wished. To make this statement more precise, we observe that substituting equations \eqref{solu} into equation~\eqref{gauge} yields the following form of the metric:
\begin{equation}\label{geom}
	\mathrm{d}s^2 = \left( b^2 + (L^2 - b^2) r^2 \right) \mathrm{d}\tau^2 + \frac{L^2 - b^2}{b^2 + (L^2 - b^2) r^2} \, \mathrm{d}r^2,
\end{equation}
and that, by introducing global coordinates via
\begin{equation}
	r = \sqrt{\frac{b^2}{L^2 - b^2}} \sinh \rho,
\end{equation}
it can be recast as
\begin{equation}\label{tru}
	\mathrm{d}s^2 =\mathrm{d}\rho^2+ b^2 \cosh^2 \rho \, \mathrm{d}\tau^2 =\mathrm{d}\rho^2+H(\rho) \mathrm{d}\tau^2,
\end{equation}
which is the metric of a hyperbolic spacetime with a geodesic hole of length $b$ located at $\rho = 0$. We will see later that the radial coordinate $\rho$ is canonically conjugate to the WdW operator, which in fact generates translations in the Euclidean "radial time" coordinate $\rho$.\footnote{Indeed the form \eqref{tru} of the metric resembles the standard ADM decomposition (with $\mathcal{N}_{\perp}=0$ and $\mathcal{N}=1$) $\mathrm{d}s^2=-\mathcal{N} \mathrm{dt}^2 +a^2(t) \left(\mathrm{d}x+\mathcal{N}_{\perp}\mathrm{d}t\right)^2$ where time and space are swapped and we are working with an Euclidean radial time $t=i\rho$.}
The saddle-point action, obtained by evaluating the action \eqref{actionn} on the classical solutions \eqref{solu}, is  
\begin{equation}\label{saddles}
S_{\mathrm{cl}}^{\pm} = S[H_{\mathrm{cl}}, \Phi_{\mathrm{cl}}; \mathcal{N}_{\mathrm{cl}}^{\pm}] 
= \phi \, \mathcal{N}_{\mathrm{cl}}^{\pm} 
= \pm \phi \sqrt{L^2-b^2},
\end{equation}

where we have included both solutions $\mathcal{N}_{\mathrm{cl}}^{\pm}$, corresponding to the positive and negative branches of the classical relation $\mathcal{N}_{\mathrm{cl}}^2 = L^2-b^2$ in \eqref{solu}. These two branches, and the associated Euclidean actions, will reappear in Section~\ref{quantum}. In the semiclassical expansion of the path integral, we are instructed to sum over both, since they encode the same trumpet-shaped geometry \eqref{tru}.  
Having established the classical solutions, we now proceed to evaluate the path integral. Expanding the fields around their classical values, the path integral can be perturbatively organized as  
\begin{equation}\label{instanton}
\begin{split}
\Psi_{b}(\phi,L) 
= \sum_{\pm} e^{-S_{\mathrm{cl}}^{\pm}} 
\int \mathrm{d}\delta\mathcal{N} \ \mathcal{D}\delta H \ \mathcal{D}\delta \Phi \ 
\exp\!\left[-S_1 \left(\delta \mathcal{N}; \mathcal{N}_{\mathrm{cl}}^{\pm}\right) 
- S_2 \left(\delta \Phi, \delta H, \delta \mathcal{N}; \mathcal{N}_{\mathrm{cl}}^{\pm}\right)\right],
\end{split}
\end{equation}
where the sum runs over the two saddle points \eqref{saddles}. 
Here $S_1$ collects the terms depending only on $\delta \mathcal{N}$:  
\begin{equation}\label{S1}
S_1 \left(\delta \mathcal{N}; \mathcal{N}_{\mathrm{cl}}\right) 
= \left(\frac{\delta \mathcal{N}^2}{2\mathcal{N}_{\mathrm{cl}}^3} 
- \frac{\delta \mathcal{N}^3}{2\mathcal{N}_{\mathrm{cl}}^4} 
+ \cdots \right) 
\int_0^1 \mathrm{d}r \, \Phi_{\mathrm{cl}}'(r) H_{\mathrm{cl}}'(r) 
= \frac{\phi}{2} \, \frac{\delta \mathcal{N}^2}{\mathcal{N}_{\mathrm{cl}}+\delta \mathcal{N}} ,
\end{equation}
while $S_2$ collects the remaining terms in the expansion:
\begin{equation}\label{SS2}
S_2 \left(\delta \Phi, \delta H,\delta \mathcal{N}; \mathcal{N}_{\mathrm{cl}}\right) 
= \frac{1}{2\left(\mathcal{N}_{\mathrm{cl}}+\delta \mathcal{N}\right)} 
\int_{0}^{1} \mathrm{d}r \, \delta \Phi'(r) \delta H'(r)+ \delta \mathcal{N} 
\frac{2 \mathcal{N}_{\mathrm{cl}}+\delta \mathcal{N}}{\mathcal{N}_{\mathrm{cl}}+\delta \mathcal{N}} 
\int_{0}^{1} \mathrm{d}r \, \delta \Phi(r) .
\end{equation}
In writing \eqref{S1} and \eqref{SS2}, we have resummed the series in $\delta \mathcal{N}$ whenever possible. For clarity, we keep the dependence on $\mathcal{N}_{\mathrm{cl}}$ implicit, specifying $\mathcal{N}_{\mathrm{cl}}^{\pm}$ only when performing the explicit sum over the two saddles in \eqref{instanton}.  

The evaluation strategy is as follows: first perform the path integral over the field fluctuations $\delta H$ and $\delta \Phi$, and then carry out the remaining integral over $\delta \mathcal{N}$. From the structure of $S_2$ in \eqref{SS2}, it is clear that the path integral over $\delta \Phi$ and $\delta H$ is Gaussian and can therefore be evaluated exactly. The resulting one-loop determinant then yields an effective action governing the final integration over the lapse fluctuations $\delta \mathcal{N}$.  A crucial step in the computation is the projection of the quantum fluctuation fields onto the subspace orthogonal to the zero mode of the quadratic operator. This zero mode precisely parametrizes the direction along which the classical solution can be varied via the collective coordinate $\phi_0$. However, this direction in functional space is gauge-fixed by the semiclassical WdW constraint in \eqref{N} and is therefore non-physical. The explicit computation, which implements this projection, is quite technical and is presented in Appendix~\ref{gaus}. The final result is
\begin{equation}\label{One}
\mathcal{Z}_{\mathrm{1\text{-}loop}} \sim \left( \mathcal{N}_{\mathrm{cl}} + \delta \mathcal{N} \right)^{-2}.
\end{equation}
where we just extracted the dependence on $\mathcal{N}$.

\paragraph{Result} Including the one-loop factor \eqref{One}, the trumpet wavefunction \eqref{instanton} then finally becomes:
\begin{equation}\label{instanton2}
\begin{split}
\Psi_{b}(\phi,L) 
= \sum_{\pm} e^{\mp\phi \sqrt{L^2-b^2}} 
\int_{-\infty}^{+\infty} \frac{\mathrm{d}\delta\mathcal{N} }{\left(\pm \sqrt{L^2-b^2}+\delta \mathcal{N}\right)^2} 
\exp\!\left[-\frac{\phi}{2} \, \frac{\delta \mathcal{N}^2}{\pm \sqrt{L^2-b^2}+\delta \mathcal{N}}\right],
\end{split}
\end{equation}
Expanding the integrand in powers of $\delta \mathcal{N} / \sqrt{\phi}$ and performing the integral term by term, one obtains:
\begin{equation}\label{steepest22}
\Psi_{b}(\phi,L)\propto \frac{\sqrt{2\pi} }{\sqrt{L^2-b^2}} \left[e^{-z} \left(\frac{1}{z^{1/2}} + \frac{3}{8z^{3/2}}  + \cdots \right) + i e^{z} \left(\frac{1}{z^{1/2}} - \frac{3}{8z^{3/2}}  + \cdots \right)\right],
\end{equation}
where $z = \phi \sqrt{L^2 - b^2}$. 
One recognizes in the expression in square brackets in \eqref{steepest22} the asymptotic expansion of the modified Bessel function $I_1(z)$ in the limit $\phi \gg 1$. As we shall see in the following section, the boundary value of the dilaton corresponds to the coupling of the wiggly boundary action at finite cutoff, playing indeed the effective role of $1/\hbar$. The presence of the imaginary unit $i$ for the perturbative expansion around $\mathcal{N}_{\mathrm{cl}}^{-}$ in \eqref{steepest22} reflects the need for a Wick rotation of $\delta \mathcal{N}$ to make the integral convergent around the instanton $\mathcal{N}_{\mathrm{cl}}^{-}$, which is unstable. A nice feature is that by redefining 
\begin{equation}
\delta \mathcal{N}=y-\mathcal{N}_{\mathrm{cl}}^{\pm},
\end{equation}
one may rewrite the trumpet wavefunction as:
\begin{equation}\label{amplitude2}
\begin{split}
\Psi_{b}(\phi,L)&=\frac{1}{2 \pi i}\oint \frac{\mathrm{d}y}{y} \exp\left( \frac{\phi}{2} \left(y+ \frac{L^2 - b^2}{y} \right) \right) \\
&= \boxed{\frac{1}{\sqrt{L^2-b^2}}I_1 \left(\phi \sqrt{L^2 - b^2}\right)},
\end{split}
\end{equation}
where the contour is now chosen to be a closed circle around the origin.\footnote{The last identity in can be proven using residue theorem and the generating function of Bessel I function, i.e. $\exp \left(\frac{x}{2}(t+1/t)\right)=\sum_{n}I_{n}(x) t^n$.} Indeed, this contour can be deformed into the two steepest contours in \eqref{instanton2}, which yields the asymptotic expansion \eqref{steepest22}. 

In the next section we will show that the above wavefunction, on top of being relevant on its own, it is also the key ingredient to determine the disk partition function of JT gravity at finite cutoff.
 
\subsection{Imposing the no-boundary condition}\label{no}

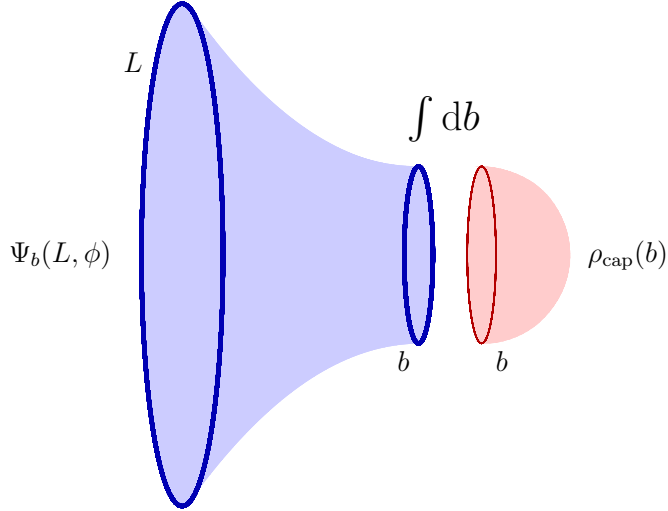
\begin{figure}
    \centering
    \begin{tikzpicture}[scale=0.9, transform shape]
        \node[inner sep=0] (img) {\includegraphics[width=0.5\textwidth, angle=90]{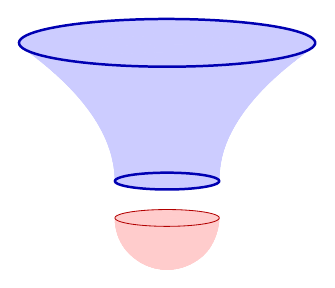}};

    \path (img.south west) -- (img.south east)
            coordinate[pos=0.48] (trumpet) 
            coordinate[pos=0.83] (cap);    

        \node at (-4.3, 0) {$\Psi_b(L,\phi)$};
        \node at (4, 0) {$\rho_{\text{cap}}(b)$}; 
        \node[font= \Large] at (1.3, 2) {$\int \mathrm{d}b$};

    \path (img.north west) -- (img.north east)
            coordinate[pos=0.05] (Lcircle)   
            coordinate[pos=0.60] (bblue)     
            coordinate[pos=0.80] (bred);     

        \node[below=30pt of Lcircle] {$L$};
        \node[below=155pt of bblue] {$b$};
        \node[below=155pt of bred] {$b$};
    \end{tikzpicture}
    \caption{In this figure we show the process of constructing the disk partition function by gluing together the trumpet wavefunction $\Psi_b(L,\phi)$ and the cap amplitude $\rho_{\text{cap}}(b)$. }

    \label{trumpetsc}
\end{figure}

In the previous section we derived the JT gravity trumpet wavefunction $\Psi_{b}(L,\phi)$ from a path-integral computation. As a consistency check, we should ask whether this wavefunction lies in the gravitational Hilbert space $\mathcal{H}_{\text{grav}}$. In a diffeomorphism-invariant theory of gravity, admissible states are annihilated by the Wheeler--DeWitt  and momentum constraints,
\[
\hat{\mathcal H}\ket{\psi}=0,
\qquad
\hat{\mathcal P}\ket{\psi}=0.
\]
The momentum constraint $\mathcal{P}$ is trivially satisfied in the minisuperspace approximation that we adopt in this paper, therefore we will only focus on the WdW constraint.
In the conventions of the previous section and from the classical point of view, the latter is obtained from the variation of the JT action with respect to the lapse function $\mathcal{N}$ and reads:
\begin{equation}\label{mini2}
\mathcal{H}_{\mathrm{WdW}} = \pi_{h}\pi_{\phi} - \frac{1}{2} \phi=0,
\end{equation}
where the above expression is written in terms of the momenta $\pi_{\phi} = H'/2\mathcal{N}$ and $\pi_{h} = \Phi'/2\mathcal{N}$ which are respectively conjugated to the fields $\phi$ and $h$. It is actually convenient to work with $h=L^2$ and rewrite \eqref{mini2} as:
\begin{equation}\label{mini3}
\mathcal{H}_{\mathrm{WdW}} = \frac{1}{2L}\left(\pi_{L}\pi_{\phi} - L\phi \right)=0.
\end{equation}
At the quantum level the idea is to promote the momenta to functional derivatives with respects to the fields $\phi(x)$ and $L(x)$ and to construct the Hilbert space of the theory with vectors that are annihilated by the WdW operator $\hat{\mathcal{H}}_{\text{WdW}}$. As stated earlier, in this paper we choose to work in the minisuperspace approximation, which for two-dimensional dilaton gravities is actually an exact treatment ~\cite{Iliesiu:2020zld}. Therefore we only retain the zero modes of the fields and the functional derivatives actually reduce to ordinary derivatives.
In a slightly more formal way we say that we impose the Euclidean commutation relations
\begin{equation}\label{comm}
\left[L,\pi_L\right]=1 \qquad \left[\phi,\pi_\phi\right]=1,
\end{equation}
such that the WdW operator \eqref{mini3} when acting on functions in the Hilbert space becomes
\begin{equation}\label{mini}
\hat{\mathcal{H}}_{\mathrm{WdW}}^{\mathrm{flat}} = \frac{\partial}{\partial \phi} \frac{\partial}{\partial L} - L \phi.
\end{equation}
As usual when promoting the phase-space variables to operators, a quantum ordering ambiguity arises in formulating the WdW equation and in particular equation~\eqref{mini} corresponds to the flat ordering choice.  
The finite-cutoff trumpet wavefunctions $\Psi_{b}(L,\phi)$ that we computed in \eqref{amplitude2}, instead can be shown to satisfy a WdW equation with non-flat ordering of the form:  
\begin{equation}
\hat{\mathcal{H}}_{\mathrm{WdW}} = \phi^{-1}\frac{\partial}{\partial \phi}\,\phi\,\frac{\partial}{\partial L} - L\,\phi, 
\qquad 
\hat{\mathcal{H}}_{\mathrm{WdW}}\,\Psi_{b}(L,\phi)=0.
\end{equation}
We notice that the label $b$ of the trumpet wave function does not enter in the above computation and  therefore we found a one parameter family of states in the physical Hilbert space, each one labeled by a different real value of $b$. This comes with no surprise as the path integral is known to prepare states that satisfy the hamiltonian constraint \cite{PhysRevD.38.2468}.
Assuming that $\{\Psi_{b}(L,\phi) \}_{b \in\mathbb{R}}$ are a basis in the physical Hilbert space, a generic physical state 
$\Psi(L,\phi)$ can be expressed as:  
\begin{equation}
\Psi(L,\phi)=\int \! \mathrm{d}b \ \rho(b)\,\Psi_{b}(L,\phi) \, ,
\end{equation}  
with some weight function $\rho(b)$ to be determined. Such a state automatically satisfies the 
WdW equation. In this instance we are interested in the Hartle–Hawking wavefunction, which is prepared by a path integration over a regular Euclidean spacetime that smoothly caps off without additional boundaries in the interior. Considering only genus zero surfaces in two dimensions, the Hartle–Hawking wavefunction, or equivalently the disk partition function, can be obtained as an integral over trumpet states,  
\begin{equation}\label{sum}
Z_{\mathrm{disk}}(L,\phi) = \Psi_{\mathrm{HH}}(L,\phi) 
= \int \! \mathrm{d}b \ \rho_{\mathrm{cap}}(b)\, \Psi_{b}(L,\phi) \, ,
\end{equation}  
where the weight $\rho_{\mathrm{cap}}(b)$ is given by the cap amplitude showed in Figure \ref{trumpetsc}. The latter is indeed given by the Euclidean path integral over a disk geometry that smoothly caps off in the bulk and ends on a geodesic 
boundary of length $b$. In infinite-cutoff JT gravity, the cap amplitude is given by  
\begin{equation}\label{cap}
\rho_{\rm cap}(b) \sim \delta'(b - 2\pi i) \, .
\end{equation}  

Since the only sensitivity to the boundary being at finite cutoff is encoded in the finite-cutoff trumpet geometries appearing in \eqref{sum}, we expect the cap amplitude to remain unchanged. The localization of the geodesic length at $b = 2\pi i$ implied by \eqref{cap} can be understood by viewing the geodesic hole as a defect of imaginary strength $\gamma = i b$ \cite{Mertens:2019tcm,Maxfield:2020ale}, where smoothness of the bulk geometry requires setting $\gamma = 2\pi$ to eliminate the conical defect. An equivalent conclusion is reached by analyzing the analytically continued trumpet metric \eqref{tru} when set $\rho \to i\theta$: as $\theta \to \pi/2$, i.e. as the Euclidean time circle shrinks, the geometry develops a conical singularity unless $b = 2\pi$ \cite{Blommaert:2021fob}. The corresponding metric contour, which runs from $\rho = +\infty$ to $\rho = 0$ and then to $\rho = i\pi/2$, is in fact equivalent to the contour connecting a Lorentzian de Sitter expanding universe to its Euclidean half-sphere preparation in the original no-boundary proposal \cite{Hartle:1983ai}.  
  
A simple way to appreciate the appearance of a derivative of delta in the form of the cap amplitude \eqref{cap} is to view JT gravity as embedded 
in sine dilaton gravity, a two-dimensional dilaton gravity theory with a sine-shaped dilaton potential \cite{Blommaert:2024ymv}. 
This theory can be regarded as a UV completion of JT gravity, due to its duality with the double-scaled 
SYK model \cite{Blommaert:2024ymv}. The cap amplitude in sine dilaton gravity/DSSYK is known 
\cite{Blommaert:2025avl,Blommaert:2025rgw,Blommaert:2025eps,Okuyama:2024eyf} and takes the form  
\begin{equation}\label{capsine}
\rho_{\rm cap}(b,q)
= e^{-i\pi b/2}\, q^{b^2/4}\left(q^{b/2}-q^{-b/2}\right) 
\simeq e^{-\frac{(b_{\rm JT}-2i\pi+2|\log q|)^2}{16|\log q|}}
- e^{-\frac{(b_{\rm JT}-2i\pi-2|\log q|)^2}{16|\log q|}} \, ,
\end{equation}  
where in the second step we discarded overall $q$-dependent prefactors and introduced the standard rescaling \cite{Okuyama:2023byh,Okuyama:2023kdo}  
$$b = 2|\log q|\, b_{\mathrm{JT}} \, .$$  
Taking the limit $|\log q|\to 0$ reduces the theory back to JT gravity. In this limit, the Gaussian factors 
in \eqref{capsine} become infinitely sharp, turning into delta functions. More precisely the difference of 
the two Gaussians approaches the derivative of a delta function, which is exactly the structure \eqref{cap} reproduced by \eqref{capsine} in the JT limit.\footnote{ Indeed  
\begin{equation}
\delta'(x) \;\simeq\; \delta(x+\epsilon)-\delta(x-\epsilon) 
\;\simeq\; e^{-(x+\epsilon)^2/4\epsilon^2} - e^{-(x-\epsilon)^2/4\epsilon^2} \, ,
\end{equation} }

Plugging the trumpet wavefunction \eqref{amplitude2} together with the cap amplitude \eqref{cap} into \eqref{sum}, we arrive at the JT gravity disk partition function at finite cutoff:
\begin{equation}\label{disk}
Z_{\mathrm{disk}}(L,\phi) = \Psi_{\mathrm{HH}}(L,\phi) \sim \frac{2 i \pi \phi I_2\left(\sqrt{L^2+4 \pi ^2} \phi \right)}{L^2+4 \pi ^2}. 
\end{equation}
In Section \ref{quantum}, we will verify this result perturbatively in $\phi$ by adopting a complementary approach based on the path integral over fluctuating boundary curves at finite distance in the bulk.

\subsection{A reduced path integral: preparing eigenstates $\ket{b}$ of the geodesic length operator}\label{reduced}

In the previous section we briefly recalled the definition of the gravitational Hilbert space and discussed how the states $\Psi(L,\phi)$ form a basis of such a space. While the discussion that follows is tangential to the main purpose of the paper, it is nevertheless useful to pause and examine how an alternative treatment of the path integral from the previous section can connect our framework to recent developments in the study of the gravitational Hilbert space of JT gravity. Several different basis of $\mathcal{H}_{\text{grav}}$ has been recently discussed in the literature \cite{Held:2024rmg,Blommaert:2025bgd} and it was suggested that
a sensible strategy to classify the admissible states is to study the eigenvectors of an operator that commutes with the WdW Hamiltonian. An operator considered in \cite{Held:2024rmg,Blommaert:2025bgd} is
\begin{equation}\label{geode}
b^2 = L^2 - \pi_{\phi}^2.
\end{equation}
This choice is interesting because from a classical view point $b$ measures the length of the geodesic defect in a trumpet spacetime \eqref{tru} as we now show. To see this, we should interpret \eqref{mini3} as the hamiltonian of a physical system and study its hamiltonian flow. The equations of motion for $L$ and $\pi_{\phi}$ are
\begin{equation}
\partial_{\rho} L = \frac{\partial \mathcal{H}_{\mathrm{WdW}}}{\partial \pi_{L}} = \pi_{\phi} ,
\qquad
\partial_{\rho} \pi_{\phi} = -\frac{\partial \mathcal{H}_{\mathrm{WdW}}}{\partial \phi} = L
\end{equation}
where $\rho$ is the Euclidean time coordinate in \eqref{mini3}, conjugate to the WdW Hamiltonian. 
As we are interested in spacetime with a geodesic boundary we impose the conditions $L(0) = b$, $L'(0) = 0$, which coincide with those used in the path integral \eqref{bcc}. The unique solutions of the equations of motion obeying these conditions are:
\begin{equation}\label{change}
L(\rho) = b \cosh \rho ,
\qquad
\pi_{\phi}(\rho) = b \sinh \rho  .
\end{equation}
The first solution coincides with the one appearing in the trumpet spacetime metric \eqref{tru}. We notice that inserting these results in \eqref{geode} the equality is immediately satisfied, therefore providing a classical interpretation for the operator \eqref{geode}, and showing that it indeed measures the length of the geodesic defect in the bulk.

One can also use the expression in \eqref{change} to perform a canonical change of variable in the classical phase space of the theory and write everything in terms of the variables $(\rho,b)$. In these new variables the hamiltonian becomes
\begin{equation}\label{change2}
\mathcal{H}_{\mathrm{WdW}}
= \pi_{\phi}\frac{\partial}{\partial L} + L \frac{\partial}{\partial \pi_{\phi}}
= \frac{\partial}{\partial \rho}  .
\end{equation}
In other words, as we expected, the WdW Hamiltonian generates translations along the radial coordinate $\rho$, while $b$ remains constant along the hamiltonian flow and is therefore conserved. This provides classical perspective on why the geodesic length operator \eqref{geode} commutes with the WdW Hamiltonian, a fact that can also be verified by a direct computation.

An obvious solution to the constraint \eqref{geode} in the coordinate space $\left(L,\pi_{\phi}\right)$ is given by 
\begin{equation}
\psi_{b}(L,\pi_{\phi})=\delta(L^2-\pi_{\phi}^2-b^2)=\frac{\delta\left(\pi_{\phi}-\sqrt{L^2-b^2}\right)+\delta\left(\pi_{\phi}+\sqrt{L^2-b^2}\right)}{2\sqrt{L^2-b^2}}
\end{equation}
One can then perform a Fourier transform to find the corresponding solution in the $\left(L,\phi\right)$ basis \cite{Held:2024rmg}: \footnote{The absence of the $i$ factor in the Fourier weight is due to Euclidean commutation relations between $\phi$ and $\pi_{\phi}$.}
\begin{equation}\label{b}
\braket{L\,\phi|b}\equiv \psi_{b}(L,\phi)=\int_{-\infty}^{+\infty} \mathrm{d}\pi_{\phi} e^{- \phi \pi_{\phi}} \psi_{b}(L,\pi_{\phi})=\boxed{\frac{\cosh \left( \phi\sqrt{L^2-b^2}\right)}{\sqrt{L^2-b^2}}}
\end{equation}
We can now ask whether there exists a version of the path integral that directly prepares the states  
$\braket{L\,\phi|b} \equiv \psi_{b}(L,\phi)$ of \eqref{b}, rather than producing the trumpet wavefunction 
$\Psi_{b}(L,\phi)$ obtained in \eqref{amplitude2}. The answer is affirmative and amounts to a different treatment 
of the zero mode of the fluctuation field $\delta \Phi$, or equivalently of the collective coordinate $\phi_{0}$ 
associated with the classical solution \eqref{collective}. 

Now we allow the path integral to also integrate over the zero mode of $\delta \Phi$ disregarding the semiclassical considerations of the previous section. Basically after the mode expansion \eqref{fluc}, and writing the action as in \eqref{SS2}, we end up with one more integral to perform resulting in the expression:
\begin{equation}\label{instanton22}
\begin{split}
\braket{L\,\phi |b}\equiv \psi_{b}(L,\phi) 
= \sum_{\pm} e^{-\phi \mathcal{N}_{\mathrm{cl}}} 
&\int_{-\infty}^{+\infty} \frac{\mathrm{d}\delta\mathcal{N} }{\left(\det^{'}(\mathcal{O})\right)^{1/2}} 
\exp\!\left[-\frac{\phi}{2} \, \frac{\delta \mathcal{N}^2}{\mathcal{N}_{\mathrm{cl}}+\delta \mathcal{N}}\right] \times \\
&\times \int \mathrm{d}c_0 \ \exp \left[-\frac{c_0 \delta \mathcal{N}}{2 \sqrt{\delta \mathcal{N}+ \mathcal{N}_{\mathrm{cl}}}}\left(\frac{\delta \mathcal{N}+2 \mathcal{N}_{\mathrm{cl}}}{\delta \mathcal{N}+ \mathcal{N}_{\mathrm{cl}}}\right)\right].
\end{split}
\end{equation}
Integrating over $c_0$ along the imaginary direction produces the sum of delta functions
\begin{equation}\label{delt}
 \delta \left( \frac{ \delta \mathcal{N}}{2 \sqrt{\delta \mathcal{N}+ \mathcal{N}_{\mathrm{cl}}}}\left(\frac{\delta \mathcal{N}+2 \mathcal{N}_{\mathrm{cl}}}{\delta \mathcal{N}+ \mathcal{N}_{\mathrm{cl}}}\right) \right) = 2 \sqrt{\delta \mathcal{N}+ \mathcal{N}_{\mathrm{cl} }} \left[ \delta (\delta \mathcal{N})+ \delta(\delta \mathcal{N} + 2 \mathcal{N}_{\mathrm{cl}} )\right].
\end{equation}
Consequently, the integral over $\delta \mathcal{N}$ localizes to the classically allowed values and the quantum dynamics of the theory disappears.
The one-loop determinant over the fluctuation fields $\delta H(r)$ and $\delta \Phi(r)$ is evaluated as in Appendix \ref{gaus}, and gives
\begin{equation}\label{onel} \left(\det (\mathcal{O})\right)^{-1/2} \propto |\delta \mathcal{N}+\mathcal{N}_{\mathrm{cl}}|^{-\frac32}, 
\end{equation}
so that when we also take into accounts the factors appearing in \eqref{delt} we obtain
\begin{equation}
\braket{L\,\phi |b}\equiv \psi_{b}(L,\phi) = \sum_{\pm} e^{-\phi \mathcal{N}_{\mathrm{cl}}} 
\int_{-\infty}^{+\infty} \frac{\mathrm{d}\delta\mathcal{N}}{\delta \mathcal{N}+\mathcal{N}_{\mathrm{cl}}} \left[ \delta (\delta \mathcal{N})+ \delta(\delta \mathcal{N} + 2 \mathcal{N}_{\mathrm{cl}} ) \right] \exp\!\left[-\frac{\phi}{2} \, \frac{\delta \mathcal{N}^2}{\mathcal{N}_{\mathrm{cl}}+\delta \mathcal{N}}\right].
\end{equation}
This path integral, where $\mathcal{N}$ is localized on its classical value, directly prepares the states $\braket{L \, \phi |b}$. Indeed, accounting for both instantons in \eqref{instanton22}, we obtain the result \eqref{b} as we wished. 

Moreover, we observe that the localization of the path integral on the classical value 
$\mathcal{N}_{\mathrm{cl}}^2 = L^2 - b^2$ is fully consistent with the expectation that the 
path integral prepares an eigenstate of the operator \eqref{geode}, since in our gauge choice 
one has $\pi_{\phi} = \mathcal{N}$. To make this explicit, consider working directly with the JT 
action in the form \eqref{JT}. Integrating out the dilaton enforces $R = -2$, which in the gauge 
\eqref{gauge} implies the quantum condition  
\begin{equation}\label{R}
H''(r) = 2 \mathcal{N}^2 \, .
\end{equation}  
The conjugate momentum to the boundary value $\phi$ of the dilaton can then be read off from the 
boundary term in \eqref{JT},  
\begin{equation}
\pi_{\phi} = \sqrt{h}\,\kappa = \frac{H'(1)}{2\mathcal{N}} = \mathcal{N} \, ,
\end{equation}  
where in the second step we used \eqref{R} together with the geodesic condition $H'(0)=0$.\footnote{Indeed, 
integrating $\int_{0}^{1} \mathrm{d}r$ over both sides of \eqref{R} yields 
$H'(1) - H'(0) = 2\mathcal{N}^2$.}  
In this formulation, the JT action reduces entirely to the boundary term in \eqref{JT}, which directly 
produces the Fourier factor appearing in \eqref{b}. To enforce the condition \eqref{geode}, one can insert 
in the path integral the Fourier representation of the delta function,  
\begin{equation}\label{delta}
\braket{L \, \phi |b} \equiv \psi_{b}(L,\phi) 
= \int \mathrm{d}\mathcal{N} \int_{-i\infty}^{+i\infty} \mathrm{d}\lambda \,
\exp\!\Big[-\lambda \left(\pi_{\phi}^2 - L^2 + b^2\right) - \phi \mathcal{N}\Big] \,.
\end{equation}  
This reproduces exactly the same structure as in \eqref{b}. In this picture, the correct dependence from $\mathcal{N}$ that previously arose from the one loop determinant and the integral over the zero mode of the fluctuations \eqref{instanton22},
now arises naturally from the Jacobian associated with the delta-function insertion \eqref{delta}. 

Another natural perspective on the result in \eqref{b}  is offered by the analysis of the amplitude $\braket{L\, \phi | L'\, \phi'}$ that was computed in \cite{Buchmuller:2024ksd}. As it is shown in Appendix \ref{fourier} the wave function $\braket{L \, \phi | b}$ is nothing but the Fourier transform of the above amplitude, meaning that:
\begin{align}\label{cosh}
    \int_{- \infty}^{+ \infty} e^{- i k x} \braket{L\, \phi| L' x} \mathrm{d}x 
    =\frac{2\,\cosh\!\big(\phi\sqrt{L^2- b^2-k^2}\,\big)}{\sqrt{L^2- b^2-k^2}}\;\mathbf{1}_{\{|k|<L^2 - b^2\}}.
\end{align}
Performing a Fourier transform in this context basically amounts to a change of basis in the Hilbert space, where instead of working with the position basis labeled by $\phi'$ we express the result in terms of the momentum basis with label $k \equiv \pi_{\phi '}$. Therefore the result that we computed from the path integral should be interpreted as the amplitude $\braket{L \, \phi | b \, \pi_{\phi '}=0}$. This is in spirit different from the computation of the trumpet partition function at finite cutoff and offers a perspective of why it is so. 

In the subsequent sections \ref{wiggly} and \ref{quantum}, we study the same finite cutoff theory from a complementary viewpoint that is entirely boundary-based, and we compare observables such as the partition function with those obtained from the bulk analysis in the preceding sections.
In particular, we will show that the disk amplitude \eqref{disk}, constructed by gluing the trumpet wavefunction $\Psi_b(L,\phi)$ to the cap amplitude, matches the result of the finite cutoff Schwarzian path integral at one loop level, thereby justifying our interpretation.

\section{The wiggly boundary curve at finite cutoff}\label{wiggly}
We now adopt a different approach to studying JT gravity at finite cutoff. Specifically, we analyze the boundary dynamics of Euclidean JT gravity on a disk topology from a semiclassical perspective, which gives a finite cutoff deformation of the standard Schwarzian description \cite{Engelsoy:2016xyb,Maldacena:2016upp}. In this section we present the form of the emergent finite cutoff boundary action,\footnote{Some of the results in this section already appeared in \cite{Iliesiu:2020zld}.} and we study its classical solutions, while in Section \ref{quantum} we turn to the quantum theory and perform a one loop path integral computation. 

The action of interest is \eqref{JT}, we write it down again for reference:
\begin{equation}\label{cla}
S_{\mathrm{JT}} = \frac{1}{2} \int_{\mathcal{M}} \mathrm{d}^2 x \sqrt{g} \, \Phi\left(R + 2\right) + \int_{\partial \mathcal{M}} \mathrm{d}\tau \sqrt{h} \, \Phi \left(\kappa-1 \right).
\end{equation}
As it is standard practice, we begin by path integrating out the dilaton over an imaginary contour. This procedure restricts the functional integral over the metrics to include only the locally hyperbolic ones. Basically, the gravitational path integral reduces to integrating over all possible ways of cutting out a portion of the hyperbolic disk with a definite boundary curve. The cost of each of these configurations is given by the Gibbons-Hawking-York boundary term in \eqref{cla}. The action of the theory that we aim study is therefore:
\begin{equation}\label{finite}
I=\int_{\partial M} \mathrm{d}u \Phi \left(\kappa-1\right),
\end{equation}
where the boundary is a curve at finite distance inside of the bulk.

To proceed further we find convenient to parametrize the hyperbolic disk as follows:
\begin{equation}\label{hyper}
\mathrm{d} s^2=\frac{\mathrm{d}\tau^2+\mathrm{d}x^2}{x^2}=-4 \frac{\mathrm{d}z \mathrm{d}\bar{z}}{(z-\bar{z})^2},
\end{equation}
where we have introduced the complex coordinates $z=\tau-i x$, $\bar{z}=\tau+i x$, which correspond to the Euclidean version of the lightcone coordinates. In this coordinate chart the boundary curve can be parametrized as  $\gamma(u)=\left(z(u),\bar{z}(u)\right)$, with $u$ the boundary proper time. What's more, the extrinsic curvature of the boundary assumes the following form:
\begin{equation}\label{extri}
\kappa=\frac{1}{4 \left(z'(u)\bar{z}'(u)\right)^{3/2}} \left[\left(\bar{z}''(u) z'(u)-z''(u)\bar{z}'(u)\right)\left(z(u)-\bar{z}(u)\right)+2\left(\bar{z}'(u)+z'(u)\right)\bar{z}'(u)z'(u)\right].
\end{equation} 
To properly define the theory both at the classical and at the quantum level we still need to impose suitable boundary conditions on the fields. In the following we will always work with Dirichlet boundary conditions both on the metric and on the dilaton. In more detail, we require that the boundary metric is constant and equal to $g_{uu}= \frac{1}{\varepsilon^2}$. When expressed in the coordinates of \eqref{hyper}, this implies the condition:
\begin{equation}\label{con}
-4 \frac{z'(u)\bar{z}'(u)}{\left(z(u)-\bar{z}(u)\right)^2}=\frac{1}{\varepsilon^2}.
\end{equation}
As a consistency check, we notice that at leading order in $\varepsilon$, equation \eqref{con} can be used to express $z(u)$ in terms of $\bar{z}(u)$ as
\begin{equation}\label{zz}
\bar{z}(u)=z(u)+2i \varepsilon z'(u) + O(\varepsilon^2),
\end{equation}
and that plugging this value in  \eqref{extri}, gives:
\begin{equation}\label{extri2}
\kappa=1+\varepsilon^2 \{z(u),u \}+O(\varepsilon^3),  \quad \text{with} \quad \{z(u),u \}=\frac{z'''(u)}{z'(u)}-\frac32 \left(\frac{z''(u)}{z'(u)}\right)^2
\end{equation}
Imposing the boundary value of the dilaton to be $\Phi=\phi_{r}/\varepsilon$, in the limit $\varepsilon \rightarrow 0$ the action \eqref{finite} reduces to the ordinary Schwarzian action, whose quantization is known \cite{Stanford:2017thb} and leads to the partition function of JT gravity. 

In the following we will be interested in keeping $\varepsilon$ finite and in quantizing \eqref{finite} at finite cutoff.

\subsection{The Riccati equation for the extrinsic curvature}\label{ric}
The first ingredient that is required to begin investigating the theory at finite cutoff is a solution of \eqref{con} for a generic value of $\varepsilon$. In order to achieve that, we introduce a new variable $\psi(u)$ related to $z(u)$ and $\bar{z}(u)$ by the following relation:
\begin{equation}\label{ree}
\bar{z}(u)=z(u)-\frac{2z'(u)^2 \psi(u)}{z''(u)\psi(u)+2\psi'(u)z'(u)}.
\end{equation}
The reasoning that lead us to \eqref{ree} is explained in Appendix \ref{ricc}.
Utilizing this new variable in equation \eqref{con}, we find that the constraint assumes a much more pleasant form in terms of a Schrodinger-like problem:
\begin{equation}\label{eq5}
\psi''+\left[\frac12 \{z(u),u \}+\frac{1}{4 \varepsilon^2}\right]\psi=0.
\end{equation}
In particular the above equation makes clear that  $\psi$ is an $\mathrm{SL}(2,\mathbb{R})$ invariant, since $\psi$ satisfies an ODE whose coefficients are invariant under $\mathrm{SL}(2,\mathbb{R})$. As a consequence, $\bar{z}(\psi,z)$ transforms correctly under $\mathrm{SL}(2,\mathbb{R})$.

We can therefore use this new variable to express the extrinsic curvature $\kappa$ in manageable way. A priori, since the extrinsic curvature is an $\mathrm{SL}(2,\mathbb{R})$ invariant, we expect it to be a function of $\psi$ only. As a matter of fact, by plugging \eqref{ree} in the definition \eqref{extri} and making use of \eqref{eq5} whenever possible, we find the following compact relation:
\begin{equation}\label{relazione}
\kappa (u)=2i \varepsilon \frac{\psi'(u)}{\psi(u)}.
\end{equation}
Remarkably, given this simple expression, it is possible to find a differential equation obeyed by $\kappa$. In particular, it is immediate to show that the extrinsic curvature satisfies a Riccati equation:
\begin{equation}\label{extra}
\kappa'(u)-\frac{i}{2\varepsilon} \kappa^2 (u)+i\left(\varepsilon  \{z(u),u \}+\frac{1}{2\varepsilon}\right)=0.
\end{equation}
Using \eqref{relazione}, we can rewrite the relation~\eqref{ree} between $z(u)$ and $\bar{z}(u)$ in terms of the extrinsic curvature as
\begin{equation}\label{rela2}
 \bar{z}(u) = z(u) - \frac{2i \varepsilon z'(u)}{\kappa(u) + i\varepsilon \dfrac{z''(u)}{z'(u)}}.
\end{equation}
As a side note, we observe that in the limit $\varepsilon \to 0$, using \eqref{extra}, we correctly recover the expected relation~\eqref{zz} between $z$ and $\bar{z}$.

Interestingly, Riccati equations of the above form are known to be solvable in a perturbative fashion order by order is some small parameter, which in our case is identified by $\varepsilon$. Therefore, \eqref{extra} will allow us to obtain a small $\varepsilon$ expansion for $\kappa$ to whatever order we are interested into. 
In order to do that we write $\kappa$ in the formal expansion
\begin{equation}
\kappa(u) = \sum_{n=0} \kappa_n(u) \varepsilon^{n},
\end{equation}
and we insert this ansatz in the Riccati equation above. Proceeding in this way, we immediately read the recursive relation that the functions $\kappa_n(u)$ have to satisfy:
\begin{equation}\label{recur}
    \begin{cases}
        \kappa_0^2(u)=1, \qquad &n=1 \\
        2  \kappa'_{n-1}(u) - i \sum_{m=0}^{n}\kappa_{n-m}(u) \kappa_m(u)  + 2 i  \delta_{n, 2}  \{z(u), u \}  =0 \qquad &n>1.
    \end{cases}
\end{equation}
The first few orders obtained with the above method selecting $\kappa_0(u)=1$ are reported below.
\begin{align}\label{kn}
&\kappa_{0}(u)=1, \qquad \kappa_{1}(u)=0, \qquad \kappa_{2}(u)=\{z(u), u \},\\
 &\kappa_{3}(u)=-i \frac{\mathrm{d}}{\mathrm{d}u}\{z(u), u \}, \qquad \kappa_{4}(u)=-\frac{\mathrm{d}^2}{\mathrm{d}u^2}\{z(u), u \}-\frac{1}{2} \{z(u), u \}^2.
\end{align}
Two comments are in order. First of all we acknowledge that  the functions in \eqref{kn} already appeared in \cite{Iliesiu:2020zld}, however it was not clear how to obtain them in a systematic way. With our treatment we made clear how to do that and we get the additional possibility of deducing some properties of the functions directly from the recursive equation \eqref{recur}. In particular, it can be shown by a direct analysis that all the odd-degree functions $k_{2n +1}(u)$ either vanish or are complex valued. At first glance this looks strange as the extrinsic curvature $\kappa$ appears in the action of the theory that should be real valued. Remarkably, it can be proved that all this terms are total derivatives, therefore when integrated in the action their contribution vanishes making the action real and perfectly legitimate. The proofs of the above statements are reported in Appendix \ref{odd}. Secondly, we notice that with our treatment we also proved that the extrinsic curvature of a closed curve embedded in the hyperbolic disk is a function of the Schwarzian and its derivatives only. This information will be important in the following section where we will study the classical equations of motion of the theory.

As a consistency check, we notice that when including up to $\kappa_2(u)$ in the perturbative expansion we recover the usual Schwarzian action.

\subsection{Classical solutions}\label{cla}
In this section we begin the investigation of the theory at the classical level aiming to write the correct equations of motion from the action.
Although we do not have an exact closed-form expression for the extrinsic curvature, knowing its functional form, at least in its perturbative expansion in the cutoff, allows us to study the classical solutions of the theory. In particular, these solutions are extrema of the action:
\begin{equation}\label{action}
I = \frac{\phi_r}{\varepsilon^2} \int_{0}^{\beta} \left(\kappa(u)-1\right)  \ d u=\phi_r \int_{0}^{\beta} \kappa_{\mathrm{reg}}(u)  \ d u.
\end{equation}
As already noticed at the end of the previous section, the extrinsic curvature is a function of the Schwarzian derivative and its derivatives only, meaning that 
\begin{equation}
\kappa(u)=\kappa \left(\{z(u),u \},\partial_u \{z(u),u \},\cdots,\partial_{u}^{k}\{z(u),u \}, \cdots \right).
\end{equation}
For this reason, in order to get the equations of motion from the action \eqref{action}, it is useful to organize its variation in the following way:
\begin{equation}
\delta I=\int_{0}^{\beta} \left[\frac{\delta \kappa}{\delta \{z(u),u \} } \delta \{z(u),u \}+\sum_{k} \frac{\delta \kappa}{\delta \partial_{u}^{k} \{z(u),u \} } \delta \partial_{u}^{k}\{z(u),u \}\right] \ \mathrm{d} u.
\end{equation}
If we now commute the variation with the derivatives and we integrate by parts a suitable number of times, the above expression assumes the form:
\begin{equation}
\delta I=\int_{0}^{\beta} g(u)  \  \delta \{z(u),u \} \ \mathrm{d} u ,
\end{equation}
where we defined the function $g(u)$ as
\begin{equation}
g(u)=\frac{\delta \kappa}{\delta \{z(u),u \} } + \sum_{k}(-1)^k \partial_{u}^{k} \frac{\delta \kappa}{\delta \partial_{u}^{k} \{z(u),u \} }.
\end{equation}
Since at the end we will be interested in the equations of motion for the field $z(u)$, it is sensible to express the variation of the Schwarzian explicitly as:
\begin{equation}
\delta \{z(u),u \}=\frac{\left(\delta z\right)'''}{z'}-\frac{z''' \left(\delta z\right)'}{z'^2}-3 \frac{z'' \left(\delta z\right)''}{z'^2}+3 \frac{z''^2 \left(\delta z\right)'}{z'^3},
\end{equation}
and by integrating by parts each derivative, we finally arrive to:
\begin{equation}
\delta I=-\int_{0}^{\beta} \bigg(g'''(u)+2g'(u) \{z(u),u \}+g(u) \{z(u),u \}'\bigg)\frac{\delta z(u)}{z'(u)} \ \mathrm{d} u.
\end{equation}
From the above expression it is immediate to read out the classical equations of motion for the theory, that are:
\begin{equation}\label{finiteEOM}
g'''(u)+2g'(u) \{z(u),u \}+g(u) \{z(u),u \}'=0
\end{equation}
This equation once again clearly exhibits the $\mathrm{SL}(2,\mathbb{R})$ invariance, as both $g(u)$ is invariant and each coefficient appearing in \eqref{finiteEOM} respects this symmetry.

First of all, we notice that if the action under exam would only be given by the Schwarzian derivative as it happens in the infinite cutoff regime, then $g(u)=1$ and \eqref{finiteEOM} would reduce to the standard equation of motion of the Schwarzian theory, i.e. $\{z(u),u \}'=0$. At finite cutoff, instead, the equations of motion become much more complicated and it seems to be very difficult to be able to find a general solution. A viable strategy could be to solve the equations in perturbation theory in $\varepsilon$, by only retaining few orders. For instance, if we truncate the perturbative expansion \eqref{recur} for the extrinsic curvature at order $\varepsilon^4$, considering only the first correction to the Schwarzian theory, then $g(u)=1-\varepsilon^2 \{z(u),u \}+O(\varepsilon^4)$ and the equations of motion \eqref{finiteEOM} would read:
\begin{equation}\label{finiteEOM2}
\frac{\mathrm{d}}{\mathrm{d}u} \left(\{z(u),u \}-\frac32 \varepsilon^2 \{z(u),u \}^2-\varepsilon^2 \{z(u),u \}''\right)+O(\varepsilon^4)=0
\end{equation}
We were not able to find a general solution to this equation and its higher order deformations in $\varepsilon$. Nonetheless, we notice that $\left\{z(u), u \right\}=\alpha$, with $\alpha$ a constant, is a solution of \eqref{finiteEOM2}. Actually, by a simple argument it remains a solution to any order. Indeed, as we have already shown, perturbatively the extrinsic curvature takes the form of a multipolynomial in the Schwarzian and its derivatives, meaning that $\frac{\partial^n}{\partial u^{n}}g(u)=0$ for any positive integer $n$ on configurations such that $\{z(u),u \}'=0$. This implies that the finite cutoff equation of motion \eqref{finiteEOM} is always satisfied on such configurations. 

In the following, we will only consider constant-Schwarzian configurations as solutions to the equations of motion, leaving as an open question the possibility of having more general solutions to \eqref{finiteEOM}.

Specializing on such configurations we notice that many properties of the solutions are easy to deduce. In particular, by requiring that $\left\{z(u), u \right\}=\alpha$, we obtain the relation
\begin{equation}
\kappa'=\frac{\delta \kappa}{\delta \left\{z(u), u \right\}} \left\{z(u), u \right\}'+\sum_{k \geq 1} \frac{\delta \kappa}{\delta \partial_{u}^{k}\left\{z(u), u \right\}} \partial_{u}^{k+1}\left\{z(u), u \right\}=0,
\end{equation}
which implies a constant extrinsic curvature. By plugging this condition inside the Riccati equation \eqref{extra}, we obtain a relation between the value of the constant $\alpha$ and the value of the extrinsic curvature on each solution:
\begin{equation}\label{alpha}
\alpha=\frac{\bar\kappa^{2}-1}{2 \varepsilon^2}
\end{equation}
where we denoted with $\bar\kappa$ the value of the extrinsic curvature. Therefore we find that there are two admissible values $\pm |\bar\kappa|$ associated with the same classical solution. We will refer to these as the perturbative and non-perturbative branches respectively.

On top of these properties we can also ask about the classical geometry of the spacetime, and in particular about the shape of the boundary curve. In order to address that we need to understand what sort of configurations are admissible for $z(u)$ by the differential equation:
\begin{equation}
    \left\{z(u), u \right\}=\alpha.
\end{equation}
This equation is well studied in the literature \cite{KRYNSKI2022104665} and its solutions are classified depending on the sign of $\alpha$: 
\begin{equation}\label{cases}
\begin{split}
    (i)\,\, \text{if}\,\, \alpha<0 &: z(u)= \frac{A e^{\sqrt{-\alpha/2} \ u}+B e^{-\sqrt{-\alpha/2} \ u}}{C e^{\sqrt{-\alpha/2} \ u}+D e^{-\sqrt{-\alpha/2} \ u}},\\
    (ii)\,\, \text{if}\,\, \alpha=0 &: z(u)= \frac{A u +B}{Cu +D},\\
    (iii)\,\, \text{if}\,\, \alpha>0 &: z(u)= \frac{A \sin(\sqrt{\alpha/2}u)+B \cos(\sqrt{\alpha/2}u)}{C \sin(\sqrt{\alpha/2}u)+D \cos(\sqrt{\alpha/2}u)}.
\end{split}
\end{equation}
Here $A$, $B$, $C$, $D$ are arbitrary constants such that $AD-BC\neq0$. 

In this instance, since we are working on the thermal circle, our focus will be on periodic solutions, therefore we will consider the case $\alpha>0$, where the solution is Mobius equivalent to a $\tan$ function, as can be seen from \eqref{cases}.\footnote{We can choose $A=D=1, \quad B=C=0$.} Specifically, we choose the solution as:
\begin{equation}\label{sol_trumpet}
	z(u)=  \tan \left(\sqrt{\alpha/2}  \ u\right) =\tan \left(\sqrt{\frac{\bar\kappa^{2}-1}{4 \varepsilon^2}}  \ u\right) = \tan \left(\frac{\pi}{\beta} u\right),
\end{equation}
where we used \eqref{alpha} and in the last equality we imposed the fact that the solution must be periodic along the thermal circle under $u\rightarrow u+\beta$. This last observation relates the value of the extrinsic curvature on-shell with the length of the boundary.

The geometry associated with this solution can be obtained by analytically continuing \( z(u) \) from~\eqref{sol_trumpet} to a holomorphic function in the bulk as a function of $\omega \in \mathbb{C}$, namely
\begin{equation}\label{qw}
z(\omega) = \tan \left( \frac{\pi \omega}{\beta} \right), \quad \bar{z}(\bar \omega) = \tan \left( \frac{\pi  \bar{\omega}}{\beta} \right).
\end{equation}
Applying this coordinate transformation to the original Poincaré patch metric~\eqref{hyper}, and reintroducing time and spatial coordinates in the new frame, yields:
\begin{equation}
\mathrm{d}s^2 = - 4 \left( \frac{\pi }{\beta} \right)^2 \frac{\mathrm{d}\tau^2 + \mathrm{d}x^2}{\sinh^2 \left( \frac{\pi}{\beta} x \right)}.
\end{equation}
We recognize this metric as the standard (Euclidean) JT gravity black hole patch at inverse temperarure $\beta$. From \eqref{sol_trumpet} and \eqref{action} we can then infer the on-shell action for this geometry at finite cutoff:
\begin{equation}
\begin{split}
I_{\mathrm{class.}}^{\pm} &= \frac{\beta \phi_r}{\varepsilon^2}\left(\pm \sqrt{1+  \frac{4 \pi^2\varepsilon^2}{\beta^2} }-1\right).
\end{split}
\end{equation}
where the $\pm$ denotes the perturbative and non-perturbative branch respectively.
As a final remark, one may wonder why the solution \eqref{sol_trumpet} does not depend on $\varepsilon$. Actually, all the dependence on $\varepsilon$ is encoded into $\bar{z}$, which is determined through \eqref{rela2}:
\begin{equation}\label{z z}
\begin{split}
\bar{z}(u)&=  \tan \left(\frac{\pi u}{ \beta }\right)-\frac{2 \pi  \, \varepsilon  \text{sec} ^2\left(\frac{\pi u}{ \beta }\right)}{\pm i \beta  \sqrt{1+\frac{4 \pi^2 \varepsilon ^2}{\beta ^2}}- 2 \pi  \varepsilon  \tan \left(\frac{\pi u}{ \beta }\right)} \\
&= \frac{2 \pi  \epsilon  \pm i \beta  \sqrt{1+\frac{4 \pi ^2 \epsilon ^2}{\beta ^2}} \tan \left(\frac{\pi  u}{\beta }\right)}{\pm i \beta  \sqrt{1+\frac{4 \pi ^2 \epsilon ^2}{\beta ^2}} -2 \pi  \epsilon  \tan \left(\frac{\pi  u}{\beta }\right)}
\end{split}
\end{equation}
This solution thus encodes a finite cutoff version of the JT gravity black hole.

It is important to notice that because of the $\text{SL}(2, \mathbb{R})$ gauge symmetry of the theory, we should not consider the functional expression of $z(u)$ and $\bar{z}(u)$ at face value, but we should understand them as representative of some equivalence class. More precisely we should declare equivalent $z(u) \sim z'(u)$ if they are related through a non-singular Mobius transformation with real coefficients. Analogously for $\bar{z}(u)$. In this context what really matters are the equivalence classes $[z(u)]$ and $[\bar{z}(u)]$ since each representative would give a physically equivalent description of the theory. Following this reasoning, we observe that the expression of $\bar{z}(u)$ in \eqref{z z} is actually equivalent to 
\begin{equation}
    \bar{z}'(u) = \tan \left(\frac{\pi}{\beta}u\right),
\end{equation}
making the result consistent with \eqref{qw}.
We point out that by choosing the classical solution~\eqref{sol_trumpet} with \( z(u) \) taken to be real, we are effectively violating the condition \( z(u) = \bar{z}^{*}(u) \). Nevertheless, as we have just shown, this relation can still be viewed as an equality between equivalence classes, so the classical analysis remains consistent.

When we pass to the quantum theory, the violation of the reality condition can be reinterpreted as a contour deformation in the space of complexified fields. We begin with complex conjugate fields \( z \) and \( \bar{z} \) subject to the boundary condition~\eqref{rela2}, with the original contour defined by the reality condition. In the quantum path integral, however, we may treat \( z \) and \( \bar{z} \) as independent variables and relax the reality condition by choosing a different integration contour on which \( z(u) \) is real, as long as~\eqref{rela2} continues to hold. We expect that the value of the path integral remains unchanged under this contour deformation.

\section{The quantum theory: boundary approach}\label{quantum}
Having analyzed the JT gravity action with a finite cutoff and its classical solutions, we now turn to the quantization of the theory. Specifically, we perform a one-loop computation and compare the result with that obtained in Section~\ref{tt}.

\subsection{Integrating out $\bar{z}$}
To properly quantize the theory at finite cutoff, we must address the measure that arises from integrating out the dilaton and localizing on hyperbolic metrics. While this procedure is well understood in the infinite cutoff case~\cite{Moitra:2021uiv, Saad:2019lba}, certain subtleties arise when attempting to extend it to finite cutoff~\cite{Moitra:2021uiv}. In the following, we adopt a minimal approach by introducing only the essential ingredients required to define the measure. We remain, for now, agnostic about any additional dependence of the measure on the cutoff parameter, and will justify our choice \emph{a posteriori} by interpreting the resulting expressions.

As the starting point for our finite cutoff path integral, we propose the following expression:
\begin{equation}\label{partition_function}
Z(\beta, \phi_r; \varepsilon) = \int \mathcal{D}z(u) \, \mathcal{D}\bar{z}(u) \, \exp \left( -\frac{\phi_r}{\varepsilon^2} \int_{0}^{\beta} \left( \kappa(u) - 1 \right) \mathrm{d}u \right) \, \delta \left( g_{uu}(z(u), \bar{z}(u)) - \frac{1}{\varepsilon^2} \right),
\end{equation}
where we integrate over boundary paths, subject to the delta functional that enforces the constraint~\eqref{con}, corresponding to the boundary condition on the induced metric. We have assumed a flat measure for \( z(u) \) and \( \bar{z}(u) \).
An important requirement on \eqref{partition_function} is that it must reproduce the Schwarzian partition function in the $\varepsilon\rightarrow 0$ limit: 
\begin{equation}
\lim_{\varepsilon \rightarrow 0}Z(\beta, \phi_r; \varepsilon)=Z_{\mathrm{Sch}}(\beta, \phi_r). 
\end{equation}
Where $Z_{\mathrm{Sch}}(\beta, \phi_r)$ has the following expression:
\begin{equation}\label{partition_function2}
Z_{\mathrm{Sch}}(\beta, \phi_r)=\int \frac{\mathcal{D} \mu (z)}{SL(2,\mathbb{R})}\  \exp \left(-\phi_r \int_{0}^{\beta} \{z(u),u \}  \ \mathrm{d}u\right) , 
\end{equation}
and $\mathcal{D} \mu (z)$ is the measure that reads:
\begin{equation}\label{measure}
\mathcal{D} \mu (z)=\prod_{u} \frac{\mathcal{D} z(u)}{z'(u)}.
\end{equation}
This measure, which is invariant under reparametrizations of the Schwarzian mode, can be derived from several independent approaches and provides the correct measure for the Schwarzian path integral \cite{Stanford:2017thb,Mertens:2017mtv,Saad:2019lba}.

Given that we have already shown in the previous section that
\begin{equation}
\kappa(u)=1+\varepsilon^2 \{z(u),u \}+\mathcal{O}(\varepsilon^3) \quad \text{which implies} \quad \lim_{\varepsilon \rightarrow 0} \kappa_{\mathrm{reg}}(u)=\{z(u),u \},
\end{equation}
our goal is to make sure that the procedure of integrating out $\bar{z}$ yields the correct measure for the remaining path integral over $z$. It turns out the simplest and most suitable choice for this purpose is to implement the constraint $g_{uu}=\frac{1}{\varepsilon^2}$ as a delta functional in the path integral with the following form:
\begin{equation}\label{const}
\delta \left(G_{\varepsilon}(z,\bar{z})\right)=\delta \left(i \varepsilon z'(u)\bar{z}'(u)-\frac{(z(u)-\bar{z}(u))^2}{4 i \varepsilon}\right).
\end{equation}
As we will check soon, integrating out $\bar{z}$ through \eqref{const} will determine the measure \eqref{measure} for the Schwarzian theory in the vanishing-$\varepsilon$ limit.
To see that explicitly, we recall that in \eqref{ree} we were able to write an expression for $\bar{z}^{*}(u)$ in terms of $z(u)$ and the extrinsic curvature of the boundary $\kappa(u)$. Using that relation inside of the delta functional and utilizing one of its usual properties we can write:
\begin{equation}\label{const2}
\delta \left(G_{\varepsilon}(z,\bar{z})\right)=\frac{\delta \left(\bar{z}(u)-\bar{z}^{*}(u)\right)}{\mathrm{det}\left(\left.\frac{\delta G_{\varepsilon}}{\delta \bar{z}}\right|_{\bar{z}(u)=\bar{z}^{*}(u)}\right)}, \quad \text{where} \quad \bar{z}^{*}(u)=z(u)-\frac{2i \varepsilon z'(u)}{\kappa(u)+i\varepsilon \frac{z''(u)}{z'(u)}}.
\end{equation}
Analyzing the determinant in the denominator we see that we need the variation of $G_{\varepsilon}$ with respect to $\bar{z}$, evaluated on the solution $\bar{z}^{*}$. With simple steps we obtain:
\begin{equation}\label{determinant}
\left.\frac{\delta G_{\varepsilon}}{\delta \bar{z}}\right|_{\bar{z}(u)=\bar{z}^{*}(u)}=i \varepsilon z'(u) \frac{\mathrm{d}}{\mathrm{d}u}+\frac{z'(u)}{\kappa(u)+i\varepsilon \frac{z''(u)}{z'(u)}} 
\end{equation}
Interestingly, if we analyze the $\varepsilon \to 0$ limit and we recall that $\lim_{\varepsilon \rightarrow 0}\kappa(u)=1$, we immediately arrive at the result
\begin{equation}
\lim_{\varepsilon \rightarrow 0}\left.\frac{\delta G_{\varepsilon}}{\delta \bar{z}}\right|_{\bar{z}(u)=\bar{z}^{*}(u)}=z'(u).
\end{equation}
This is exactly what we wanted. Indeed $1/\mathrm{det}(z'(u))$ will exactly reproduce the measure \eqref{measure} once we integrate out $\bar{z}$ using \eqref{const2}. This shows that our treatement of the constraint is consistent with result without the cutoff.

With this in mind we now leave $\varepsilon$ finite and we investigate where the path integral defined above leads us. We begin by noticing that since \( z'(u) \) factorizes in equation ~\eqref{determinant}, we can rewrite the partition function \eqref{partition_function} in the following form:
\begin{equation}
Z(\beta, \phi_r; \varepsilon) = \int \frac{\mathcal{D}z(u)}{z'(u)} \ 
\exp\left( -\frac{\phi_r}{\varepsilon^2} \int_{0}^{\beta} \left( \kappa\left(z(u)\right) - 1 \right) \mathrm{d}u \right) \ 
\left[ \det\left( i \varepsilon \frac{\mathrm{d}}{\mathrm{d}u} + \frac{1}{\kappa(u) + i \varepsilon \, \frac{z''(u)}{z'(u)}} \right) \right]^{-1},
\end{equation}
where we already integrated out \( \bar{z}(u) \). To proceed further in the evaluation, inspired by the infinite cutoff case, it is sensible to perform the following field redefinition by introducing the $\eta(u)$ field:
\begin{equation}\label{redef}
z(u) = \tan \left( \frac{\pi}{\beta} \eta(u)\right).
\end{equation}
In this way the path integral that we aim to compute takes the following form:  
\begin{equation}\label{path}
Z(\beta, \phi_r; \varepsilon) = \int \frac{\mathcal{D}\eta(u)}{\eta'(u)} \ 
\exp\left( -\frac{\phi_r}{\varepsilon^2} \int_{0}^{\beta} \left( \kappa\left(\eta(u)\right) - 1 \right) \mathrm{d}u \right) \ 
\det\left( \Delta\left(\eta(u)\right) \right)^{-1},
\end{equation}
with the differential operator $\Delta$ defined as  
\begin{equation}
\Delta\left(\eta(u)\right) = i \varepsilon  \frac{\mathrm{d}}{\mathrm{d}u}+\frac{1}{\kappa(u)+i\varepsilon \left(\frac{2\pi}{\beta} \tan\left( \frac{\pi}{\beta}\eta(u)\right)  \ \eta ' (u)  +  \frac{\eta '' (u)}{\eta ' (u)}\right)} .
\end{equation}
We observe that the parametrization \eqref{redef} is chosen such that on classical solution \eqref{sol_trumpet} we have $\eta(u) = u$.

As it would be extremely challenging to exactly evaluate the path integral following this route, in the following section  we expand around the classical configuration and we obtain the result at one-loop computation in the limit $\phi_r \gg 1$. In this regime, $\phi_r$ plays the role of an inverse Planck constant ($1/\hbar$), effectively localizing the path integral around the classical saddle point. We therefore consider small fluctuations around the classical solution of the form:
\begin{equation}\label{grav}
\eta(u) = u + \phi_r^{-1/2} \, \delta\eta(u).
\end{equation}
As anticipated at the end of the previous section, we will choose a real contour of integration for $\delta \eta$.

\subsection{The one loop path integral}
In order to compute the  partition function at one loop, we first deal with the measure factors appearing in \eqref{path}.
In particular we expand the functional determinant $\det \Delta$ around the classical solution $\eta(u)=u$:
\begin{equation}\label{higher}
	\det \Delta=\det\left( i \varepsilon  \frac{\mathrm{d}}{\mathrm{d}u}+\frac{1}{\kappa_0+\frac{2\pi i \varepsilon}{\beta} \text{tan}\left( \frac{\pi u}{\beta}\right) }  \right)+\mathcal{O}\left(\phi_r^{-\frac12}\right),
\end{equation}
suppressing the higher order corrections in \eqref{higher} that do not affect the result at one-loop order. To evaluate the determinant it is now crucial to characterize the spectrum $\{\lambda_n \}$ of the operator in \eqref{higher}, defined on a $S^{1}$ of length $\beta$.
The eigenfunctions of the operator under exam are of the following form:
\begin{equation}
\Psi(u;\lambda)= \frac{ \exp \left[i \left(u \left(\kappa_0-\lambda \right)/\varepsilon-2 \arctan\left(\frac{2 \pi \varepsilon  \tan \left(\pi u/\beta\right)}{\beta \kappa_0}\right)\right)\right]}{2 \kappa_{0}^2+\cos (\pi u/\beta)-1},
\end{equation}
with $\lambda$ the corresponding real eigenvalue. Upon imposing the periodicity condition $\Psi(u+ \beta) = \Psi(u)$ we find that the spectrum discretizes and the only allowed eigenvalues become:
\begin{equation}
	\lambda_n=  \sqrt{1+ \frac{4 \pi ^2 \varepsilon ^2}{\beta ^2}}-\frac{2 \pi  \varepsilon}{\beta } n = \kappa_0 - \frac{2 \pi  \varepsilon}{\beta } n.
\end{equation}
Therefore, at one loop, the path integral that we need to compute becomes:
\begin{equation}
Z(\beta, \phi_r; \varepsilon)_{1\text{loop}}=\frac{1}{\kappa_0 \, \prod_{n=2}^{+ \infty} \left( 1 - \left(\frac{2\pi}{\beta}\right)^2 \varepsilon^2 (n^2 -1) \right) }\int \frac{\mathcal{D} \eta(u)}{\eta ' (u)}    \exp \left(-\frac{\phi_r}{\varepsilon^2} \int_{0}^{\beta} \left(\kappa(u)-1\right) \mathrm{d}u\right).
\end{equation}
To proceed further, we need to make sense of the path integral measure $\mathcal{D} \eta(u)/\eta'(u)$. As explained in \cite{Stanford_2017}, the correct approach is to identify the path integral measure with the Pfaffian of the symplectic measure associated with the Schwarzian theory
\begin{equation}
\frac{\mathcal{D} \eta(u)}{\eta'(u)} = \text{Pf}(\omega),
\end{equation}
where $\omega$ is defined as
\begin{equation}
\omega = \int_{0}^{\beta} \mathrm{d}u \left( \frac{\mathrm{d} \eta'(u) \wedge \mathrm{d} \eta''(u)}{\eta'(u)^2} - \mathrm{d} \eta(u) \wedge \mathrm{d} \eta'(u) \right).
\end{equation}
In general the Pfaffian $\text{Pf}(\omega)$ admits a useful representation in terms of fermionic variables by identifying $\mathrm{d} \eta(u) = \psi(u)$ \cite{Stanford_2017}:
\begin{equation}\label{fermion}
\text{Pf}(\omega) = \int \mathcal{D} \psi \ 
\exp \left( -\frac{1}{2} \int_0^\beta \mathrm{d}u \left( \frac{\psi'' \psi'}{\psi'^2} - \psi' \psi \right) \right),
\end{equation}
For our purposes it is sufficient to expand around the classical configuration $\psi(u) = u$, and by performing the functional integral \eqref{fermion} decomposing the fluctuations into Fourier modes, we obtain:
\begin{equation}
\text{Pf}(\omega) = \prod_{n=2}^{\infty} 2\pi \left( \frac{2\pi}{\beta} \right)^2 n(n^2 - 1),
\end{equation}
where the product excludes the $n=0,1$ modes, corresponding to the gauge redundancy associated with the $\text{PSL}(2,\mathbb{R})$ symmetry.

Having discussed the measure, we now focus on expanding the action up to second order in the fluctuation of the fields. In order to do that we need to find a way of expanding the extrinsic curvature.  The sensible way of doing it is to recall that the extrinsic curvature is bound to obey the following Riccati equation
\begin{equation}\label{ri}
2 \varepsilon \kappa'(\eta(u)) - i \kappa^2(\eta(u)) + 2 \varepsilon i \left( \varepsilon \{z(\eta(u)), u \} + \frac{1}{2 \varepsilon} \right) =0.
\end{equation}
The strategy is then to formally expand the extrinsic curvature and the Schwarzian derivative $\{z(\eta(u)),u \}= \{z(u + \phi_r^{-\frac{1}{2}}\delta \eta(u)),u \}$ in powers of $\phi_{r}^{-\frac{1}{2}}$. In detail, for the Schwarzian derivative, recalling that $z(u) = \tan \left( \frac{\pi}{\beta} u \right)$ we obtain the expansion:
\begin{equation}\label{Sc}
\begin{split}
\{z(u + \phi_r^{-\frac{1}{2}}\delta \eta(u)),u \}=&\frac{2 \pi^2}{\beta^2}+\phi_{r}^{-1/2}\left( \frac{4 \pi^2}{\beta^2} \delta \eta'(u)+\delta \eta^{(3)}(u)\right)\\
+&\phi_{r}^{-1}\left(\frac{2 \pi^2}{\beta^2} \delta \eta'(u)^2-\frac32 \delta \eta''(u)^2- \delta \eta ^{(3)}(u) \delta \eta '(u)\right)+\mathcal{O}\left(\phi_r^{-\frac32}\right),
\end{split}
\end{equation}
and for the extrinsic curvature $\kappa\left(\eta(u)\right)$ we use the expression
\begin{equation}\label{k}
\kappa\left(\eta(u)\right)=\kappa_{0}(\delta \eta(u))+\phi_r^{-1/2} \kappa_1(\delta \eta(u))+\phi_r^{-1} \kappa_2(\delta \eta(u))+\mathcal{O}\left(\phi_r^{-\frac32}\right).
\end{equation}
By inserting \eqref{Sc} and \eqref{k} in \eqref{ri} we now solve perturbatively in $\phi_{r}^{- \frac{1}{2}}$ the Riccati equation, and we find up to second order the equalities
\begin{equation}\label{system}
\begin{split}
\kappa_0^{2}=&1+ \frac{4 \pi^2}{\beta^2} \varepsilon^2, \\
2 i \varepsilon \kappa_1'+2 \kappa_0 \kappa_1=& 2\varepsilon^2 \left( \frac{4 \pi^2}{\beta^2} \delta \eta'+\delta \eta^{(3)}\right), \\
2 i \varepsilon \kappa_2'+2 \kappa_0 \kappa_2+\kappa_1^2=& 2\varepsilon^2 \left( \frac{2 \pi^2}{\beta^2} \delta \eta'^2-\frac32 \delta \eta ''^2- \delta \eta ^{(3)} \delta \eta '\right).
\end{split}
\end{equation}
 To proceed further it is now sensible to Fourier expand along the thermal circle so that we can convert the differential equations in \eqref{system} in algebraic ones. Explicitly, we write
\begin{equation}
\delta \eta(u)=\sum_{n=-\infty}^{+\infty} e^{-i \frac{2 \pi}{\beta} n u} \delta \eta_{n}, \qquad \kappa_i (u)=\sum_{n=-\infty}^{+\infty} e^{-i \frac{2 \pi}{\beta} n u} \kappa_{i,n} \quad \text{for $i=0,1,2$}.
\end{equation}
We can now solve the equations in \eqref{system} one after the other. In particular the second equation yields the following value for the coefficient $\kappa_{1,n}$:
\begin{equation}
\kappa_{1,n}=\frac{i \left( \frac{2 \pi}{\beta} \right)^3 n \varepsilon^2\left(n^2-1\right)}{\frac{2 \pi}{\beta} n \varepsilon+\kappa_0} \ \delta \eta_{n},
\end{equation}
that we can use to solve for $\kappa_{2, n}$. In hindsight, we will only need to know the expression of the coefficient $\kappa_{2, 0}$, therefore we specialize the third equation in \eqref{system} to only include the zero mode and we obtain the following expression
\begin{equation}\label{kappa2}
\kappa_{2,0}=-\sum_{n=-\infty}^{+\infty} \frac{\left( \frac{2 \pi}{\beta} \right)^4 \varepsilon^2 n^2 \left(n^2-1\right)}{2 \kappa_0\left(\kappa_{0}^2-\left( \frac{2 \pi}{\beta} \right)^2 n^2 \varepsilon^2\right)} \ \delta \eta_{-n} \delta \eta_{n}.
\end{equation}
Using \eqref{kappa2}, we can now write down the form of the action of the theory expanded up to the second order in the fluctuation on the fields, $I^{\pm}[\delta \eta;\phi_{r}]= I^{\pm}_0+ I^{\pm}_2[\delta \eta;\phi_{r}]$:
\begin{equation}\label{qq}
\begin{split}
&I^{\pm}_0=\frac{\phi_r \beta}{  \varepsilon^2}\left(\pm\sqrt{1+\left( \frac{2 \pi}{\beta} \right)^2 \varepsilon ^2}-1\right),\\
&I^{\pm}_2[\delta \eta;\phi_{r}] = \mp\frac{ 2\pi \left( \frac{2 \pi}{\beta} \right)^3 }{\sqrt{1+\left( \frac{2 \pi}{\beta} \right)^2 \varepsilon ^2} } \sum_{n=2}^{+\infty} \ \frac{n^2 \left(n^2-1\right)}{1-\left( \frac{2 \pi}{\beta} \right)^2 \varepsilon ^2\left(n^2-1\right)} \left|\delta \eta_{n}\right|^2.
\end{split}\end{equation}
The plus or minus signs correspond to the two different the signs of $\kappa_0$ which are determined by this procedure, as both the positive and the negative sign are on equal footing. This is reminiscent of the fact that there exist two admissible values for the extrinsic curvature corresponding to the classical solution of the theory. 

With all the necessary ingredients we now detail the procedure for determining the one-loop correction to the partition function in the case of positive $\kappa_0$. Analogous results apply for the other case.

Using the correct path integral measure, and including the contribution coming from the determinant \eqref{higher}, the expression takes the form
\begin{equation}
\begin{split}
Z(\beta, \phi_r; \varepsilon)_{1\text{loop}}^{+}&= \frac{e^{I_0^+}}{\kappa_0}\prod_{n=2}^{\infty}\frac{1}{\phi_r}\frac{ 2 \pi \left(\frac{2 \pi}{\beta}\right)^2 n (n^2-1) }{  1 - \left(\frac{2 \pi}{\beta}\right)^2 \varepsilon^2 (n^2 -1) }\int_{- \infty}^{+ \infty} \mathrm{d} \mathrm{Re} \delta \eta_n \ \mathrm{d} \mathrm{Im} \delta \eta_n \exp \left( -I^{+}_2[\delta \eta; \phi_r]\right)
\end{split}
\end{equation}
At face value some of the integrals above are divergent, as they involve gaussian integrals with the wrong sign. In particular we notice that as long as $n> s^* = \left( 1 + \left( \frac{\beta}{2 \pi \varepsilon}  \right)^2 \right)^{1/2} $ the integrals are well defined, therefore we assume we first compute the amplitude for a value of $\varepsilon$ such that all modes with $n \geq 2$ yield convergent integrals, and subsequently perform an analytic continuation in $\varepsilon$. Proceeding in this fashion we obtain the regularized result:
\begin{equation}
\begin{split}
  Z(\beta, \phi_r; \varepsilon)_{1\text{loop}}^{+} &= \frac{e^{I_0^+}}{\kappa_0}\prod_{n=2}^{\infty} \frac{1}{\phi_r} \frac{ 2 \pi \left(\frac{2 \pi}{\beta}\right)^2 n (n^2-1) }{  1 - \left(\frac{2 \pi}{\beta}\right)^2 \varepsilon^2 (n^2 -1) } \prod_{n=2}^{+\infty} \frac{\sqrt{1+\left(\frac{2 \pi}{\beta}\right)^2 \varepsilon ^2} }{2\pi \left(\frac{2 \pi}{\beta}\right)^3} \frac{1-\left(\frac{2 \pi}{\beta}\right)^2 \varepsilon ^2\left(n^2-1\right)}{n^2 \left(n^2-1\right)} \\
  &=  \frac{\left( \frac{2 \pi \phi_r}{\beta}\right)^{\frac{3}{2}}}{\sqrt{2 \pi} \left(1+ \left(\frac{2 \pi}{\beta}\right)^2 \varepsilon^2\right)^{\frac{5}{4}}} e^{I_0^+},
\end{split}
\end{equation}
where in the second step we used the following identities valid in the sense of the $\zeta$ function regularization:
\begin{equation}
\prod_{n=2}^{\infty} \frac{1}{n}= \frac{1}{\sqrt{2 \pi}}, \qquad \prod_{n=2}^{\infty} c = c^{-\frac{3}{2}}.
\end{equation}

As we have previously mentioned, the result that we have just obtained is valid in the case of the expanding branch partition function, an analogous result can be also obtained in the case of the contracting branch. Indeed, performing the same steps as before, we obtain
\begin{equation}\label{unstable}
\begin{split}
Z(\beta, \phi_r; \varepsilon)_{1-\text{loop}}^{-}
= i \frac{\left( \frac{2 \pi \phi_r}{\beta}\right)^{\frac{3}{2}}}
{\sqrt{2 \pi} \left(1 + \left(\frac{2 \pi}{\beta}\right)^2 \varepsilon^2\right)^{\frac{5}{4}}}
e^{I_0^-}.
\end{split}
\end{equation}
The prefactor of $i$ in \eqref{unstable} arises from the regularization of the infinite product
\begin{equation}\label{unstable_prod}
\begin{split}
\prod_{n=2}^{\infty} (-1),
\end{split}
\end{equation}
which originates from the Wick rotation of both $\mathrm{Re}\delta\eta_n$ and $\mathrm{Im}\delta\eta_n$ required to render the Gaussian integrals convergent.
Comparison of the contribution of $Z_{\mathrm{1loop}}^{+}$ and $Z_{\mathrm{1loop}}^{-}$ reveals that the main difference in the one-loop correction between the two branches comes from a relative phase. This in turn indicates that the complete result for the partition function at this order has the following structure:
\begin{equation}
    \boxed{Z(\beta, \phi_r; \varepsilon)_{1\text{loop}}=\frac{(\phi_r \frac{2 \pi}{\beta})^{\frac{3}{2}}}{\sqrt{2 \pi} \left(1+ \left(\frac{2 \pi}{\beta}\right)^2 \varepsilon^2\right)^{\frac{5}{4}}} \left( e^{-I_0^{+}} +  ie^{-I_0^{-}}\right)}.
\end{equation}
Interestingly, we emphasize that this result captures only the leading term in the $\hbar = 1/\phi_r$ expansion, while remaining exact to all orders in the cutoff $\varepsilon$.

\section{One loop vs exact results}\label{onevs}
Before comparing with previous results, we first perform an internal consistency check between the two main results of this paper:
\begin{equation}\label{sum}
\begin{split}
Z_{\mathrm{disk}}(L,\phi) &\sim 
\frac{2 i \pi \phi \, I_2\!\left(\sqrt{L^2+4 \pi ^2}\, \phi \right)}{L^2+4 \pi ^2} \\ 
Z(\beta, \phi_r; \varepsilon)_{1-\text{loop}}
&= \frac{\left(\phi_r \frac{2 \pi}{\beta}\right)^{\frac{3}{2}}}{\sqrt{2 \pi}\,\left(1 + \left(\frac{2 \pi}{\beta}\right)^2 \varepsilon^2\right)^{\frac{5}{4}}}
\left( e^{-I_0^{+}} - i\, e^{-I_0^{-}} \right).
\end{split}
\end{equation}
with $I_0^{\pm}$ given in \eqref{summary}.
Under the usual mapping
\begin{equation}
\phi = \frac{\phi_r}{\varepsilon}, 
\qquad 
L = \frac{\beta}{\varepsilon},
\end{equation}
and performing the asymptotic expansion for large $\phi_r$, the semiclassical expansion parameter used in Section~\ref{quantum}, 
\begin{equation}
I_2(z) \simeq \frac{1}{\sqrt{2 \pi z}} \left( e^{z} + i\, e^{-z} \right),
\qquad 
z \rightarrow \infty,
\end{equation}
of the first exact result, we find agreement between the two expressions at this order.  
The match is up to an overall factor of $L\,\phi$ multiplying $Z_{\mathrm{disk}}(L,\phi)$, which would produce perfect agreement between the results.\footnote{%
$e^{I_0^{\pm}}$ contains an additional factor $e^{-\frac{\phi_r \beta}{\varepsilon^2}}$, originating from a boundary counterterm that we did not include in Section~\ref{tt}. This factor simply corresponds to an extra $e^{-L \phi}$ contribution.
}  
As discussed in Section~\ref{no}, these factors can be naturally attributed to different operator orderings in the WdW equation and therefore, within the path integral formulation of Section~\ref{tt}, we argue that they are encoded in different choices for the path integral measure.

The semiclassical and one-loop calculation carried out in Section \ref{quantum} can be viewed as the finite-\(\varepsilon\) generalization of the computation performed in \cite{Stanford:2017thb} for the ordinary Schwarzian theory. However, while in \cite{Stanford:2017thb} the partition function is one-loop exact due to a localization argument, here we expect the result to admit a full asymptotic expansion in \(1/\phi_r\), the boundary gravitational coupling constant, reflecting the nontrivial structure of the asymptotic series. It would be interesting to compute, along the lines of Section \ref{quantum}, higher-loop corrections to the finite-\(\varepsilon\) Schwarzian path integral and check whether the subleading terms in the asymptotic expansion of \(I_{2}(z)\) are reproduced.

From the integral representation \eqref{instanton2}, we observe that the asymptotic series is naturally organized in powers of \(\delta \mathcal{N}/\sqrt{\phi}\). This suggests that, from the bulk perspective, it is the integration over \(\delta \mathcal{N}\) that generates the full boundary Schwarzian perturbation theory. We regard this as a nontrivial manifestation of the bulk/boundary duality in this theory. Indeed, it is straightforward to verify that, in the gauge \eqref{gauge} where the boundary is fixed at \(r = 1\), no Schwarzian dynamics can arise on the boundary in the first bulk path integral approach. Rather than integrating over boundary reparametrizations, this approach integrates over bulk fluctuations, while boundary fluctuations remain frozen. By contrast, in the computation of Section \ref{quantum}, the bulk dynamics is frozen and all nontrivial dynamics is encoded in the boundary mode. This explains why the two computations are genuinely independent, and why their agreement provides a nontrivial consistency check.
A remarkable feature of both approaches is the presence of two instantons with a relative phase of \(i\), which precisely accounts for the instability of the nonperturbative saddle in \(\varepsilon\).



\subsection{Comparison with earlier work}\label{compa}
\paragraph{Semiclassical analysis} In order to compare our results with existing semiclassical studies on this topic, let us first take a closer look at the leading and subleading terms in the semiclassical expansion of the boundary Schwarzian action at finite cutoff derived in Section~\ref{quantum}, which we reproduce here:
\begin{equation}\label{summary}
\begin{split}
&I^{\pm}_0=\frac{\phi_r \beta}{  \varepsilon^2}\left(\pm\sqrt{1+\left( \frac{2 \pi}{\beta} \right)^2 \varepsilon ^2}-1\right),\\
&I^{\pm}_2[\delta \eta;\phi_{r}] = \mp\frac{ 2\pi \left( \frac{2 \pi}{\beta} \right)^3 }{\sqrt{1+\left( \frac{2 \pi}{\beta} \right)^2 \varepsilon ^2} } \sum_{n=2}^{+\infty} \ \frac{n^2 \left(n^2-1\right)}{1-\left( \frac{2 \pi}{\beta} \right)^2 \varepsilon ^2\left(n^2-1\right)} \left|\delta \eta_{n}\right|^2.
\end{split}\end{equation}

We first focus on the leading on-shell action $I^{\pm}_0$ and compare it with the $T\bar{T}$ flow of the Schwarzian theory. This can be achieved by performing a dimensional reduction of the flow of two-dimensional CFTs under the $T\bar{T}$ operator \cite{Gross:2019ach}. Specifically, one can write a flow equation for the energy eigenvalues
\begin{equation}\label{flow}
\begin{split}
\frac{\partial H}{\partial \lambda}=\frac{H^2}{\frac12-2\lambda H_{0}}, 
\end{split}
\end{equation}
where $H_{0}$ is the undeformed Hamiltonian, and the $T\bar{T}$ deformation parameter $\lambda$ is related to the bulk cutoff in our conventions as $\lambda=\frac{\varepsilon^2}{4 \phi_r}$. This equation admits two solutions, referred to as the perturbative and nonperturbative branches of the $T\bar{T}$ spectrum, namely
\begin{equation}\label{HTT}
\begin{split}
H_{\mathrm{T\bar{T}}}^{\pm}=\frac{\phi_r}{\varepsilon^2}\left(1\mp\sqrt{1-\frac{2 \varepsilon^2}{\phi_r} H_{0}}\right),
\end{split}
\end{equation}
where the perturbative branch is the one smoothly connected to the undeformed theory in the $\lambda \rightarrow 0$ limit. By performing a Legendre transform of the perturbative branch of the $T\bar{T}$-deformed Hamiltonian, a $T\bar{T}$-deformed version of the Schwarzian theory was obtained in \cite{Gross:2019ach}, which in our conventions reads
\begin{equation}\label{TT}
\begin{split}
S_{\mathrm{T\bar{T}}}=\phi_r \int_{0}^{\beta} \mathrm{d} u\left[-\frac{\left(\tau'-e^{\rho}\right)^2}{2 \tau' e^{\rho} \varepsilon^2}+\frac{e^{\rho}}{2 \tau'}\left(\rho'^2-\left(\frac{2\pi}{\beta}\right)^2 \tau'^2 \right)\right],
\end{split}
\end{equation}
expressed in terms of the original reparametrization mode $\tau(u)$ and a new radial degree of freedom $\rho(u)$.\footnote{Sending $\varepsilon$ to zero amounts to fixing $e^{\rho}=\tau'$, i.e.\ freezing the radial degree of freedom. Plugging this condition back into the action, we find $\lim_{\varepsilon\rightarrow 0}L_{\varepsilon}=-\phi_r \left(-\frac12 \left(\frac{\tau''}{\tau'}\right)^2+\frac12(\frac{2\pi}{\beta})^2 \tau'^2\right)$, which corresponds to the Schwarzian action up to a total derivative term.}

Assuming that $\tau_0(u)=u$ remains a solution for the deformed theory \cite{Gross:2019ach}, varying with respect to $\rho(u)$ gives the classical equation of motion
\begin{equation}\label{rho}
\begin{split}
2 \varepsilon^2\left(\rho''+\frac12 \rho'^2\right)+1+\varepsilon^2 \left(\frac{2\pi}{\beta}\right)^2=e^{-2\rho} \qquad \Longrightarrow e^{-\rho_{0}^{\pm}}=\pm\left(1+\left(\frac{2\pi}{\beta}\right)^2 \varepsilon^2\right)^{1/2},
\end{split}
\end{equation}
where in the second step we have assumed $\rho_{0}^{'}(u)=\rho_{0}^{''}(u)=0$. Evaluating the action \eqref{TT} on these solutions yields an exact match with the on-shell bulk action \eqref{summary}, namely
\begin{equation}\label{match}
\begin{split}
I^{\pm}_0=S_{\mathrm{T\bar{T}}}\left[\tau_0,\rho_{0}^{\pm}\right].
\end{split}
\end{equation}
The on-shell equivalence between finite cutoff JT gravity and the $T\bar{T}$-deformed boundary theory was already shown in \cite{Gross:2019ach} for the first orders in the perturbative expansion in $\varepsilon$. The match \eqref{match} holds for arbitrary values of $\varepsilon$, and moreover, we see explicitly from the classical bulk solutions of the Riccati equation the emergence of both the perturbative and nonperturbative branches of the $T\bar{T}$-deformed spectrum. Although a full quantum identification between the two theories has been suggested, and we will return to this point shortly, we first pause to check whether our one-loop analysis in Section~\ref{quantum} remains consistent with the $T\bar{T}$ flow.
 
Expanding around the background configuration $(\tau_0, \rho_0)$, the $T\bar{T}$-deformed Schwarzian action \eqref{TT} at quadratic order in the fluctuations reads
\begin{equation}\label{quadratic}
\begin{split}
S_{\mathrm{T\bar{T}}}^{(2)}=\frac{1}{2 \varepsilon ^2 \sqrt{1+\left(\frac{2\pi}{\beta}\right)^2 \varepsilon ^2}} 
\int_{0}^{\beta} \mathrm{d}u \,
\left[\varepsilon ^2 \left(\delta \rho '(u)^2-\left(\frac{2\pi}{\beta}\right)^2 \delta \rho (u)^2\right)
-\left(\delta \tau '(u)- \delta \rho (u)\right)^2 \right].
\end{split}
\end{equation}
Remarkably, when expanding \eqref{quadratic} in Fourier modes on the thermal circle \cite{Gross:2019ach}, the resulting expression takes precisely the same form as $I^{\pm}_2[\delta \eta;\phi_{r}]$ in \eqref{summary} (see in particular Eqs.~(4.18) and (4.19) of \cite{Gross:2019ach}). This is a significant observation, as our quadratic action $I^{\pm}_2[\delta \eta;\phi_{r}]$ was derived purely from a bulk geometric computation, and this comparison indicates that the correspondence extends beyond the semiclassical regime.
The precise agreement provides strong evidence for the identification of finite cutoff JT gravity with the \(T\bar{T}\) deformation of the Schwarzian theory, serving as a non-trivial bulk/boundary consistency check.

Before commenting on the full quantum extension of our results, some comments are in order.
\begin{itemize}
\item In Section~\ref{cla}, we identified the solutions with constant Schwarzian, $\{z(u),u\}=\alpha$, as those producing the on-shell action $I^{\pm}_0$. From the previous analysis, these solutions correspond to the $(\tau_0, \rho_0)$ configurations on the $T\bar{T}$ side. However, as discussed in Section~\ref{cla}, more general solutions to \eqref{finiteEOM} appear to be possible, which could exhibit non-vanishing derivatives of the Schwarzian and thereby modify the value of $I^{\pm}_0$. Similarly, on the $T\bar{T}$ side, more general solutions to \eqref{rho} may exist.\footnote{It is noteworthy that, once one sets $f'(u)=e^{-\rho(u)}$, the equation takes the suggestive form
$$2 \varepsilon^2\left\{f(u),u\right\}=1+\left(\frac{2\pi \varepsilon}{\beta}\right)^2-f'(u)^2,$$
which bears a striking resemblance to the Riccati equation.} The role of these more general solutions, both in the bulk and on the $T\bar{T}$ side, as well as their potential correspondence, is left for future investigation.
\item It is important to emphasize that if the action terms involving derivatives of the Schwarzian in \eqref{kn} were neglected, the quadratic action \(I^{\pm}_2[\delta \eta;\phi_{r}]\) would take a different form than that shown in \eqref{summary}. This can be verified explicitly by expanding the various terms in the perturbative \(\varepsilon\) expansion of \eqref{kn} to quadratic order in the quantum fluctuations. Hence, our results suggest that the derivatives of the Schwarzian play an essential role in contributing to the first quantum correction of the bulk action. 

However, as we explain below, one of the bulk approaches to obtaining the full partition function of finite cutoff JT gravity relies on the argument that derivatives of the Schwarzian do not contribute to the boundary gravitational path integral \cite{Iliesiu:2020zld}. To reconcile these statements, we conjecture that possible contributions of derivatives of the Schwarzian in the semiclassical expansion of the boundary action, which are manifest in our analysis and consistent with $T\bar{T}$ at one-loop order, may be precisely canceled by measure factors in the path integral at higher loops. Nonetheless, further investigation is required.
 
\item While our results naturally align with the \(T\bar{T}\) ones, other relevant frameworks for comparison include those considered in \cite{Stanford:2020qhm,Chaudhuri:2024yau}. We did not find any illuminating match; however, we would like to point out an observation that appears interesting to us. In \cite{Stanford:2020qhm}, an intermediate-size regime was considered for loops of the boundary curve in the bulk, where an entropic effective force was included, leading to an effective action for the system. The quadratic action (equation 5.21 of \cite{Stanford:2020qhm}) takes the form
\begin{equation}
\begin{split}
I_{\mathrm{eff}} = I_{0} + \frac{\sinh^2(\rho_0)}{\cosh(\rho_0)} \int \mathrm{d}u \left[ \frac{\delta \rho'^2 - \delta \rho^2}{\sinh^2(\rho_0)} + 3 \left( \coth(\rho_0)\, \delta \rho + \delta \tau' \right)^2 \right] ,
\end{split}
\end{equation}
which, under the identification
\begin{equation}
\sinh(\rho_0) \simeq \frac{\beta}{2 \pi \varepsilon},
\end{equation}
shows a structure quite similar to the $T\bar{T}$ quadratic action \eqref{quadratic}. Clearly, further investigation is required.

\end{itemize}
\paragraph{Exact result}
We now turn to compare the exact result \eqref{sum}, which we computed using the bulk JT path integral in Section \ref{tt}, with existing exact proposals for the partition function of finite cutoff JT gravity. By computing the Laplace transform of the Schwarzian theory spectral density, weighted by the perturbative branch of the \(T\bar{T}\)-deformed spectrum, the full \(T\bar{T}\)-deformed partition function for the Schwarzian theory was obtained in \cite{Gross:2019ach}, yielding
\begin{equation}\label{TTexact}
Z(\beta) = \int_{0}^{\infty} \mathrm{d}E \, \phi_r \sinh(2 \pi \sqrt{2 \phi_r E}) \, e^{-\beta H_{\mathrm{TT}}^{+}} 
= \frac{\beta \phi_r^{2} e^{-\frac{\beta \phi_r}{\varepsilon^2}}}{\varepsilon (\beta^2 + 4 \pi^2 \varepsilon^2)} K_2\Bigg(-\frac{\phi_r}{\varepsilon^2} \sqrt{\beta^2 + 4 \pi^2 \varepsilon^2} \Bigg).
\end{equation}
From the gravity side, the same result for the full quantum partition function of JT gravity at finite cutoff was obtained in \cite{Iliesiu:2020zld} using two distinct methods. Both methods led to the integral representation on the left-hand side of \eqref{TTexact}. In the boundary path integral approach, the precise structure \eqref{TTexact} was reproduced by treating the finite-cutoff action as an operator insertion inside the infinite-cutoff path integral, and by arguing that the expectation value of derivatives of the Schwarzian vanishes identically. We have already commented on the role of these derivatives above.  
The second approach interprets the \(T\bar{T}\)-deformed Boltzmann factor in \eqref{TTexact} as arising from an energy-dependent basis of solutions to the Wheeler–DeWitt equation. The Hartle-Hawking wavefunction is then constructed as a superposition of these solutions, where the weight $\rho(E)$, found to be given by the Schwarzian spectral density, is isolated by taking the large $L$ (or small \(\varepsilon\)) limit, in which the Boltzmann factor reduces to the undeformed one.\footnote{Related relevant works that studied and found basis of solutions of the JT gravity Wheeler–DeWitt equation in different contexts and purposes are for instance \cite{Fanaras:2021awm}. }

An unsatisfactory aspect of the partition function \eqref{TTexact} is that for \(E = H_{0} > \phi_r / 2\varepsilon^2\), the spectrum becomes complex, implying a loss of unitarity in the theory. To address this issue, the partition function on the right-hand side of \eqref{TTexact} is typically evaluated after performing the analytic continuation \(\varepsilon \rightarrow i \varepsilon\), where no complex energies appear, and is then continued back. 
A more formal treatment of the result \eqref{TTexact} was implemented in \cite{Griguolo:2021wgy}, where the integral representation was regarded as a formally ill-defined expression encoding its perturbative expansion in \(\varepsilon\). By applying resurgence theory, the resulting asymptotic series was Borel resummed, leading to a unique nonperturbative completion of \eqref{TTexact}, given by\footnote{A nonperturbative completion of the form \eqref{after_resurgence} was previously proposed in \cite{Iliesiu:2020zld} based on the requirement of reality, although its uniqueness was not established.
}
\begin{equation}\label{after_resurgence}
Z_{\mathrm{nonpert.}}(\beta) = \frac{\beta \phi_r^{2} e^{-\frac{\beta \phi_r}{\varepsilon^2}}}{\varepsilon \left(\beta^2 + 4 \pi^2 \varepsilon^2\right)} 
I_{2}\!\left(\frac{\phi_r}{\varepsilon^2} \sqrt{\beta^2 + 4 \pi^2 \varepsilon^2}\right).
\end{equation}
This nonperturbative completion precisely accounts for the resurgence of the nonperturbative branch of the $T\bar{T}$-spectrum \cite{Griguolo:2021wgy}, which does not appear in the original starting expression \eqref{TTexact}.

We note that the bulk path integral computation presented in this paper, up to the overall factors of \(L \phi\) already discussed, precisely matches the result of the disk partition function \eqref{after_resurgence} obtained via resurgence.
Although the final result \eqref{after_resurgence} was already reported in the literature, the existence of other, differing results in the subject suggested that certain aspects remained unclear. The advantage of the present approach lies in providing two completely independent derivations of the same result, one from the bulk and one from the boundary, thereby strengthening the connection with the \(T\bar{T}\) deformation and placing both the correspondence and the result \eqref{after_resurgence} on much firmer ground.
The nonperturbative branch naturally emerges as related to one of the two instantons appearing in both the bulk and boundary formulations discussed in this paper and, importantly, it is automatically included in our computation, without the need for any ad hoc reinstatement at the end of the analysis.

Furthermore, a clear advantage of our approach, specifically the one proposed in Section \ref{tt}, is that we impose the Hartle–Hawking condition directly in the limit \(L \rightarrow 0\), by requiring that the geodesic defect of the trumpet spacetime becomes regular as it shrinks to zero size, as discussed in Section \ref{no}. This contrasts with the more indirect implementation of the no-boundary condition through the \(L \rightarrow +\infty\) Schwarzian limit.

In the next section, leveraging on the lessons learned from JT gravity, we begin the investigation of a general dilaton gravity at finite cutoff.

\section{A general dilaton gravity at finite cutoff}\label{gene}
Having discussed JT gravity at finite cutoff and its relation to the $T\bar T$ deformation of the Schwarzian theory, it is natural to ask whether similar statements can be made for a more general class of dilaton gravities with arbitrary potential $V(\phi)$, as already observed for JT-like dilaton potentials in \cite{Aguilar-Gutierrez:2024nst}. The analysis presented below should not be regarded as conclusive, but it does provide several indications that a finite cutoff in the bulk gravity theory is closely related to a $T\bar T$--type deformation of the putative dual boundary theory\footnote{With "$T\bar T$--type" we mean that the putative Hamiltonian of the theory obeys a differential equation of the form of \eqref{flow}. }.
One of the hallmark signatures of a $T\bar T$ deformation is the specific relation between the deformed Hamiltonian and its undeformed counterpart \eqref{HTT}. As a first step, we would like to derive the structure of the Hamiltonian from a gravitational computation at finite cutoff. This analysis is very similar in spirit to the one of \cite{McGough:2016lol}, where the energy of the BTZ black hole in the presence of a finite cutoff Dirichlet wall was matched with the energy of the dual CFT under the  $T\bar{T}$ flow.

In our case, one possible approach is to analyze a general dilaton gravity model at the classical level and compute the on-shell action of the theory without taking the asymptotic boundary limit. By identifying the on-shell action with the free energy, one can then deduce the quasi-local energy \(E_{\varepsilon}\) \cite{Aguilar-Gutierrez:2024nst}.
Indeed the entropy term in the on-shell action via the Bekenstein--Hawking formula reads $S_E \propto \phi_h$, where $\phi_h$ is the value of the dilaton at the black hole horizon, and the remaining part of the expression should be interpreted as $\text{"length of thermal circle "} \times E_{\varepsilon}$. The action that we are referring to including the appropriate boundary terms introduced in \cite{Belaey:2025kiu} is the following:
\begin{equation}\label{dila1}
    S_E = -\frac{1}{2} \int_{\mathcal{M}} \mathrm{d}^2 x \sqrt{g}\,\bigl(\Phi R + V(\Phi)\bigr)
      - \oint_{\partial \mathcal{M}} \mathrm{d}\tau \sqrt{h}\left(
        \Phi K - \sqrt{\int_{c}^{\Phi_{\partial}} V(\Phi)\, d\Phi}
      \right),
\end{equation}
and in the gauge where the dilaton parametrizes the radial coordinate its classical solutions read \cite{Witten:2020ert}:
\begin{equation}\label{dila}
    \mathrm{d}s^2_{\text{cl}} = \left(W(r)-W(r_h)\right) \mathrm{d} \tau^2 + \frac{\mathrm{d}r^2}{W(r)-W(r_h)} , \qquad \phi_{\text{cl}} = r,
\end{equation}
where we defined $W(y) = \int V(y) dy$ to be the prepotential. The following thermodynamic properties of a general dilaton gravity black hole \eqref{dila} can then be easily derived \cite{Mertens:2022irh}:
\begin{equation}\label{termo}
\beta_{\mathrm{BH}}= \frac{4\pi}{\lvert V(\phi_h) \rvert}
\qquad 
S = 2\pi \phi_h
\qquad 
E_{\mathrm{ADM}} = \frac12 W(\phi_h)
\end{equation}
where the Hawking inverse temperature follows from demanding Euclidean smoothness of \eqref{dila} as we expand around the black hole horizon $r = \phi_h$, the entropy is given by the usual Bekenstein Hawking formula\footnote{$S \propto \phi_h$ is the two dimensional version of the area law, since the dilaton parametrizes the size of the sphere of the higher dimensional black hole after dimensional reduction.}, and the ADM energy is obtained by requiring that the first law of black hole thermodynamics $\beta_{\mathrm{BH}} = \frac{\mathrm{d}S}{\mathrm{d}E_{\mathrm{ADM}}}$ be satisfied.

The on-shell action \eqref{dila1} on the solutions \eqref{dila} can be computed in terms of the boundary length $L=\int \mathrm{d}\tau \sqrt{h}=\int \mathrm{d}\tau \sqrt{W(\phi_b)-W(\phi_h)}$ and yields:\footnote{The computation is carried out in detail in Appendix B1 of \cite{Belaey:2025kiu} following \cite{Witten:2020ert} and uses that the scalar curvature on the solutions \eqref{dila} is $R(r)=-W''(r)=-V'(r)$ and the extrinsic curvature is $\kappa(r)=\frac12 \frac{W'(r)}{\left(W(r)-W(r_h)\right)^{1/2}}$.}
\begin{equation}\label{dila2}
S_E^{\mathrm{on-shell}}=-2\pi \phi_h-L \left( \sqrt{W(\phi_b)-W(\phi_h)}- \sqrt{W(\phi_b)} \right).
\end{equation}
Under the identification 
\begin{equation}\label{ide}
W(\phi_b)=1/\varepsilon^2 \qquad L = \frac{\beta}{\varepsilon}
\end{equation}
we can then read off from \eqref{dila2} the finite cutoff energy of the theory:
\begin{equation}\label{adm_finite}
E_{\varepsilon}= \frac{1}{\varepsilon^2}\left(1-\sqrt{1-\varepsilon^2 W(\phi_h)}\right)=H_\varepsilon(\phi_h).
\end{equation}
In the standard infinite cutoff framework, one usually identifies a location $\phi_b$ where $W(\phi_b)\rightarrow +\infty$, i.e. the metric blows up as we approach the asymptotic boundary $r\rightarrow \phi_b$. In this limit, the gravitational energy is given by the $\varepsilon \rightarrow 0$ limit of \eqref{adm_finite} and yields:
\begin{equation}
  E_{\varepsilon\to 0}(\phi_h)
  \;=\; \frac{1}{2}\,W(\phi_h)
  \;\equiv\; H_0(\phi_h)\,.
\end{equation}
In this regime, the quasi-local energy corresponds to the background energy of the dilaton gravity black hole, the ADM energy we identified in \eqref{termo}. 
If we instead we keep $\varepsilon$ finite, $E_{\varepsilon}$ in \eqref{adm_finite} is the finite cutoff energy we need to consider. Specializing to JT gravity, one has
$W(\phi_b) = \phi_b^2=1/\varepsilon^2$: the mapping \eqref{ide} is thus the generalization to generic $V(\phi)$ of the bulk-boundary identification performed in \cite{Iliesiu:2020zld}.\footnote{Notice that $\exp (-\beta E^{\varepsilon}_{\mathrm{ADM}})$ is precisely the expanding branch solution of the WdW equation in the minisuperspace approximation and for constant boundary dilaton profile, presented in \cite{Iliesiu:2020zld}.}

Having discussed the energy of the gravitational theory, we now turn our attention to the expected structure of a partition function for a general dilaton gravity at finite cutoff. Starting from the on shell action \eqref{dila}, the semiclassical disk partition function of the finite cutoff theory is expected to take the form
\begin{equation}
\begin{split}
  Z_{\mathrm{semi}}^{V(\phi)}(\beta;\varepsilon)
  \;\simeq\;
  \int d\phi_h\,
    V(\phi_h)\,e^{2\pi \phi_h}\,
    \exp\!\left[
      -\beta\,H_\varepsilon(\phi_h)
    \right]
  \label{eq:Z_semi_phi},
\end{split}
\end{equation}
where the integral over $\phi_h$ is interpreted as an integration over conical saddle solutions \eqref{dila1}, in which the horizon value of the dilaton is not fine tuned so that the associated Hawking temperature matches the Euclidean time periodicity $\beta$ \cite{Carlip:1993sa}.
Equivalently, in terms of the undeformed energy
$E_0 \equiv H_0 = \tfrac12 W(\phi_h)$, the semiclassical density of
states of the infinite cutoff theory is
\begin{equation}
  \rho_{\mathrm{semi}}(E_0)
  \;\sim\;
  \exp\!\left[ 2\pi\,W^{-1}(2E_0) \right],
  \label{eq:rho_semi}
\end{equation}
so that\footnote{The factor $V(\phi_h)$ was included in \eqref{eq:Z_semi_phi} to account for the Jacobian between $E$ and $\phi_h$. }
\begin{equation}
  Z_{\mathrm{semi}}^{V(\phi)}(\beta;\varepsilon)
  \;\simeq\;
  \int dE_0\,
    \rho_{\mathrm{semi}}(E_0)\,
    \exp\!\left[
      -\beta\,H_\varepsilon(E_0)
    \right].
  \label{eq:Z_semi_E}
\end{equation}
Up to overall normalization factors, \eqref{eq:rho_semi} reproduces the
standard semiclassical result for general dilaton potentials, and reduces to the well–known JT and Liouville densities in the appropriate limits \cite{Blommaert:2023wad}.

The idea is to improve the result above in two natural ways. The first improvement consists of replacing the semiclassical density of states with the exact one. To do this, we can exploit the same trick performed in \cite{Iliesiu:2020zld,Blommaert:2025rgw}, where one takes the infinite cutoff limit and fixes the exact spectral density in that regime:
\begin{equation}
  Z_{\mathrm{exact},\,\varepsilon\to 0}^{V(\phi)}(\beta)
  \;=\;
  \int \mathrm{d}\phi_h\,V(\phi_h)\,
    \rho_{\mathrm{exact}}(\phi_h)\,
    e^{-\beta W(\phi_h)/2}\,,
  \label{eq:Z_exact_eps0}
\end{equation}
and then extrapolate the same form at every value of $\varepsilon$, assuming that the Hartle--Hawking condition can be consistently fixed at $L\propto W(\phi_b)^{\frac12}\rightarrow +\infty$ \cite{Iliesiu:2020zld}.

Leveraging this observation, one can then improve the form of the partition function by employing the proposed gas of defects representation of the spectral density \cite{Maxfield:2020ale,Turiaci:2020fjj,Kruthoff:2024gxc} for a general dilaton gravity model. In this approach, the dilaton potential is expanded as a series of defect-type insertions, and the perturbative series can be exactly resummed \cite{Turiaci:2020fjj} , yielding
\begin{equation}
  \rho_{\mathrm{exact}}(\phi_h)
  \;=\;
  \frac{1}{8\pi^2}
  \int_{\mathcal C}\frac{\mathrm{d}\Phi}{2\pi i}\,
    e^{2\pi\Phi}\,
    \frac{W'(\Phi)}{W(\Phi)-W(\phi_h)}\,
    \sqrt{\frac{W(\phi_h)}{W(\Phi)}}\,.
  \label{eq:rho_exact_Phi}
\end{equation}
where $\mathcal C$ is the inverse Laplace transform contour.\footnote{Compared with the result presented in \cite{Turiaci:2020fjj}, we performed an additional integration by parts in \eqref{eq:rho_exact_Phi} and we set $E_{0}=0$ for simplicity. }
An interesting manipulation of this formula was performed in \cite{Blommaert:2024whf}, and we report it here because it connects naturally with the semiclassical analysis in \eqref{eq:Z_semi_phi}.
By changing variables to the energy variable
$M = W(\Phi)$, one can rewrite \eqref{eq:rho_exact_Phi} in the form
\begin{equation}
  \rho_{\mathrm{exact}}(\phi_h)
  \;=\;
  \frac{1}{8\pi^2}
  \int_{\mathcal C'}
    \frac{\mathrm{d}M}{2\pi i}\,
    \frac{e^{2\pi W^{-1}(M)}}{M - W(\phi_h)}\,
    \sqrt{\frac{W(\phi_h)}{M}}\,.
  \label{eq:rho_exact_M}
\end{equation}
Picking the pole at $M = W(\phi_h)$ reproduces the semiclassical
contribution \eqref{eq:Z_semi_phi}, while additional poles associated to other branches of
$W^{-1}$ produce some modifications. Remarkably, summing over all such
branches, one finds the following proposed form for the exact density:\footnote{We should stress that possible additional contributions to this formula may arise because of the presence of a branch cut in \eqref{eq:rho_exact_M}. }
\begin{equation}
  \rho_{\mathrm{exact}}(E)
  \;=\;
  \sum_{\substack{E_i \\ W(E_i)=W(E)}}
    (-1)^{\sigma_i}\,
    \frac{1}{8\pi^2}\,
    \exp\!\bigl[2\pi W^{-1}(E_i)\bigr]\,,
  \label{eq:rho_exact_images}
\end{equation}
where the sum runs over all images $E_i$ with the same value of the
prepotential, and $(-1)^{\sigma_i}$ encodes the sign inherited from the
Jacobian.\footnote{For monotonic $W(\phi)$ this sum reduces to the
single semiclassical branch. For JT and Liouville/sinh dilaton gravity,
the two–branch structure reproduces the standard
$\sinh$–type spectral densities.}

A second way to improve the expression of the partition function is to implement a truncation of the physical spectrum given in \eqref{adm_finite}.
The need for this modification follows from a simple observation: the finite cutoff energy \eqref{eq:H_eps_general} develops a branch point singularity when
\begin{equation}
  1 - \varepsilon^{2} W(\phi_h) = 0
  \qquad \Longleftrightarrow \qquad
  W(\phi_h) = \frac{1}{\varepsilon^{2}} = W(\phi_{b})\,,
  \label{eq:threshold}
\end{equation}
which renders the integral over the radial location of the black hole horizon $\phi_h$ in \eqref{eq:Z_semi_phi} ill defined. Beyond this threshold the Hamiltonian becomes complex, reflecting the familiar singular behaviour of $T\bar T$--deformed spectrum \cite{Smirnov:2016lqw,Cavaglia:2016oda}.

This branch point singularity has a clear geometric interpretation in our setting.
If we gradually increase $\varepsilon$, thereby pushing the physical boundary deeper into the bulk, then for fixed black hole energy, and hence fixed horizon radius, the condition \eqref{eq:threshold} corresponds exactly to the moment when the finite cutoff boundary crosses the black hole horizon. This can be seen explicitly from the classical solutions \eqref{dila}.
Conversely, if we keep $\varepsilon$ fixed and instead increase the energy, enlarging the black hole horizon, the singularity \eqref{eq:threshold} is reached precisely when the dilaton gravity black hole \eqref{dila} fills out the entire spacetime region inside the finite cutoff boundary.

Moreover, as the black hole becomes larger and its horizon approaches the fixed cutoff surface at 
$r = \phi_b$, the boundary observer will therefore sit at an increasingly large
gravitational redshift relative to the horizon, which corresponds to a growing
proper acceleration required to remain static. As a result, the local temperature, or Tolman temperature, measured by the boundary observer increases and should finally diverge. This is indeed what can be found by using the first law to compute the effective temperature associated to the finite cutoff energy \eqref{adm_finite} and Bekenstein-Hawking entropy \eqref{termo}, which yields:
\begin{equation}\label{tolman}
T_{\varepsilon}^{\mathrm{T}}(\phi_h)
= \frac{V(\phi_h)}{4\pi \sqrt{1 - \varepsilon^{2} W(\phi_h)}}.
\end{equation}
This realizes within general two dimensional dilaton gravity the same phenomenon observed earlier in the AdS$_3$/CFT$_2$ context \cite{McGough:2016lol}.

Once a physical intuition has been established for the need to truncate the radial integral over the horizon position $\phi_h$ at the bound\footnote{Notice that in the form \eqref{eq:Z_semi_phi} of the partition function, the truncation of the integral variable directly encodes the truncation of spacetime at $r=\phi_b$, or equivalently a truncation of the physical phase space of conical solutions \eqref{dila} labeled by $\phi_h$.}
\[
\phi_h = W^{-1}\!\left(\frac{1}{\varepsilon^{2}}\right),
\]
the lesson from JT gravity is that such a truncation necessarily brings into the game the nonperturbative branch of the $T\bar T$ deformation. The emergence of this branch appears to be inherently tied to the appearance of a spectral cutoff.
Motivated by the JT gravity analysis, in which a well defined partition function such as \eqref{sum} or \eqref{after_resurgence} is obtained only after selecting an appropriate contour prescription that effectively incorporates the nonperturbative branch, as discussed in Section~\ref{compa} and in \cite{Griguolo:2021wgy}, we now attempt to apply the same logic to a general dilaton gravity theory. The most natural generalization of the JT result for arbitrary
$W(\phi)$ is to choose for the partition function an integration contour $\gamma$ that wraps the branch cut of
the finite cutoff energy spectrum. Concretely, we take
\begin{equation}
  Z_{\mathrm{exact},\,\varepsilon}^{V(\phi)}(\beta)
  \;=\;
  \int_{\gamma} \mathrm{d}\phi_h\,
    V(\phi_h)\,\rho_{\mathrm{exact}}(\phi_h)\,
    \exp\!\left[
      -\frac{\beta}{\varepsilon^2}
       \Bigl(
         1 - \sqrt{1 - \varepsilon^2 W(\phi_h)}
       \Bigr)
    \right],
  \label{eq:Z_exact_contour}
\end{equation}
where $\gamma$ encircles the branch cut starting at
$W(\phi_h) = 1/\varepsilon^2$.\footnote{This is indeed the same type of contour that, starting from the ill defined integral representation \eqref{TTexact}, leads to the exact result \eqref{disk}. } After accounting for the discontinuity
across the cut, this localizes to a real integral over the range where
the square root is real,
\begin{equation}
  \boxed{Z_{\mathrm{exact},\,\varepsilon}^{V(\phi)}(\beta)
  \;=\; e^{- \frac{\beta}{\varepsilon^2}}
  \int_{0}^{W^{-1}\!\left(1/\varepsilon^2\right)} \mathrm{d}\phi_h\,
    V(\phi_h)\,\rho_{\mathrm{exact}}(\phi_h)\,
    \sinh\!\left(
      \frac{\beta}{\varepsilon^2}
         \sqrt{1 - \varepsilon^2 W(\phi_h)}  
    \right)}.
  \label{eq:Z_exact_sinh}
\end{equation}
We argue this result has the structure that we expect for the exact partition function of a general dilaton gravity theory at finite cutoff.

As an example, consider Liouville (or sinh) dilaton gravity, for
which the exact disk spectral density  \cite{Mertens:2020hbs,Fan:2021bwt} and prepotential are
\begin{equation}
  \rho_{\mathrm{exact}}(\phi_h)
  \;\propto\;
  \sinh(2\pi b\,\phi_h)\,
  \sinh\!\left(2\pi\,\frac{\phi_h}{b}\right) \qquad W(\phi_h) = \cosh(2\pi b\,\phi_h)\,.
\end{equation}
Plugging this into \eqref{eq:Z_exact_sinh} (and absorbing
overall constants into the normalization) we obtain the fully
nonperturbative finite cutoff partition function
\begin{equation}
  Z_{\mathrm{exact},\,\varepsilon}^{\sinh(\phi)}(\beta)
  \;=\; e^{- \frac{\beta}{\varepsilon^2}}
  \int_{0}^{\phi_*} \mathrm{d}\phi_h\,
    \sinh(2\pi b\,\phi_h)\,
    \sinh\!\left(2\pi\,\frac{\phi_h}{b}\right)\,
    \sinh\!\left(
      \frac{\beta}{\varepsilon^2} \sqrt{1 - \varepsilon^2 \cosh(2\pi b\,\phi_h)} \right),
\end{equation}
with $2 \pi b \,\phi_* = \cosh^{-1}\!\left(1/\varepsilon^2 \right)$.

An important remark concerning the results of this section is the class of dilaton potentials to which we expect them to apply. Strictly speaking, the gas of defects formula has been derived only for a family of potentials that can be understood as deformations of the JT potential \cite{Turiaci:2020fjj}. Nevertheless, it is natural to conjecture that its validity extends beyond this restricted class. In fact, using Eq.~\eqref{eq:rho_exact_images}, one can also reproduce the correct exact spectral density of sinh dilaton gravity, which provides nontrivial evidence for this extrapolation. Moreover, Eq.~\eqref{eq:Z_exact_sinh} has been written under the assumption that $W(\phi)$ is a convex function, so that the square root $\sqrt{1 - \varepsilon^2 W(\phi_h)}$ has a single branch cut. For more general prepotentials we expect additional contributions, obtained by appropriately accounting for each branch cut in the complex plane.

Another very interesting class of potentials $V(\phi)$ that has recently attracted considerable attention is the class of periodic potentials. In this case, in order to find a boundary location where $W(\phi_b)$ diverges, one has in general to continue $\phi_h$ along a complex contour \cite{Blommaert:2024whf}. This makes the considerations about the contour $\gamma$ more subtle in these periodic cases. 

Before concluding this section, we comment on the relationship between the finite cutoff ADM energy \eqref{adm_finite} and a possible $T\bar{T}$ interpretation.
As already proven by \cite{Aguilar-Gutierrez:2024nst}, for a generic dilaton gravity sharing the same asymptotics as JT gravity, the finite cutoff black hole energy \eqref{adm_finite} takes exactly the form of the perturbative branch of the $T\bar T$–deformed spectrum, that solves the dimensionally reduced flow equation \eqref{flow} governing the flow of a CFT.
More precisely, identifying the deformation parameter with the bulk cutoff via $\lambda \;=\; \frac{\varepsilon^2}{4}\,$,
and using $H_0(\phi_h)=\tfrac12 W(\phi_h)$, we obtain in our conventions 
\begin{equation}
  E_{\varepsilon}
  \;=\; H_{\lambda=\varepsilon^2/4}^{T\bar{T}}\!\left(\tfrac12 W(\phi_h)\right)
  \;=\; \frac{1}{\varepsilon^2}\left(
        1 - \sqrt{1 - \varepsilon^2 W(\phi_h)}
      \right).
  \label{eq:H_eps_general}
\end{equation}
Our analysis shows that this \(T\bar{T}\)-like structure persists for general dilaton potentials with different asymptotics, provided that the more general boundary counterterm in \eqref{dila1} is included. We aknowledge that similar results have already been found in \cite{Aguilar-Gutierrez:2024oea} using different techniques.. However, for dilaton gravity theories with asymptotics different from those of JT gravity, one expects a nontrivial boundary dynamics generally distinct from the Schwarzian, and the dimensional reduction of the flow equation inherited from the standard AdS\(_3\)/CFT\(_2\) framework should no longer hold. A more precise characterization of this persisting \(T\bar{T}\)-like structure in such general settings therefore remains an interesting open problem.

\section{Toward UV completeness in 2d finite cutoff holography}\label{toward}
The content of this section is inherently more speculative. Indeed, while the results we present are not conclusive, they are intended to open new directions for future investigation. The overarching goal is to explore how introducing a finite cutoff in the bulk may lead to a better behaved ultraviolet completion of the theory. Throughout, we focus primarily on JT gravity, while indicating when extensions to more general dilaton potentials appear feasible.

In \ref{discrete} we investigate possible signatures of discreteness in finite cutoff holography, both in the energy spectrum and in the underlying spacetime structure. Specifically, we propose an extension to higher topologies, present a nonperturbative completion of finite cutoff JT gravity in terms of a matrix model, and try to explain the finiteness of the number of states of the theory within the open-channel quantization framework.
In \ref{corre}, we study from the bulk perspective the coupling of a matter scalar field to a finite cutoff disk geometry and, from the boundary viewpoint, we then propose an exact expression for the boundary to boundary correlator, for which the usual UV singularity at coincident operator insertions appears to be resolved.
\subsection{Higher topologies and open channel quantization}\label{discrete}
Up to this point, our analysis has focused exclusively on the disk topology.
It's well known that standard JT gravity at infinite cutoff admits a full topological expansion on higher genus surfaces, captured by the loop equations of a double scaled matrix model \cite{Saad:2019lba}. This naturally raises the question of how JT gravity at \emph{finite} cutoff should be defined on higher topologies, and whether a dual matrix model description exists that reduces to the usual Saad–Shenker–Stanford duality in the infinite cutoff limit. This direction was initiated in \cite{Griguolo:2021wgy}, with several encouraging results, although a number of conceptual and technical issues remain unresolved.
Assuming that the standard Weil–Petersson measure
\[
\int \mathrm{d}b\, b
\]
continues to hold in the finite cutoff setting, in \cite{Griguolo:2021wgy} the partition function of the wormhole geometry was computed  by gluing together two finite cutoff
trumpets ~\eqref{amplitude2} with the above measure. This yields
\begin{equation}\label{wormhole}
Z_{\varepsilon}^{\mathrm{wormhole}}(\beta_1,\beta_2)
=\frac{\beta_1 \beta_2}{\varepsilon^2(\beta_{1}^2-\beta_{2}^{2})}
\left[
\beta_1 I_{0}(\beta_2/\varepsilon^2)\, I_{1}(\beta_1/\varepsilon^2)
-
\beta_2 I_{0}(\beta_1/\varepsilon^2)\, I_{1}(\beta_2/\varepsilon^2)
\right].
\end{equation}
Furthermore, by combining the standard Weil–Petersson volumes with the finite cutoff trumpet~\eqref{amplitude2}, \cite{Griguolo:2021wgy} were able to compute amplitudes for certain higher genus geometries and showed that these results are consistent with loop equations generated by a deformation of the SSS spectral curve.

This deformed spectral curve can be extracted by taking the inverse Laplace transform of the finite cutoff JT partition function~\eqref{disk}. One obtains the finite cutoff spectral density
\begin{equation}\label{spectral}
\rho(E;\phi_r,\varepsilon)
=
\phi_r\!\left(1-\frac{\varepsilon^2}{\phi_r}\,E\right)
\sinh\!\left(
2\pi\sqrt{\phi_r E\,\left(2-\frac{\varepsilon^2}{\phi_r}\,E\right)}
\right),
\end{equation}
which has compact support,
\begin{equation}\label{compact}
E \in \left[\,0,\; \frac{2\phi_r}{\varepsilon^2}\,\right].
\end{equation}

Some comments are in order:
\paragraph{The role of the nonperturbative saddle in the topological gluing}
A fundamental obstacle in the topological gluing procedure carried out in \cite{Griguolo:2021wgy} is the physical intuition underlying the integration over the neck $b$ of the trumpet: since it ranges from $0$ to $\infty$, it necessarily includes regions in which the geodesic boundary exceeds the size of the finite cutoff boundary, for any finite $\varepsilon$. However, the inclusion of the nonperturbative saddle in the trumpet wavefunction ~\eqref{amplitude2}, whose relevance is substantiated by the calculations in this paper, renders the integration mathematically well defined.
\paragraph{UV completeness}The emergence of a dynamical cutoff in the spectrum, which the bulk computation carried out in this paper confirms, is quite remarlable and signals a notion of UV completeness for the gravity theory at finite cutoff. This feature is strongly reminiscent of what happens in periodic dilaton gravities \cite{Blommaert:2024whf}, particularly in the sine dilaton gravity dual to DSSYK \cite{Blommaert:2024ymv}, where the spectrum is bounded and the theory can therefore be viewed as a UV completion of JT gravity. Because of the above considerations, the UV completeness at the bulk level appears here to be closely tied with the dual $T\bar{T}$ flow of the boundary theory, whose integrable structure enables one to follow the deformation into the UV regime in a controlled way. See also the analysis in \ref{corre} on the finite cutoff correlator for another possible manifestation of UV completeness.
\paragraph{Finite cutoff JT as a finite cut matrix integral?} Since, unlike in the infiite cutoff case, the spectrum naturally develops a cutoff \eqref{compact}, one may wonder whether it is actually more natural to describe the dual matrix model of JT gravity at finite cutoff as a finite matrix integral, analogous to the case of periodic dilaton gravities \cite{Blommaert:2024whf}, rather than as a deformation of the double scaled model. It was already observed in \cite{Blommaert:2025avl} that the wormhole amplitude \eqref{wormhole} bears a striking resemblance to the corresponding amplitude in sine dilaton gravity. Moreover, the finite cut matrix model amplitudes exhibit similarities with the JT finite cutoff amplitudes on higher topologies found in \cite{Griguolo:2021wgy}.\footnote{Compare for instance the similarity between the amplitudes of the Gaussian matrix model \cite{Okuyama:2018aij} or ETH matrix model \cite{Okuyama:2023kdo} with the JT finite cutoff amplitudes presented in Section 5 of \cite{Griguolo:2021wgy}. } However, in a finite cut matrix model the universal trumpet contribution entering the topological decomposition would feature a discrete value for $b$ \cite{Okuyama:2023kdo}. Can we find some signatures of discreteness in the finite cutoff geometry? The canonical quantization framework presented in the following might suggest so.
\paragraph{Bulk length discretization?}
An unsatisfactory aspect of the spectral density \label{spectral} is that it becomes negative in the range $\left[\frac{\phi_r}{\varepsilon^2},\,\frac{2\phi_r}{\varepsilon^2}\right]$. An interpretation of this negativity is still missing, but we point out that it occurs precisely in the range of energies associated with the nonperturbative saddle encountered in this work. 
Moreover, as anticipated above, the finite support of the spectral density may hint at an underlying spacetime
discretization, in line with a possible dual description in terms of a finite cut matrix
integral. In this work we have focused on a path integral approach to dilaton gravities at
finite cutoff. It is therefore natural to ask whether these features admit an alternative
explanation within a canonical framework, similar to that employed in the quantization of
periodic dilaton gravities \cite{Blommaert:2024whf}, which is based on a two-sided open channel canonical quantization of the
system.

As is well known, the classical phase space of two-dimensional dilaton gravity can be parametrized by the ADM energy and the boundary time. For simplicity, let us specialize to JT gravity. We can attempt to generalize the two sided construction of \cite{Harlow:2018tqv} to a finite cutoff by introducing the symplectic form
\begin{equation}\label{simple}
\omega = \mathrm{d}E_{\varepsilon} \wedge \mathrm{d}t , \qquad
E_{\varepsilon}= r_b-\sqrt{r_b^2-r_h^2},.
\end{equation}
Here $E_{\varepsilon}$ is the finite cutoff energy, obtained for general dilaton potentials in \eqref{adm_finite} and specialized here to JT gravity, with $r_b=1/\varepsilon$.
One can then perform a canonical change of variables to a more geometrical pair $(\ell_{\varepsilon},p_{\varepsilon})$, where $\ell_{\varepsilon}$ is the proper length of the spacelike geodesic connecting the two finite cutoff boundaries at fixed global time in the Kruskal two sided extension of the Schwarzschild patch \eqref{dila}. The length $\ell_{\varepsilon}$ of the Einstein–Rosen bridge in the finite-cutoff geometry is related to the Schwarzschild time by
\begin{equation}
\sinh\!\left(\frac{\ell_{\varepsilon}}{2}\right) = \sqrt{\frac{r_b^2}{r_h^2}-1}\; \cosh\!\left(r_h\, t\right), \label{eq:length-relation} 
\end{equation}
which is precisely the two-sided analytic continuation through
$\tau=\beta_{\mathrm{JT}}/2 + i t = \pi/r_h + i t$
of the Euclidean calculation performed in Appendix~\ref{finite_geo}.
Equation \eqref{eq:length-relation} allows one to trade the boundary time for the geometric variable $\ell_{\varepsilon}$ and to rewrite the symplectic form as $\omega = \mathrm{d}p_{\varepsilon}\wedge \mathrm{d}\ell_{\varepsilon}$, thus identifying $(\ell_{\varepsilon},p_{\varepsilon})$ as canonical variables at finite cutoff.

In terms of these variables, the finite cutoff JT Hamiltonian takes the compact form
\begin{equation}\label{finham}
  H(\ell_{\varepsilon}, p_{\varepsilon})
  = r_b\!\bigg[
    1-\tanh (\ell_{\varepsilon}/2)\,
    \cos\!\left(\,p_{\varepsilon}\right)
  \bigg],
\end{equation}
In the infinite cutoff limit, after the standard renormalization of the length, this Hamiltonian reduces to the familiar Liouville-type Hamiltonian of JT gravity \cite{Harlow:2018tqv}. This occurs upon taking $\ell_{\varepsilon} \gg 1$ and, consequently, $p_{\varepsilon} \ll 1$. Unlike the infinite cutoff case, where holographic renormalization of the bare length drives the renormalized length $\ell$ to $-\infty$, the finite cutoff length $\ell_{\varepsilon}$ is, by construction, always positive. It is tempting to interpret the lower bound $\ell_{\varepsilon} \ge 0$ as a canonical
counterpart of a truncation of spacetime degrees of freedom. In particular, in periodic
dilaton gravity the positivity constraint on the geodesic length can be viewed as
integrating out a portion of the disk. From this
perspective, the finite cutoff boundary may play a role similar to that of a cosmological
horizon, cutting out a portion of spacetime and thereby reducing the number of available states.

Upon canonical quantization, the Hamiltonian \eqref{finham} becomes a non-local operator in the $\ell$ representation, leading to a
finite-difference eigenvalue problem. Moreover, the
periodicity of the Hamiltonian under shifts
$p_{\varepsilon} \to p_{\varepsilon} + 2\pi$ may suggest a discretization of the spectrum of the length
operator, in close analogy with what occurs in periodic dilaton gravity \cite{Blommaert:2024whf}.

We also point out that, while from the boundary perspective it is more natural to work in a fixed inverse-temperature ensemble $\beta_{\mathrm{BH}}$, corresponding to the periodicity associated with the asymptotic Schwarzschild time \(t = t_{\infty}\) in \eqref{simple}, from the bulk perspective it appears more natural to work with the proper time in \eqref{dila}, which we denote by \(t_{\varepsilon}\). The latter is instead associated with the redshifted Tolman temperature \eqref{tolman}, which depends explicitly on the cutoff. Using \(\omega = \mathrm{d} t_{\varepsilon} \wedge \mathrm{d} E_{\varepsilon}\) then leads to a different form of the Hamiltonian, namely, \footnote{We acknowledge that similar results were independently found by Adam Levine and Andreas Blommaert.}
\begin{equation}\label{ham2}
  \tilde{H}_{\varepsilon}(\tilde{\ell}_{\varepsilon},\tilde{p}_{\varepsilon})
  = r_b\left[
    1-\sqrt{1-\frac{1}{r_{b}^2}\left(\tilde{p}_{\varepsilon}^2+\frac{r_{b}^2}{\cosh^2(\tilde{\ell}_{\varepsilon}/2)}\right)}
  \right].
\end{equation}
A thorough investigation of the quantization of the Hamiltonians \eqref{finham} and \eqref{ham2}, including their different origins and interpretations, as well as the computation of the resulting spectra and observables, will be presented in a future paper \cite{newpaper}. 

Finally, we note that an extension of the open-channel Hamiltonian framework to general dilaton gravity models appears feasible. This could be achieved by using the effective AdS$_2$ geometry introduced in \cite{Blommaert:2024whf}, performing canonical quantization in the finite cutoff version of this effective geometry, and subsequently mapping the results back to the finite cutoff original geometry.
In particular, a formulation of sine dilaton gravity at a finite cutoff\footnote{An interesting investigation in this direction was initiated in \cite{Aguilar-Gutierrez:2024oea}.} appears highly desirable, as it is expected to yield a $T\bar{T}$-like deformation of DSSYK or, equivalently, of its q-Schwarzian incarnation \cite{Blommaert:2023opb}.\footnote{As the flow of the Schwarzian can be related to the dimensional reduction of the flow of a CFT, one might expect the flow of the q-Schwarzian to be given by the dimensional reduction of the flow of a q-CFT.} As a related remark, one may envisage computing the Krylov complexity in (double-scaled) SYK \cite{Rabinovici:2023yex,Heller:2024ldz,Ambrosini:2024sre,Xu:2024gfm,Aguilar-Gutierrez:2025mxf, Aguilar-Gutierrez:2025pqp, Aguilar-Gutierrez:2025sqh}, deforming it along a \(T\bar T\)-like flow \cite{Anninos:2022qgy,Gross:2019uxi}, and then taking the JT limit. This procedure would realize a holographic notion of length as a function of time whose semiclassical expectation value is expected to be encoded in the finite cutoff length formula \eqref{eq:length-relation}.\footnote{A holographic notion of complexity=length at finite cutoff has been studied in \cite{Bhattacharyya:2025gvd}, based on the wormhole computation of \cite{Griguolo:2021wgy}.}

\subsection{Boundary to boundary correlators at finite cutoff}\label{corre}

An interesting open direction is the study of boundary to boundary correlators on the disk at finite cutoff. To the best of our knowledge, this problem has not been fully addressed from the bulk perspective within the existing JT gravity literature, particularly with the aim of connecting it directly to the boundary.\footnote{A related analysis has been conducted in \cite{Hartman:2018tkw}.} The purpose of this section is not to provide a complete solution, but to outline the progress we have made and the steps required to move forward.

On the bulk side, mirroring the infinite cutoff analysis, we introduce a free scalar matter field $\chi$, covariantly coupled to the AdS$_2$ background. The field obeys the equations of motion in Poincaré coordinates \eqref{hyper}, supplemented by the boundary conditions:
\begin{equation}\label{wave}
\partial_{x}^2 \chi-\partial_{t}^2 \chi=\frac{1}{x^2}m^2 \chi \qquad \chi(\varepsilon,t)=\varepsilon^{1-\Delta} \tilde{\chi}_b(t).
\end{equation}
Adapting the standard infinite cutoff procedure to the finite cutoff case reveals three distinct sources of corrections:
\paragraph{Dirichlet to Neumann kernel at finite cutoff}
After integrating the action by parts, the Gaussian path integral over $\chi$ is exact and produces the one-loop determinant of the Klein--Gordon operator on AdS$_2$, which can be absorbed into the overall normalization. The remaining boundary term contains no dynamical degrees of freedom and is fixed entirely by the boundary conditions of the fields at $x=\varepsilon$:
\begin{equation}\label{bounda}
I_{m}=-\frac12 \int \! \mathrm{d}t \, \chi(\varepsilon,t)\,
\left.\partial_{x} \chi(x,t)\right|_{x=\varepsilon}.
\end{equation}

Solving the wave equation \eqref{wave} in Fourier space for $\chi(x,p)$ with boundary condition $\chi(\varepsilon,p)=\varepsilon^{1-\Delta}\tilde{\chi}_{b}(p)$, one finds a unique exponentially damped bulk solution compatible with this boundary condition:
\begin{equation}\label{four}
\chi(x,p)=\frac{x^{1/2} K_{\nu}(p x)}{\varepsilon^{1/2} K_{\nu}(p\varepsilon)} \,
\varepsilon^{1-\Delta}\, \tilde{\chi}_{b}(p),
\end{equation}
where $\nu=\Delta-\tfrac12$.
Substituting this solution into the boundary term \eqref{bounda}, one obtains
\begin{equation}
\begin{split}
I_{m}^{\varepsilon}
&=-\frac{\varepsilon^{2-2\Delta}}{2}
\int \frac{\mathrm{d}p}{2\pi}\,
\tilde{\chi}_{b}(-p)\,
\frac{\mathrm{d}}{\mathrm{d}\varepsilon}
\bigl(\varepsilon^{1/2} K_{\nu}(p\varepsilon)\bigr)\,
\tilde{\chi}_{b}(p)\\[4pt]
&=-\frac{\varepsilon^{2-2\Delta}}{2}
\int\! \mathrm{d}t_1 \int\! \mathrm{d}t_2 \,
\tilde{\chi}_{b}(t_1)\tilde{\chi}_{b}(t_2)\,
\mathcal{A}_{\varepsilon}(t_1,t_2).
\end{split}
\end{equation}

Transforming back to position space, the key quantity to evaluate at finite cutoff $\varepsilon$ is the Dirichlet-to-Neumann kernel
\begin{equation}\label{fini}
\mathcal{A}_{\varepsilon}(t_1,t_2)
=\int\frac{\mathrm{d}p}{2\pi}\,
e^{ip(t_1-t_2)}
\left(
\frac{1-2\nu}{2\varepsilon}
-\frac{p\, K_{\nu-1}(p\varepsilon)}{K_{\nu}(p\varepsilon)}
\right).
\end{equation}

Taking the limit $\varepsilon\to 0$ and dropping contact terms, one recovers the standard conformal expression for the boundary two-point function:\footnote{The expansion of $\frac{\mathrm{d}}{\mathrm{d}\varepsilon} \left(\varepsilon^{1/2}K_{\nu}(p\varepsilon)\right)$ organizes as
\begin{equation}
\frac{\mathrm{d}}{\mathrm{d}\varepsilon}\left(\varepsilon^{1/2}K_{\nu}(p\varepsilon)\right)
=\frac{1}{\varepsilon}\!\left[
\frac12
-\nu\!\left(1+c_2 (\varepsilon p)^2+\cdots\right)
-2^{1-2\nu}\frac{\Gamma(1-\nu)}{\Gamma(\nu)}
(\varepsilon p)^{2\nu}\!\left(1+d_2 (\varepsilon p)^2+\cdots\right)
\right].
\end{equation}
The terms proportional to $p^2$ and higher generate derivatives of $\delta(t_1-t_2)$, which correspond to contact terms that we omit.}
\begin{equation}\label{confo}
\lim_{\varepsilon\to 0}\mathcal{A}_{\varepsilon}(t_1,t_2)
= C_{\nu} \int\! \mathrm{d}p \, e^{ip(t_1-t_2)} p^{2\nu}
= C_{\nu} \frac{2^{2\nu}\Gamma(\Delta)}{\pi^{1/2}\Gamma(-\nu)}
\frac{1}{|t_1-t_2|^{2\Delta}},
\end{equation}
with $C_{\nu}=\dfrac{\nu\, \Gamma(1-\nu)}{\Gamma(1+\nu)}$.
At finite $\varepsilon$, one should evaluate the full expression in \eqref{fini}. We have not found a closed-form expression for the Fourier transform, although it can be computed perturbatively in $\varepsilon$. The short-distance behaviour of the correlator \eqref{fini} is governed by the large-$p$ behaviour of the integrand, which differs qualitatively when $\varepsilon$ is kept finite. We will return to this point later.

\paragraph{Transformation of the sources at finite cutoff}
The standard next step, in order to incorporate the effects of quantum gravity, is to couple the matter fields to the boundary reparametrization modes. A convenient way to achieve this is by considering the transformation of the renormalized scalar field $\tilde{\chi}_b$ under the wiggly boundary curve in Poincaré coordinates, $(t(u),x(u))$. This transformation can be inferred by studying the behavior of the scalar field near the boundary:
\begin{equation}
\chi(x,\tau)\simeq x^{1-\Delta} \tilde{\chi}_b(t)=\varepsilon^{1-\Delta}t'^{1-\Delta}(u)\tilde{\chi}_b (t)=\varepsilon^{1-\Delta} \tilde{\chi}_b (u)
\end{equation}
where we use $x \simeq \varepsilon t'(u)$ in the limit $\varepsilon \ll 1$. In the standard infinite-cutoff framework, this bulk transformation leads to the familiar transformation law $\tilde{\chi}_b (u)=t'^{1-\Delta}(u)\tilde{\chi}_b \left(t(u)\right) $, which is the usual behavior of sources dual to operators of dimension $\Delta$ on the boundary, in accordance with the standard AdS/CFT dictionary.

Generalizing this construction to finite $\varepsilon$ is considerably more complicated. The leading near-boundary behavior $\chi \sim z^{1-\Delta}$ no longer suffices, and the simple relation between $x$ and $t$ along the boundary curve no longer holds. To infer the transformation of the source at finite cutoff, one can consider the full solution for $\chi(x,t)$:
\begin{equation}\label{convo}
\chi(x,t)
=\frac{x^{1/2}}{\varepsilon^{1/2}}\,
\varepsilon^{\,1-\Delta}
\int_{-\infty}^{\infty}ds\;
\tilde{\chi}(t-s)\,
g_x(s)
=\frac{x^{1/2}}{\varepsilon^{1/2}}\,
\varepsilon^{\,1-\Delta}\,
(\tilde{\chi}*g_x)(t),
\end{equation}
by interpreting the Fourier transform of \eqref{four} as a convolution of the source with $g_x(t)=\int_{-\infty}^{\infty}\frac{dp}{2\pi}\,e^{ip t}\,\frac{K_{\nu}(p x)}{K_{\nu}(p\varepsilon)}$. One can then study the transformation of \eqref{convo} under a boundary reparametrization $(x(u), t(u)) \to (z(u), \bar{z}(u))$, which is constrained by the condition derived in \ref{ric}:
\begin{equation}
\bar{z}(u)=z(u)-\frac{2z'(u)^2 \psi(u)}{z''(u)\psi(u)+2\psi'(u)z'(u)} \qquad \psi''+\left[\frac12 \{z(u),u \}+\frac{1}{4 \varepsilon^2}\right]\psi=0.
\end{equation}
This constraint ensures that we remain on the boundary curve at finite cutoff. From the transformed field $\chi(x(u), t(u))$, one can then extract the transformation of the boundary source $\tilde{\chi}_b(t)$, at least perturbatively in $\varepsilon$.

Reversing the logic, if the correspondence between finite-cutoff bulk physics and the $T\bar{T}$ flow on the boundary holds, this transformation can be viewed as a bulk manifestation of how a dimensionally reduced CFT primary operator transforms under a $T\bar{T}$ deformation. While the transformation of a $T\bar{T}$-deformed primary is expected to be highly nontrivial, it can be analyzed perturbatively using the flow equation, to verify whether it aligns with the perturbative expansion in $\varepsilon$ from the bulk perspective.
\paragraph{Coupling to the boundary mode at finite cutoff}
Coupling to the boundary mode is implemented by inserting the matter observable into the finite-$\varepsilon$ Schwarzian path integral introduced in Section \ref{quantum}:
\begin{equation}\label{deformed}
\int \frac{\mathcal{D}z(u)}{z'(u)} \ 
\exp\!\left[ -\frac{\phi_r}{\varepsilon^2}
\int_{0}^{\beta} \big( \kappa_{\varepsilon}(z(u)) - 1 \big)\,du \right]
\left[\det(\Delta_\varepsilon)\right]^{-1}
\frac{\delta^2 I_m^\varepsilon}{\delta \tilde{\chi}_b(u_1)\,\delta \tilde{\chi}_b(u_2)} ,
\end{equation}
where $I_m^\varepsilon$ incorporates both the finite-$\varepsilon$ kernel $\mathcal{A}_\varepsilon(t_1,t_2)$ and the finite-$\varepsilon$ transformation of the sources discussed above.
In the $\varepsilon\to 0$ limit, this expression reduces to the standard Schwarzian expectation value of the boundary two-point function:
\begin{equation}\label{undeformed}
\int \frac{\mathcal{D}f(u)}{f'(u)} \ 
\exp\!\left[ -\phi_r \int_{0}^{\beta} \{f(u),u\}\,du \right]
\left(\frac{f'(u_1)f'(u_2)}{(f(u_1)-f(u_2))^{2}}\right)^{\Delta}.
\end{equation}
Assuming that the operator insertion in \eqref{deformed} were known, at least as a perturbative expansion in $\varepsilon$ around the conformal correlator in \eqref{undeformed}, one could study the path integral \eqref{deformed} semiclassically, in analogy with the infinite-cutoff case. This involves expanding around the classical saddle \eqref{redef}, \eqref{grav} as we did in Section \ref{quantum} and treating fluctuations of the boundary mode $\delta\eta(u)$.
The finite cutoff graviton propagator $\langle\delta\eta(0)\delta\eta(u)\rangle_\varepsilon$ is obtained by inverting the quadratic action \eqref{qq}:
\begin{equation}
\begin{split}
\langle\delta \eta(0)\delta \eta(u)\rangle_{\varepsilon}
= \frac{\sqrt{1+\left(\frac{2\pi}{\beta}\right)^2 \varepsilon^2}}{2\pi}
\left(\frac{\beta}{2\pi}\right)^{3}
\sum_{n\neq 0,\pm1} e^{2\pi i n u/\beta}
\frac{1-\left(\frac{2\pi}{\beta}\right)^2 (n^{2}-1)\varepsilon^{2}}
     {n^{2}(n^{2}-1)} .
\end{split}
\end{equation}
Using the standard contour integral method of \cite{Maldacena:2016upp,Sarosi:2017ykf}, the sum can be evaluated by picking up residues at $n=0,\pm 1$. Writing $\omega=2\pi/\beta$, one obtains the closed form result
\begin{equation}\label{p}
\begin{split}
\langle \delta\eta(0)\,\delta\eta(u) \rangle_{\varepsilon}
= \frac{\sqrt{\omega^{2}\varepsilon^{2}+1}}{12\pi\,\omega^{3}}
\Big[
(3\omega^{2}u^{2}-6\pi\omega u+2\pi^{2})(\omega^{2}\varepsilon^{2}+1)
\\[1mm]
-3(4\omega^{2}\varepsilon^{2}+5)\cos(\omega u)
+6(\pi-\omega u)\sin(\omega u)
-6
\Big].
\end{split}
\end{equation}
In a semiclassical evaluation of \eqref{deformed}, any quadratic appearance of the graviton fluctuation must then be replaced by this propagator.

As this discussion illustrates, coupling matter to JT gravity at finite cutoff is considerably more intricate than in the standard $\varepsilon\to 0$ setup. Nevertheless, the main ingredients required for such a computation have been identified. An appealing alternative viewpoint is to examine the boundary dual: the corresponding observable should be the $T\bar{T}$-deformed two-point function of the boundary theory, which may offer a simpler and more tractable formulation of the problem.

Quite surprinsingly, the only instance where this object was studied is in \cite{Ebert:2022ehb}, where they considered expectation value of a Wilson line in the BF first order formulation of JT, in the presence of the $T\bar{T}$ deformation. The result is actually quite natural and equals basically the undeformed two point function \cite{Blommaert:2018oro,Iliesiu:2019xuh,Mertens:2018fds} where the Boltzmann factors are replaced by the perturbative branch of the $T\bar{T}$:\footnote{With $N_d$ a normalization chosen as $N_{d}=\phi_r^{2\Delta}Z_{\mathrm{disk}}^{-1}$.}
\begin{equation}\label{nonimpro}
\begin{split}
\langle \mathcal{O}(\tau) \mathcal{O}(\beta-\tau)\rangle_{\Delta}=N_{d}\int_{0}^{+\infty}\int_{0}^{+\infty} &\mathrm{d}s_1 \mathrm{d}s_2 \  s_1 s_2 \ \sinh(2\pi s_1)\sinh(2\pi s_2) \ e^{-\frac{\phi_r}{2}\tau  H_{\mathrm{TT}}^{+}(s_1)} \times \\
&\times e^{-\frac{\phi_r}{2}(\beta-\tau) H_{\mathrm{TT}}^{+}(s_2)} \frac{\Gamma \left(\Delta \pm i s_1 \pm i s_2\right)}{\Gamma (2\Delta)}
\end{split}
\end{equation}
If it is not interpreted as an asymptotic expansion of the exact result, the expression \eqref{nonimpro} exhibits the same pathologies as the partition function \eqref{TTexact}. Our proposed improvement consists of introducing a cutoff and adding the nonperturbative term (shown in red), in accordance with the discussion above and the results presented in this paper:\footnote{From the BF perspective that led to the derivation of \eqref{nonimpro}, a cutoff on the integral over $s$ corresponds to a cutoff on the label $s$ of the principal series representations of $SL(2,\mathbb{R})$. We note a clear analogy with the truncation of physical representations for $SU(2)$ and for more general compact gauge groups, as proven in \cite{Griguolo:2022xcj,Griguolo:2022hek} for Yang-Mills theory deformed by the $T\bar{T}$ flow. Since the BF formulation of JT gravity can be basically viewed as a Yang-Mills action in which the non-topological piece has support only on the boundary rather than the full surface, it would be very interesting to understand the physical meaning of this truncation of the representation label in the noncompact gauge group case.}
\begin{equation}\label{improvement}
\begin{split}
&\langle \mathcal{O}(\tau)\mathcal{O}(\beta-\tau)\rangle_{\Delta}^{\text{nonpert.}}
= \frac{N_{d}}{4}\, e^{-2\beta \phi_r/\varepsilon^2} 
\int_{0}^{\phi_r/\varepsilon} \int_{0}^{\phi_r/\varepsilon} \mathrm{d}s_1\,\mathrm{d}s_2\; s_1 s_2\, \sinh(2\pi s_1)\sinh(2\pi s_2) \\
&\quad \times 
\Big(e^{\frac{\tau\phi_r}{\varepsilon}\sqrt{1-\frac{\varepsilon^2}{\phi_r^2}s_1^2}}
- \textcolor{red}{e^{-\frac{\tau\phi_r}{\varepsilon}\sqrt{1-\frac{\varepsilon^2}{\phi_r^2}s_1^2}}}\Big)
\Big(e^{\frac{(\beta-\tau)\phi_r}{\varepsilon}\sqrt{1-\frac{\varepsilon^2}{\phi_r^2}s_2^2}}
- \textcolor{red}{e^{-\frac{(\beta-\tau)\phi_r}{\varepsilon}\sqrt{1-\frac{\varepsilon^2}{\phi_r^2}s_2^2}}}\Big)
\frac{\Gamma(\Delta \pm i s_1 \pm i s_2)}{\Gamma(2\Delta)}.
\end{split}
\end{equation}
Indeed, in the infinite $\Delta$ limit, where the matter particle becomes infinitely heavy and effectively splits the disk into two components, the two integrals decouple, giving the product of two disk partition functions with boundary lengths length $\tau$ and $\beta-\tau$ respectively, such that \eqref{improvement} appears to be the correct improvement of \eqref{nonimpro} to match with the above results \eqref{sum} or \eqref{after_resurgence}.
\begin{figure}[ht]
    \centering
    \begin{tikzpicture}
        \node[anchor=south west, inner sep=0] (img) {\includegraphics[width=0.75\textwidth]{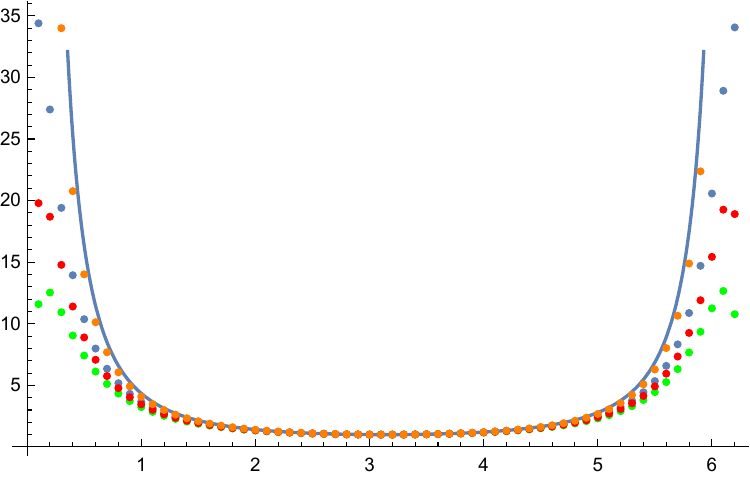}};

        \begin{scope}[x={(img.south east)}, y={(img.north west)}]

            \node at (0.5, -0.07) {$\tau$};

            \node[rotate=90] at (-0.07, 0.5) {$\langle \mathcal{O}(\tau)\mathcal{O}(\beta-\tau)\rangle$};

        \end{scope}
    \end{tikzpicture}
    \caption{Numerical evaluation of the $T\bar{T}$-deformed two point function \eqref{improvement} for $\Delta=1$, $\phi_r=20$, $\beta=2\pi$. Orange dots correspond to $\varepsilon=0.5$, blue dots to $\varepsilon=0.8$, red dots to $\varepsilon=0.9$ and green dots to $\varepsilon=1$. The line represents the semiclassical conformal answer $\langle \mathcal{O}(\tau)\mathcal{O}(\beta-\tau)\rangle=\frac{1}{\sin (\pi \tau/\beta)^{2\Delta}}$ obtained by evaluting \eqref{undeformed} on the Schwarzian thermal saddle. All data sets have been normalized such that the amplitude is equal to $1$ at $\tau=\beta/2$. }
\end{figure}\label{figureUV}

A remarkable feature of our nonperturbative improvement \eqref{improvement} is that the UV singularities at $\tau = 0$ and $\tau = \beta$ are removed. Technically, this is tied to the emergence of an effective cutoff in the spectrum. To see this, consider the infinite cutoff expression \eqref{nonimpro}, where $H^{+}_{T\bar{T}}(s) \to s^{2}$. Setting $\tau \to 0$ (or $\tau \to \beta$) in that formula eliminates one of the two Gaussian damping factors, rendering the integral divergent. By contrast, in our nonperturbative improvement \eqref{improvement} the cutoff ensures that the integral remains finite as $\tau \to 0$. Drawing once again a parallel with the behavior of the semiclassical two-point function in DSSYK/sine dilaton gravity \cite{Goel:2023svz,Blommaert:2024ymv}, the smoothing of the UV singularity in the correlator can be interpreted as a direct manifestation of the UV completeness of the cutoff theory. The consequent removal of the OPE singularity of the dual boundary operators can be appreciated very explicitly in the numerical evaluation of the amplitude shown in Figure~\ref{figureUV}.

Is this phenomenon consistent with what we expect from the bulk? Assuming that the transformation of the sources and the coupling to the boundary mode do not modify the qualitative behavior of the correlator near $t = t_1 - t_2 \to 0$, we expect this behavior to be encoded in the small-$t$ limit of $\mathcal{A}_{\varepsilon}(t)$ in \eqref{fini}. The short-time regime is governed by the large-$p$ region in Fourier space, which is controlled by the following asymptotic expansion of the integrand for large $p$:
\begin{equation}\label{asintotica}
\frac{1-2\nu}{2\varepsilon}
-\frac{p\, K_{\nu-1}(p\varepsilon)}{K_{\nu}(p\varepsilon)}
\simeq
-p+\frac{1-4\nu^2}{8\,\varepsilon^2 p}
+\frac{4\nu^2-1}{8\,\varepsilon^3 p^2}
+\cdots
\end{equation}
This asymptotic expansion remains unchanged under $p \to \pm \infty$. Since the correlator is expected to be symmetric under the exchange of the boundary points $t_1 \leftrightarrow t_2$, odd powers in the asymptotic expansion do not contribute by parity.\footnote{The Fourier trasnform can be replaced by a cosine Fourier transform.} Therefore, for $t \ll 1$ we expect a behavior of the form:\footnote{We have generally assumed the presence of a constant term in \eqref{tzero} because the asymptotic expansion in \eqref{asintotica} is valid only in a neighborhood of infinity. When we extrapolate this expansion in order to perform the full Fourier transform, additional contributions may appear that are not captured by the asymptotics. These missing contributions should, however, be regular terms, since the singular behavior as $t \to 0$ is entirely determined by the asymptotic region near $t \to \infty$.}
\begin{equation}\label{tzero}
\mathcal{A}_{\varepsilon}(t)\simeq \mathrm{const.} +\sqrt{\frac{\pi}{2}}\frac{1-4\nu^2}{8\,\varepsilon^3 } |t|+\cdots\,.
\end{equation}
In other words, this analysis shows that the leading behavior at $t \ll 1$ differs dramatically depending on whether we first take the limit $\varepsilon \to 0$ as in \eqref{confo} and then extrapolate the large-$p$ regime, or instead keep $\varepsilon$ finite in Fourier space. The two limits do not commute.

As further evidence, we can consider the geodesic approximation of the correlator, which is given by the length of a geodesic in the cutoff Euclidean geometry. This computation is performed in Appendix \ref{finite_geo} and yields:
\begin{equation}\label{ge}
\begin{split}
e^{-\Delta L_{\mathrm{geo}}}=\exp \Bigg[-\Delta \ \mathrm{arcsinh} \Big(\sqrt{\frac{\beta^2}{4 \pi^2 \varepsilon^2}-1}\,\sin \Big(\frac{\pi \tau}{\beta}\Big) \Big)\Bigg].
\end{split}
\end{equation}
Once again, \eqref{ge} exhibits a regular behavior as $\tau \rightarrow 0$. From these two bulk perspectives, we again infer that the cutoff in the spectrum is a key feature of the finite-cutoff theory that resolves the UV singularity of the correlator. Furthermore, if the formula \eqref{improvement} captures the full quantum behavior of the correlator, the inclusion of the nonperturbative (red) terms is responsible for driving the correlator to zero as $\tau \rightarrow 0, \beta$.  

While we have argued from two different perspectives for the finiteness of the correlator and the resolution of the UV singularity, we have not yet explained from a purely bulk perspective why it should vanish at coincident points. However, following the same manipulations used in \cite{Griguolo:2021zsn}, the expression \eqref{improvement} can be rewritten in the form:\footnote{where $C=\frac{\pi^2 \tau(\beta-\tau)\Gamma(2\lambda)e^{-\frac{\beta}{2\kappa \epsilon^2}} \mathcal{N}_{d}}{\varepsilon^2 (2\kappa)^4 2^{4\lambda-2}} 2i e^{-3\pi i \lambda} \sin(2 \pi \lambda)
$.}
\begin{equation}
\begin{split}
C\int_{-\infty}^{+\infty} \mathrm{d}q \int_{0}^{+\infty} \mathrm{d}t \frac{q^2-\left(t- i\pi \right)^2}{\left(\cosh \frac{q}{2}\right)^{2\Delta}\left(\sinh \frac{t}{2}\right)^{2\Delta}} \frac{I_{2}\left(\frac{\sqrt{\tau^2-\varepsilon^2 \left(t+q- \pi i\right)^2}}{\phi_r \varepsilon^2}\right)}{\tau^2-\varepsilon^2 \left(t+q- \pi i\right)^2} \frac{I_{2}\left(\frac{\sqrt{(\beta-\tau)^2-\epsilon^2 \left(t-q-\pi i\right)^2}}{\phi_r \varepsilon^2}\right)}{(\beta-\tau)^2-\varepsilon^2 \left(t-q-\pi i\right)^2}.
\end{split}
\end{equation}
This form is suitable for taking the semiclassical limit $\phi_r \rightarrow 0$ by using the asymptotic expansion of the Bessel functions and applying a steepest descent approximation for the integral. It would be very interesting to carry out this procedure to extract the expected behavior in \eqref{ge}, or even better, its nonperturbative modification that drives the correlator to zero, thus matching the numerical plots shown in Figure~\ref{figureUV}. We leave these more detailed investigations for future work.

\section{Acknowledgments}

We thank Andreas Blommaert, Thomas Mertens and Alex Tarana for useful discussions. We thank Rodolfo Panerai for having contributed to some of the early stages of this work. LG, LR, DS have been supported in part by the Italian Ministero
dell’Università e della Ricerca (MIUR), and by Istituto Nazionale di Fisica Nucleare (INFN)
through the “Gauge and String Theory” (GAST) research project. JP acknowledges financial support from the European Research Council (grant BHHQG-101040024). Funded
by the European Union. Views and opinions expressed are however those of the author(s)
only and do not necessarily reflect those of the European Union or the European Research
Council. Neither the European Union nor the granting authority can be held responsible
for them.

\appendix
\section{More details on the Riccati equation} \label{ricc}
In this Appendix we give a more detailed justification of the rationale behind the change of variable in \eqref{ree}. The idea is to begin with the relation:
\begin{equation}
\bar{z}(u)=z(u)+\frac{\varepsilon}{y(u)}
\end{equation}
which, plugged in \eqref{con}, leads to the equation
\begin{equation}\label{eq1}
y'-\frac{z'}{\varepsilon}y^2-\frac{1}{4z' \varepsilon}=0
\end{equation}
where we suppressed the dependence on $u$ for simplicity.
We now perform the further change of variables $y=- \frac{w'}{rw}$ and the equation \eqref{eq1} is rewritten as
\begin{equation}\label{eq}
w''(u)-\frac{r'}{r}w'(u)+\frac{w(u)}{4 \varepsilon^2}=0,
\end{equation} 
if we choose $r=\frac{z'}{\varepsilon}$ (for now we leave $r(u)$ generic for simplicity).
We then set $w(u)=h(u) \psi(u)$ where $h(u)$ will be determined soon. In this way \eqref{eq} becomes
\begin{equation}\label{eq2}
\psi''(u)+\left(2\frac{h'}{h}-\frac{r'}{r}\right)\psi'(u)+\left(\frac{h''}{h}-\frac{h'}{h}\frac{r'}{r}\right)\frac{1}{4 \varepsilon^2}\psi(u)=0.
\end{equation}
At this point we can choose $h$ in such a way that the coefficient of $\psi'$ vanishes, i.e.
\begin{equation}\label{eq3}
\log r=2\log h+\log c\rightarrow h=c \sqrt{r}=c\sqrt{\frac{z'}{\varepsilon}}
\end{equation}
Substituting this back in \eqref{eq2} we get
\begin{equation}\label{eq4}
\psi''+\left[\frac12 \left(\frac{z'''}{z'}-\frac32\left(\frac{z''}{z'}\right)^2\right)+\frac{1}{4 \varepsilon^2}\right]\psi=0
\end{equation}
where following all change of variables back to the origin we find the relation between $z$ and $\bar{z}$ in terms of $\psi$: 
\begin{equation}\label{re}
\bar{z}=z-\frac{2z'^2 \psi}{z''\psi+2\psi'z'}.
\end{equation}

\subsection{Odd coefficients of the extrinsic curvature $\kappa$} \label{odd}

In this Appendix we analyze the recursive relation \eqref{extra} and we deduce some important properties of the coefficients $\kappa_n(u)$. In particular, we will firstly prove by induction that all the odd terms $\kappa_{2n + 1}(u)$ are purely imaginary, and after that we will show that they are also total derivatives.

First of all we start by recalling the recursive relation of interest:
\begin{equation}
\begin{cases}
\kappa_0(u)^2=1,\\
2\,\kappa'_{n-1}(u)-i\displaystyle\sum_{m=0}^n \kappa_{n-m}(u)\,\kappa_m(u)
+2i\,\delta_{n,2}\{z(u),u\}=0, \qquad n\ge1.
\end{cases}
\end{equation}
Separating the $m=0$ and $m=n$ terms we obtain
\begin{equation}
2\kappa_0(u)\kappa_n(u)=-\sum_{m=1}^{n-1}\kappa_{n-m}(u)\kappa_m(u)+2\delta_{n,2}\{z(u),u\}
-2 i\,\kappa'_{n-1}(u).
\end{equation}
We now prove by induction that all the odd degree terms are purely imaginary and all the even ones are completely real.
In order to do that we start with the following induction hypothesis: suppose all even $\kappa_{2m}$ are real and all odd $\kappa_{2m+1}$ are purely imaginary for $m<n$. Then:

\begin{itemize}
\item If $n$ is even, the sum involves products of terms with the same parity (even-even or odd-odd), hence real; $\kappa'_{n-1}$ is purely imaginary, so $-2 i \kappa'_{n-1}$ is real; and $\delta_{n,2}\{z,u\}$ is real. Thus $\kappa_n\in\mathbb{R}$.
\item If $n$ is odd, the sum involves products of terms with opposite parity, hence purely imaginary; $\kappa'_{n-1}$ is real, so $-2 i \kappa'_{n-1}$ is purely imaginary; and $\delta_{n,2}=0$. Thus $\kappa_n\in i\mathbb{R}$.
\end{itemize}

Therefore, by induction, all even coefficients $\kappa_{2m}$ are real, while all odd coefficients $\kappa_{2m+1}$ are purely imaginary.

We now proceed to show that all the odd-degree coefficient are total derivatives. In order to do that it is convenient to go back to the Riccati equation \eqref{extra}, that we write again for reference:
\begin{equation}
2\epsilon\,\partial_u \kappa=i\left(\kappa^2-1\right)-2i\,\epsilon^2\{z(u),u\}.
\end{equation}
Since we are interested in the odd-degree terms it is now convenient to decompose $\kappa=\kappa_{\mathrm{e}}+\kappa_{\mathrm{o}}$ into even and odd parts in $\epsilon$, and set $\kappa_{\mathrm{o}}=\epsilon J$ with $J$ even. Substituting this form into the functional equation, the coefficient of $\epsilon^1$ gives directly the relation
\begin{equation}
2\,\partial_u \kappa_{\mathrm{e}}=2i\,\kappa_{\mathrm{e}}J, \quad \text{that implies} \quad
J=-i\,\kappa_{\mathrm{e}}^{-1}\partial_u \kappa_{\mathrm{e}}
=-\,i\,\partial_u(\log \kappa_{\mathrm{e}}).
\end{equation}
Therefore we proved that
\begin{equation}
\kappa_{\mathrm{o}}=-\,i\,\epsilon\,\partial_u\!\left(\log \kappa_{\mathrm{e}}\right),
\end{equation}
so that every odd coefficient is a total derivative:
\begin{equation}
\kappa_{2m+1}(u)=-\,i\,\partial_u\!\left([\epsilon^{2m}]\,\log \kappa_{\mathrm{e}}(u,\epsilon)\right),
\qquad m=0,1,2,\dots
\end{equation}
In particular, $\kappa_1=-i\,\partial_u \log\kappa_0=0$, and all higher odd coefficients are total derivatives of expressions built from the even sector.

\section{The Gaussian integral over $\delta \Phi$ and $\delta H$}\label{gaus}
In this section we aim to compute the one-loop determinant associated to the quadratic action \eqref{SS2}, which we report here:
\begin{equation}\label{sss}
\begin{split}
S_2 \left(\delta \Phi, \delta H, \delta \mathcal{N}; \mathcal{N}_{\mathrm{cl}}\right) 
&= \frac{1}{2\left(\mathcal{N}_{\mathrm{cl}}+\delta \mathcal{N}\right)} 
\int_{0}^{1} \mathrm{d}r \, \delta \Phi'(r) \, \delta H'(r)  + \, F(\delta \mathcal{N})
\int_{0}^{1} \mathrm{d}r \, \delta \Phi(r) .
\end{split}
\end{equation}
where we denoted
\begin{equation}
F(\delta \mathcal{N})=\frac{\delta \mathcal{N}(2\mathcal{N}_{\mathrm{cl}}+\delta \mathcal{N})}{\mathcal{N}_{\mathrm{cl}}+\delta \mathcal{N}}
\end{equation}
In order to do that, it is useful to analyze the following spectral problem, associated to the quadratic operator $\mathcal{O}$ associated to the first term of \eqref{sss},
\begin{equation}\label{quadra}
\mathcal{O} \ \delta \vec{\chi}(r)=\frac{1}{\left(\mathcal{N}_{\mathrm{cl}}+\delta \mathcal{N}\right)^2}\begin{pmatrix}
0 & -\partial_r^2 \\[6pt]
-\partial_r^2 & 0
\end{pmatrix}
\begin{pmatrix}
\delta \Phi(r) \\[6pt]
\delta H(r)
\end{pmatrix}
=\frac{\lambda}{\left(\mathcal{N}_{\mathrm{cl}}+\delta \mathcal{N}\right)^2} \  \delta \vec{\chi}(r),
\end{equation}
where we organized the fluctuation fields into a vector $\delta \vec{\chi}$.\footnote{The  quadratic operator $\mathcal{O}$ can be read off from \eqref{SS2} as $\int \mathrm{d}^2 x \sqrt{g} \  \delta \chi(x) \mathcal{O} \delta \chi(x)$. This ensures that the resulting eigenmodes are normalized with respect to the standard inner product $\int \mathrm{d}^2 x \sqrt{g} \ \ \delta \chi_{k}(x) \delta \chi_{k'}(x)=\delta_{k,k'}$.}
This will allow us to expand the field fluctuations in a basis of eigenfunctions of this second-order differential operator. The quadratic operator $\mathcal{O}$ in \eqref{quadra} arises from integrating by parts in \eqref{SS2}, with all boundary terms vanishing under the following boundary conditions imposed on the fluctuation fields:
\begin{equation}\label{bcc2}
\delta H(0) = 0, \qquad \delta H'(0) = 0, \qquad \delta H(1) = 0, \qquad \delta \Phi(1) = 0.
\end{equation}
These conditions descend directly from the original boundary conditions \eqref{bcc}. 
The most general solution to the differential system \eqref{quadra} is given by:
\begin{equation}
\begin{aligned}
\delta \Phi(r) = \frac{1}{2\sqrt{2\lambda}} \Big[ 
& (c_1 + c_3) \sqrt{2\lambda} \cos(\sqrt{2\lambda} \, r) 
+ (c_1 - c_3) \sqrt{2\lambda} \cosh(\sqrt{2\lambda} \, r) \\
& + (c_2 + c_4) \sin(\sqrt{2\lambda} \, r) 
+ (c_2 - c_4) \sinh(\sqrt{2\lambda} \, r) 
\Big]
\end{aligned}
\end{equation}

\begin{equation}
\begin{aligned}
\delta H(r) = \frac{1}{2\sqrt{2\lambda}} \Big[ 
& (c_1 + c_3) \sqrt{2\lambda} \cos(\sqrt{2\lambda} \, r) 
+ (c_3 - c_1) \sqrt{2\lambda} \cosh(\sqrt{2\lambda} \, r) \\
& + (c_2 + c_4) \sin(\sqrt{2\lambda} \, r) 
+ (c_4 - c_2) \sinh(\sqrt{2\lambda} \, r) 
\Big]
\end{aligned}
\end{equation}
We now impose the boundary conditions \eqref{bcc2} on the eigenfunctions. Three of the four conditions determine the values of three integration constants in the general solution, while the remaining condition quantizes the spectrum. In particular, we find the following relations among the constants:
\begin{equation}\label{cond}
c_1 = - c_2 \, \frac{\tanh\left( \sqrt{2\lambda} \right)}{\sqrt{2\lambda}}, \qquad c_3 = 0, \qquad c_4 = 0,
\end{equation}
and the allowed eigenvalues $\lambda$ must satisfy the transcendental equation:
\begin{equation}\label{trasce}
\frac{\tanh\left( \sqrt{2\lambda} \right)}{\sqrt{2\lambda}} = \frac{\tan\left( \sqrt{2\lambda} \right)}{\sqrt{2\lambda}}.
\end{equation}
The above equality discretizes the spectrum and imposes the fact that there is a one-to-one correspondence between the eigenvalues and the number of disconnected intervals where the tangent is defined. For this reason from now on we will label the eigenvalues as $\lambda_k$ with $ k \in \mathbb{Z}$.\footnote{Interestingly, for each $\lambda_n > 0$ with $n \in \mathbb{N}$ satisfying \eqref{trasce}, there exists a corresponding $\lambda_{-n} = -\lvert \lambda_n \rvert$ which also solves the same equation. This is because $\tan$ and $\tanh$ are interchanged in \eqref{trasce}, while the eigenvalue equation retains its form. Moreover,
\[
\vec{v}_{-n}(r) =
\begin{pmatrix}
\delta \Phi_{-n}(r) \\[12pt]
\delta H_{-n}(r)
\end{pmatrix}
=
\begin{pmatrix}
\delta \Phi_{n}(r) \\[12pt]
-\delta H_{n}(r)
\end{pmatrix},
\]
so the corresponding eigenvector is linearly independent from $\vec{v}_{n}(r)$. Therefore, the negative eigenvalues must also be included in the one-loop determinant as well.
}. Implementing the conditions \eqref{cond} and \eqref{trasce}, the eigenmodes of the quadratic operator take the form:
\begin{equation}\label{eigenmode}
\begin{split}
\vec{v}_{k}(r) &= A_k\begin{pmatrix}
\displaystyle \delta \Phi_{k}(r) \\[12pt]
\delta H_{k}(r)
\end{pmatrix} \\
&=A_k
\begin{pmatrix}
\displaystyle \frac{\sinh \left(\sqrt{2\lambda_k } \, r\right)+\sec \left(\sqrt{2\lambda_k }\right) \sin \left(\sqrt{2\lambda_k } (r-1)\right)-\tan \left(\sqrt{2\lambda_k }\right) \cosh \left(\sqrt{2\lambda_k } \, r\right)}{2 \sqrt{2\lambda_k }} \\[12pt]
\displaystyle \frac{-\sinh \left(\sqrt{2 \lambda_k } \, r\right)+\sec \left(\sqrt{2\lambda_k }\right) \sin \left(\sqrt{2\lambda_k } (r-1)\right)+\tan \left(\sqrt{2\lambda_k }\right) \cosh \left(\sqrt{2\lambda_k } \, r\right)}{2 \sqrt{2\lambda_k }}
\end{pmatrix}
\end{split}
\end{equation}
where $A_k=\left(\mathcal{N}_{\mathrm{cl}}+\delta \mathcal{N}\right)^{-1/2} \frac{2 \sqrt{\lambda_k}}{ \tan(\sqrt{2\lambda_k})}$ is a normalization factor which ensures that the eigenmodes are orthonormal with respect to the covariant inner product.
With these results in mind, we can now expand the vector of fluctuations of the original fields in terms of the eigenvectors that we have just found:
\begin{equation}\label{fluc}
    \delta \vec{\chi}(r) =\begin{pmatrix}
\displaystyle \delta \Phi(r) \\[12pt]
\delta H(r)
\end{pmatrix}= c_0 \vec{v}_{0}(r) + \sum_{k \neq 0} c_k \vec{v}_k(r),
\end{equation}
where $\vec{v}_{0}(r)$ is the vector of the kernel of the quadratic operator. Indeed, the eigenvalue $\lambda = 0$ is a solution of the eigenvalue equation~\eqref{trasce}. 
The corresponding zero mode can be obtained by taking the limit $\lambda \to 0$ of~\eqref{eigenmode}, yielding
\begin{equation}\label{zero}
    \vec{v}_{0}(r) = 
    \begin{pmatrix}
        r - 1 \\[6pt]
        0
    \end{pmatrix}.
\end{equation}
Its form admits a natural interpretation: it corresponds to the moduli space of classical solutions~\eqref{solu}, which depend on the collective coordinate $\phi_0$,
\begin{equation}\label{collective}
    \Phi_{\mathrm{cl}}(r) = \phi\, r + \phi_0 (1 - r).
\end{equation}
This one-parameter family of classical solutions~\eqref{collective} shares the same on-shell action~\eqref{saddles}, which is independent of $\phi_0$. 
This degeneracy of classical configurations is precisely what gives rise to the zero mode in the fluctuation spectrum. 
In fact, the variation of a classical solution along a direction parametrized by a collective coordinate always yields a zero mode of the quadratic fluctuation operator. 
In the present case,
\begin{equation}
    \delta \vec{\chi}_{\mathrm{cl}}
    = \vec{\chi}_{\mathrm{cl}} + \delta \phi_0 \, \partial_{\phi_0} \vec{\chi}_{\mathrm{cl}}
    = \vec{\chi}_{\mathrm{cl}} + \delta \phi_0 
    \begin{pmatrix}
        r - 1 \\[6pt]
        0
    \end{pmatrix}
    = \vec{\chi}_{\mathrm{cl}} + \delta \phi_0 \, \vec{v}_{0}(r),
\end{equation}
where $\vec{v}_0(r)$ is precisely the zero mode identified in~\eqref{zero}.\footnote{This correspondence between collective coordinates and zero modes is completely general. 
Differentiating the classical equation of motion, 
$\left.\frac{\delta S[\phi]}{\delta \phi(x)}\right|_{\phi_{\mathrm{cl}}}=0$, 
with respect to $\phi_0$ yields
\[
\int \mathrm{d}y \ 
\left.\frac{\delta^2 S[\phi]}{\delta \phi(x)\,\delta\phi(y)}\right|_{\phi_{\mathrm{cl}}}
\, \partial_{\phi_0} \phi_{\mathrm{cl}}(y)
= 0,
\]
showing that $\partial_{\phi_0} \phi_{\mathrm{cl}}$ lies in the kernel of the quadratic operator, i.e. 
$\mathcal{O}(\partial_{\phi_0}\phi_{\mathrm{cl}})=0$.}

However, the semiclassical Wheeler--DeWitt constraint~\eqref{wdw_contr} selects the solution with $\phi_0 = 0$, thereby gauge-fixing the freedom to deform the classical configuration along this direction in functional space. 
When defining the path-integral measure over fluctuations, we must therefore recall that the direction in field space associated with the collective coordinate, $\partial_{\phi_0}\vec{\chi}_{\mathrm{cl}}$, is unphysical. 
We thus project onto the subspace orthogonal to it, which can be implemented by inserting into the path integral a delta-functional constraint enforcing that the inner product 
\[
\langle \partial_{\phi_0}\vec{\chi}_{\mathrm{cl}}, \, \delta \vec{\chi} \rangle = 0
\]
vanishes.
The integration measure of the path integral over fluctuations thus becomes
\begin{equation}
\begin{split}
    \mathcal{D}\delta \vec{\chi}^{\perp} &= \mathrm{d}c_0\prod_{n=1}^{\infty} \mathrm{d}c_n \mathrm{d}c_{-n}  \ \delta \left(\int \mathrm{d}^2 x \sqrt{g} \ \partial_{\phi_0} \vec{\chi}_{\mathrm{cl}}(x) \cdot \left(\vec{\chi}(x)-\vec{\chi}_{\mathrm{cl}}(x)\right)\right) \\
    &= \mathrm{d}c_0\prod_{n=1}^{\infty} \mathrm{d}c_n \mathrm{d}c_{-n}  \ \delta \left(c_0 \left(\mathcal{N}_{\mathrm{cl}}+\delta \mathcal{N}\right)^{\frac12}\right)=\left(\mathcal{N}_{\mathrm{cl}}+\delta \mathcal{N}\right)^{-\frac12}\prod_{n=1}^{\infty} \mathrm{d}c_n \mathrm{d}c_{-n}
    ,
\end{split}
\end{equation} 
where in the second step we inserted the expansion \eqref{fluc} and used the orthogonality between eigenmodes.
Plugging in the expansion \eqref{fluc} into \eqref{sss}, and using the orthogonality of the eigenmodes, we obtain the following gaussian integral
\begin{equation}\label{modes}
\mathcal{Z}_{\text{1-loop}}=\left(\mathcal{N}_{\mathrm{cl}}+\delta \mathcal{N}\right)^{-\frac12}\int \prod_{n=1}^{\infty}\!\mathrm{d}c_{n}\,\mathrm{d}c_{-n}\;
\exp\!\left[-\frac{\lambda_{n}}{(\mathcal{N}_{\mathrm{cl}}+\delta \mathcal{N})^2}(c_{n}^2-c_{-n}^2) 
+ I_{n} F(\delta \mathcal{N})(c_n+c_{-n})\right],
\end{equation}
where $I_{n}$ is given by:
\begin{equation}
I_{n}=\int_{0}^{1} \mathrm{d}r  \ \delta \Phi_{n}(r)=\frac{\sech\sqrt{\lambda_n}-\sec\sqrt{\lambda_n}}{2\lambda_n}=I_{-n}
\end{equation}
which can be checked via explicit calculation using \eqref{eigenmode}. The linear shift in \eqref{modes} can be reabsorbed by completing the square and redefining 
\begin{equation}
\tilde{c}_{\pm n}=c_{\pm n}\mp \frac{(\mathcal{N}_{\mathrm{cl}}+\delta \mathcal{N})^2F(\delta \mathcal{N})I_{n}}{2\lambda_{n}},
\end{equation} with no additional terms appearing thanks to the $\pm n$ symmetry. We finally get 
\begin{equation}\label{modes2}
\begin{split}
\mathcal{Z}_{\text{1-loop}}&=(\mathcal{N}_{\mathrm{cl}}+\delta \mathcal{N})^{-\frac12}\int \prod_{n=1}^{\infty}\!\mathrm{d}\tilde{c}_{n}\,\mathrm{d}\tilde{c}_{-n}\;
\exp\!\left[-\frac{\lambda_{n}}{(\mathcal{N}_{\mathrm{cl}}+\delta \mathcal{N})^2}(\tilde{c}_{n}^2-\tilde{c}_{-n}^2) 
\right] \\
&\propto (\mathcal{N}_{\mathrm{cl}}+\delta \mathcal{N})^{-\frac12}\left(\det\mathcal{O}\right)^{-1},
\end{split}
\end{equation}
where we performed a Wick rotation of the negative modes $c_{-n}$ to render the gaussian integral over them convergent. Going to the second line of \eqref{modes2}, we dropped all proportionality constants coming from the gaussian integral (including the zeta function regularization of the infinite product of Wick rotation factors from the measure) because we are just interested in extracting the dependence on $\mathcal{N}$. Therefore, the object we need to evaluate is
\begin{equation}\label{deter}
\det\mathcal{O}\;\sim\;\prod_{n=1}^{+\infty} 
\frac{\lambda_{n}}{(\mathcal{N}_{\mathrm{cl}}+\delta \mathcal{N})^2}
\;\sim\;
\prod_{n=1}^{+\infty}\frac{k_{n}^{2}}{(\mathcal{N}_{\mathrm{cl}}+\delta \mathcal{N})^{2}},
\end{equation}
where $k_n$ satisfy 
\begin{equation}\label{eige_equation}
\tan k_n = \tanh k_n,
\end{equation}
as a consequence of \eqref{trasce}.  
The infinite product in \eqref{deter} is divergent, so we regularize it using the Riemann zeta function. In Sec.~\ref{zeta} we show that this regularization yields
\begin{equation}
\det\mathcal{O}\sim \left(\mathcal{N}_{\mathrm{cl}}+\delta \mathcal{N}\right)^{3/2}.
\end{equation}
Inserting this result into \eqref{modes2} reproduces the final expression \eqref{One} given in the main text.

\subsection{Zeta–function regularization}\label{zeta}

To make sense of the determinant \eqref{deter}, we define the zeta–regularized determinant of the operator $\mathcal{O}$ as
\begin{equation}
    \det\mathcal{O} = \exp\!\left[-\,\zeta_{\mathcal{O}}'(0)\right],
\end{equation}
where $\zeta_{\mathcal{O}}(s)$ is the analytic continuation of the operator-dependent zeta function
\begin{equation}\label{Z}
    \zeta_{\mathcal{O}}(s) \equiv   \sum_{n=1}^{\infty} 
    \left(\frac{k_{n}^{2}}{(\mathcal{N}_{\mathrm{cl}}+\delta \mathcal{N})^{2}}\right)^{-s},
\end{equation}
with $k_n$ determined by \eqref{eige_equation}.  
Taking the derivative at $s=0$ gives
\begin{equation}
    \zeta_{\mathcal{O}}'(0)
    =
    2\log\!\left(\mathcal{N}_{\mathrm{cl}}+\delta\mathcal{N}\right)\,
    \zeta_{k_n^2}(0)
    + \zeta_{k_n^2}'(0),
\end{equation}
where $\zeta_{k_n^2}(s)=\sum_{n=1}^{\infty} k_n^{-2s}$.
Our claim is that
\begin{equation}
    \zeta_{k_n^2}(0)= -\frac{3}{4}.
\end{equation}
Since we are only interested in the $\mathcal{N}$–dependence, we disregard the finite contribution $\zeta_{k_n^2}'(0)$. Hence,
\begin{equation}\label{resu}
    \det\mathcal{O}\sim 
    A\left(\mathcal{N}_{\mathrm{cl}}+\delta \mathcal{N}\right)^{3/2},
\end{equation}
where the constant prefactor is $A=\exp\!\left[-\zeta_{k_n^2}'(0)\right]$.
We now prove the above claim. Define the auxiliary sequence $y_n = \left(n+\tfrac{5}{4}\right)\pi$ and introduce the Hurwitz zeta function
\begin{equation}
    \zeta_H(s;a)=\sum_{n=0}^{\infty} (n+a)^{-s}.
\end{equation}
Then we may rewrite
\begin{equation}
    \zeta_{k_n^2}(s)
    = \pi^{-2s}\,\zeta_H\!\left(2s;\tfrac{5}{4}\right)
      +\sum_{n=0}^{\infty}\!\left(k_{n+1}^{-2s}-y_n^{-2s}\right),
\end{equation}
where we have subtracted and added the corresponding Hurwitz zeta term.  
Its analytic continuation satisfies
\begin{equation}
    \zeta_H\!\left(0;\tfrac{5}{4}\right)=-\frac{3}{4}.
\end{equation}
Therefore, to complete the proof we must show that the remainder
\begin{equation}\label{remainder}
    R(s)=\sum_{n=0}^{\infty}\!\left(k_{n+1}^{-2s}-y_n^{-2s}\right),
\end{equation}
admits an analytic continuation such that $R(0)=0$.
An important preliminary observation is the behavior of the solutions \(k_n\) of \eqref{eige_equation}. These roots accumulate precisely near the points \[ y_n = \Bigl(n+\tfrac14\Bigr)\pi, \] because \(\tan y_n = 1\), and for large argument \(\tanh x\) also approaches \(1\) extremely fast. In other words, for large \(n\) the exact \(k_n\) differ from the simple reference values \(y_n\) only by an exponentially small amount. 
To make this precise we write \[ k_n = y_n + \varepsilon_n , \qquad \varepsilon_n\ll 1 \quad (n\gg1). \] and, to estimate \(\varepsilon_n\), we expand both sides of the equation \eqref{eige_equation} around \(y_n\).
With simple manipulations, one can prove that
\begin{equation}
|\varepsilon_n| \le C e^{-2\pi n}
\end{equation}
for some constant \(C\).
Thus the deviation of \(k_n\) from \(y_n\) is exponentially small in \(n\).

We now return to the remainder \eqref{remainder}.
From the expression above for \(\varepsilon_n\), it is clear that for small \(s\) the difference in each term is tiny. It is convenient to estimate it by Taylor expanding \(t^{-2s}\) around \(t=y_n\). Writing the first-order expansion in $\varepsilon_n$ and using that \(\varepsilon_n\sim e^{-2\pi n}\), \(y_n\sim \pi n\), we obtain that each term in \eqref{remainder} behaves like
\[
k_{n+1}^{-2s}-y_n^{-2s}
\sim s\, e^{-2\pi n} \, n^{-1-2 s}.
\]
The crucial point is simply that the factor \(e^{-2\pi n}\) makes each term in the series extremely small for large \(n\). Because of this exponential falloff, in a small neighborhood of $s=0$, we can bound each term by a convergent series, which ensures that the series converges uniformly near $s=0$. Because of that, we can safely take \(s\to0\) term by term:
\[
\lim_{s\to0}\bigl(k_{n+1}^{-2s}-y_n^{-2s}\bigr)
=1-1=0,
\]
for every \(n\). Therefore $R(0)=\sum_{n=0}^\infty 0 = 0$.
This shows that the only contribution to \(\zeta_{\mathcal O}(0)\) comes from the Hurwitz zeta function we added and subtracted earlier, and therefore this completes the proof of \eqref{resu}.

\section{Fourier transform of the amplitude $\braket{L, \phi | L' \phi'}$} \label{fourier}
In this appendix we are interested in computing the Fourier transform of the amplitude $\braket{L\, \phi| L' \phi'}=I_0 \left(\alpha \sqrt{\phi^2- \phi'^2} \right)$ computed in \cite{Buchmuller:2024ksd} with respect to $\phi'$, from now on denoted by $x$. For notational simplicity let us define $\alpha^2 = L^2 - L'^2$, so that the transform that we want to compute is written as:
\begin{equation}
    \hat{f}(k)=\int_{- \infty}^{+ \infty} e^{- i k x} \braket{L\, \phi| L' x} \mathrm{d}x = \int_{- \infty}^{+ \infty} e^{- i k x} I_0 \left(\alpha \sqrt{\phi^2- x^2} \right) \mathrm{d}x,
\end{equation}
where $I_0 \left(z \right)$ is the modified Bessel function of zeroth order. In order to perform the Fourier transform, it is convenient to resort to the standard integral representation for the Bessel function inside of the integral, which is obtained starting from:
\begin{equation}
I_0(R)=\frac{1}{2\pi}\int_{0}^{2\pi}e^{R\cos \theta}\,\mathrm{d}\theta,
\label{eq:I0-standard}
\end{equation}
which is valid for every $R \in \mathbb{C}$.
If we now consider $A=\phi$ and $B=i x$, then the trigonometric identity
\begin{equation}
A\cos\theta+B\sin\theta=R\cos(\theta-\delta),\qquad
R=\sqrt{A^2+B^2}=\sqrt{\phi^2+(ix)^2}=\sqrt{\phi^2-x^2},
\label{eq:LC-identity}
\end{equation}
holds for a suitable phase shift $\delta$ (its value is irrelevant because we integrate over a full period). 
Using this identity in \eqref{eq:I0-standard} yields
\begin{align}
I_0\!\Big(\alpha\sqrt{\phi^2-x^2}\Big)
&=\frac{1}{2\pi}\int_{0}^{2\pi}e^{\alpha(\phi\cos\theta+i x\sin\theta)}\,\mathrm{d}\theta
\notag\\
&=\frac{1}{2\pi}\int_{0}^{2\pi}e^{\alpha\phi\cos\theta}\,
\Big[\cos(\alpha x\sin\theta)+i\,\sin(\alpha x\sin\theta)\Big]\,\mathrm{d}\theta.
\label{eq:I0-expand}
\end{align}
By symmetry, the integral of the sine term vanishes over $[0,2\pi]$, therefore we can retain the real part alone:
\begin{equation}
I_0\!\Big(\alpha\sqrt{\phi^2-x^2}\Big)
=\frac{1}{2\pi}\int_{0}^{2\pi}e^{\alpha\phi\cos\theta}\,\cos(\alpha x\sin\theta)\,\mathrm{d}\theta.
\label{eq:I0-real}
\end{equation}
Using the periodicity and symmetry of the integrand we can simplify further the result and get to,
\begin{equation}
\frac{1}{2\pi}\int_{0}^{2\pi}(\cdots)\,\mathrm{d}\theta
=\frac{1}{\pi}\int_{0}^{\pi}(\cdots)\,\mathrm{d}\theta
=\frac{2}{\pi}\int_{0}^{\pi/2}\cosh(\alpha\phi\cos\theta)\,\cos(\alpha x\sin\theta)\,\mathrm{d}\theta,
\label{eq:range-reduction}
\end{equation}
where we used $e^{\alpha\phi\cos\theta}=\cosh(\alpha\phi\cos\theta)+\sinh(\alpha\phi\cos\theta)$ and the $\sinh$ contribution cancels by the map $\theta\mapsto \pi-\theta$. Therefore this proves the identity
\begin{equation}
I_0\!\Big(\alpha\sqrt{\phi^2-x^2}\Big)
=\frac{2}{\pi}\int_{0}^{\pi/2}\cosh(\alpha\phi\cos\theta)\,\cos(\alpha x\sin\theta)\,\mathrm{d}\theta\;.
\label{eq:I0-angle}
\end{equation}
Using this result we can now compute the Fourier transform. indeed by interchanging the order of integrations :
\begin{align*}
\widehat f(k)
&=\frac{2}{\pi}\int_{0}^{\pi/2}\cosh(\alpha\phi\cos\theta)\,
\bigg[\int_{-\infty}^{\infty} e^{-ikx}\cos(\alpha x\sin\theta)\,\mathrm{d}x\bigg]\,\mathrm{d}\theta.
\end{align*}
In the sense of distributions,
\[
\int_{-\infty}^{\infty}e^{-ikx}\cos(px)\,\mathrm{d}x
=\pi\big[\delta(k-p)+\delta(k+p)\big].
\]
With $p=\alpha\sin\theta$ we obtain
\[
\widehat f(k)=\frac{2}{\pi}\int_{0}^{\pi/2}\cosh(\alpha\phi\cos\theta)\,
\pi\big[\delta(k-\alpha\sin\theta)+\delta(k+\alpha\sin\theta)\big]\,\mathrm{d}\theta.
\]
The $\delta$-of-a-function rule states that if $g(\theta)$ has simple zeros at $\theta_i$, then
\[
\int \delta(g(\theta))\,h(\theta)\,\mathrm{d}\theta=\sum_i \frac{h(\theta_i)}{|g'(\theta_i)|}.
\]
Here $g(\theta)=k-\alpha\sin\theta$ with $g'(\theta)=-\alpha\cos\theta$. There are real solutions $\theta_i$ if and only if $|k|<\alpha$. When $|k|<\alpha$, the unique solution on $[0,\pi/2]$ is $\theta_1=\arcsin(|k|/\alpha)$. Using $\cos\theta_1=\sqrt{1-(k/\alpha)^2}$, we get
\[
\widehat f(k)=\frac{2\,\cosh\!\big(\phi\sqrt{\alpha^2-k^2}\,\big)}{\sqrt{\alpha^2-k^2}}\;\mathbf{1}_{\{|k|<\alpha\}},
\]
which is the result quoted in \eqref{cosh}.

\section{Geodesic length in finite cutoff geometry}\label{finite_geo}
In this appendix we study spacelike geodesics in the black hole backgrond of AdS$_2$ and we derive inherent formulas that were used in the main text.

We consider the Euclidean AdS$_2$ black hole metric
\begin{equation}
  ds^2 = (r^2 - r_h^2)\,d\tau^2 + \frac{dr^2}{r^2 - r_h^2}\,,
\end{equation}
with $r \geq r_h$. We introduce a radial cutoff at $r = r_c$ and look for spacelike geodesics that connect two boundary points at $r=r_c$ separated by Euclidean time
\begin{equation}
  \tau = -\frac{\Delta\tau}{2} \quad\text{and}\quad \tau = +\frac{\Delta\tau}{2}\,.
\end{equation}
By symmetry, such a geodesic reaches a minimum radius $r=r_\ast$ at $\tau=0$, and the curve is symmetric under $\tau \to -\tau$.

We parametrize the geodesic as $x^\mu(\lambda) = (\tau(\lambda), r(\lambda))$ and we study the variational problem with the usual Lagrangian
\begin{equation}
  \mathcal{L}
  \;=\; \sqrt{g_{\mu\nu} \dot{x}^\mu \dot{x}^\nu}
  \;=\; \sqrt{(r^2 - r_h^2)\,\dot{\tau}^2 + \frac{\dot{r}^2}{r^2 - r_h^2}}\,,
\end{equation}
where dot denotes differentiation with respect to the affine parameter $\lambda$. Since the metric is $\tau$-independent, there is a conserved momentum
\begin{equation}
  p_\tau \equiv \frac{\partial \mathcal{L}}{\partial \dot{\tau}}
  = \frac{(r^2 - r_h^2)\,\dot{\tau}}{\mathcal{L}}\,.
\end{equation}
Choosing $\lambda$ to be the proper length parameter, $\mathcal{L}=1$, the conserved charge simplifies to
\begin{equation}
  E \;\equiv\; p_\tau = (r^2 - r_h^2)\,\dot{\tau}\,.
\end{equation}
The unit-speed condition $\mathcal{L}=1$ allow us to obtain a differential equation for $r(\lambda)$ in terms of the conserved quantity, that reads
\begin{equation}
  \dot{r}^2 = r^2 - r_h^2 - E^2 =  r^2 - r_\ast^2\,.
\end{equation}
where we defined $r=r_\ast$ the radial position of the turning point. The main goal of the remaining part of the Appendix is to relate the length of a geodesic $L$ to the euclidean time separation $\Delta \tau$ of its end points.

Since $\lambda$ is the proper length, the total geodesic length is
\begin{equation}
  L = 2\int_{r_\ast}^{r_c} d\lambda
    = 2\int_{r_\ast}^{r_c} \frac{dr}{\sqrt{r^2 - r_\ast^2}} = 2\,\operatorname{arcosh}\!\left(\frac{r_c}{r_\ast}\right)\,.
\end{equation}
Instead from the conserved charge, we get that the half time-separation between the turning point $r_\ast$ and the boundary $r_c$ is
\begin{equation}
  \frac{\Delta\tau}{2}
   = \sqrt{r_\ast^2 - r_h^2}\int_{r_\ast}^{r_c}
     \frac{dr}{(r^2 - r_h^2)\sqrt{r^2 - r_\ast^2}} = \frac{2}{r_h}\,\arctan\!\left(
    \frac{r_h\sqrt{r_c^2 - r_\ast^2}}{r_c\sqrt{r_\ast^2 - r_h^2}}
  \right).
\end{equation}
The equations that we just derived give a parametric representation of the geodesic length and boundary time separation in terms of the turning point $r_\ast$. For our purposes it is convenient to eliminate $r_\ast$ in favor of the length $L$. After a short algebraic manipulation we obtain the geodesic length as a function of the (finite-cutoff) boundary time separation which is
\begin{equation}
  L(\Delta\tau; r_c, r_h)
  = 2\,\operatorname{arsinh}\!\left[
    \frac{\sqrt{r_c^2 - r_h^2}}{r_h}\,
    \left|\sin\left(\frac{r_h\Delta\tau}{2}\right)\right|
  \right],
  \label{eq:L-of-Dtau-finite-rc}
\end{equation}
which is the relation used in the main text. As a non trivial check we notice that in the limit $r_c \to \infty$, we have
\begin{equation}
  L(\Delta\tau)
  \;\sim\;
  2\log\left[
    \frac{2\,\sqrt{r_c^2 - r_h^2}}{r_h}\,
    \left|\sin\left(\frac{r_h\Delta\tau}{2}\right)\right|
  \right].
\end{equation}
Subtracting the $\Delta\tau$-independent divergence, one obtains the usual renormalized length
\begin{equation}
  L_{\rm ren}(\Delta\tau)
  \propto -2\log\left|\sin\left(\frac{r_h\Delta\tau}{2}\right)\right|,
\end{equation}
which reproduces the expected finite-temperature behavior of two-point functions via the geodesic approximation.

\bibliographystyle{ourbst}
\bibliography{cutoff.bib}
\end{document}